\documentclass[12pt]{article}
\usepackage{graphicx}
\usepackage{amsmath}
\usepackage{amsfonts}
\usepackage{amssymb}

\begin{document}

\title{Calculations for Extended Thermodynamics of dense gases up to whatever order and with only some symmetries}
\author{S. Pennisi \\
Dipartimento di Matematica ed Informatica , Universit\'{a} di Cagliari, Cagliari,
Italy\\
spennisi@unica.it}
\date{}
\maketitle \vspace{0.5 cm}
 \small {\em \noindent } \\

\begin{abstract}
The 14 moments model for dense gases, introduced in the last years by Ruggeri, Sugiyama
and collaborators, is here considered. They have found the closure of the balance
equations up to second order with respect to equilibrium; subsequently, Carrisi has found
the closure up to whatever order with respect to equilibrium, but for a more constrained
system where more symmetry conditions are imposed. Here the closure is obtained up to
whatever order and without imposing the supplementary conditions. It comes out that the
first non symmetric parts appear only at third order with respect to equilibrium, even if
Ruggeri and Sugiyama found a non symmetric part proportional to an arbitrary constant
also at first order with respect to equilibrium. Consequently, this constant must be
zero, as Ruggeri, Sugiyama assumed in the applications and on an intuitive ground.
\end{abstract}

\section{Introduction}

Starting point of this research is the article  \cite{1} which belongs to the framework
of Extended Thermodynamics. Some of the original papers on this subject are \cite{2},
\cite{3} while more recent papers are \cite{4}-\cite{18} and the theory has the advantage
to furnish hyperbolic field equations, with finite
speeds of propagation of shock waves and very interesting analytical  properties. \\
It starts from a given set of balance equations where some arbitrary functions appear;
restrictions on these arbitrariness are obtained by imposing the entropy principle and
the relativity principle. \\
However, these restrictions were so strong to allow only particular state functions; for
example, the function $p=p( \rho , T)$ relating the pression $p$ with the mass density
$\rho$ and the absolute temperature $T$, was determined except for a single variable
function so that it was adapt to describe only particular gases or continuum. \\
This drawback has been overcome in \cite{1} and other articles such as
\cite{19}-\cite{34} by considering two blocks of balance equations, for example, in the
14 moments case treated in \cite{1}, they are
\begin{eqnarray}\label{2.1}
&{}&  \quad \quad \quad \, \, \partial_t F^N + \partial_k F^{kN} = P^N \quad , \quad
\partial_t G^E +
\partial_k G^{kE}
= Q^E \, ,  \\
&{}& \nonumber \\
 &{}&  \mbox{where} \quad F^N = (F, F^i, F^{ij}) \quad \, \quad , \quad G^E = (G,
G^i) \quad ,
\nonumber \\
&{}&  \quad \quad \quad \, \, F^{kN} = (F^k, F^{ki}, F^{kij}) \quad , \quad G^{kE} =
(G^k,
G^{ki}) \quad , \nonumber \\
&{}&  \quad \quad \quad  \, \, P^{N} = (0, 0, P^{ij}) \quad\quad\quad , \quad Q^{E} = (0,
Q^{i}) \quad . \nonumber
\end{eqnarray}
The first 2 components of $P^N$ are zero because the first 2 components of equations
$(\ref{2.1})_1$ are the conservation laws of mass and momentum; the first  component of
$Q^E$ is zero because the first  component of equations $(\ref{2.1})_2$ is the
conservation laws of energy. The whole block $(\ref{2.1})_2$ can be considered an "Energy
Block". \\
The equations $(\ref{2.1})$ can be written in a more compact form as
\begin{eqnarray}\label{2.2}
&{}&  \quad \quad \quad \, \, \partial_t F^A + \partial_k F^{kA} = P^A \quad ,
 \\
&{}& \nonumber \\
 &{}&  \mbox{where} \quad F^A = (F^N, G^E) \quad  , \quad F^{kA} = (F^{kN}, G^{kE}) \quad , \quad
 P^{A} = (P^N, Q^E) \quad . \nonumber
\end{eqnarray}
In the whole set (\ref{2.2}), $F^A$ are the independent variables, while $F^{kij}$,
$G^{ki}$, $P^{ij}$, $Q^{i}$ are constitutive functions. Restrictions on their
generalities are obtained by imposing
\begin{enumerate}
  \item \textbf{The Entropy Principle} which guarantees the existence of an entropy density $h$
  and an entropy flux $h^k$ such that the equation
  \begin{eqnarray}\label{2.3}
&{}&   \partial_t h + \partial_k h^{k} = \sigma \geq 0  \quad ,
\end{eqnarray}
holds whatever solution of the equations (\ref{2.2}). \\
Thanks to Liu' s Theorem \cite{35}, \cite{36}, this is equivalent to assuming the
existence of Lagrange Multipliers $\mu_A$ such that
  \begin{eqnarray}\label{2bis.1}
&{}&   d \, h= \mu_A d \, F^A  \quad , \quad d \, h^k= \mu_A d \, F^{kA} \quad , \quad
\sigma = \mu_A P^A \, .
\end{eqnarray}
An idea conceived by Ruggeri is to define the 4-potentials $h'$, $h'^k$ as
  \begin{eqnarray}\label{2bis.2}
&{}&  h'= \mu_A  F^A - h \quad , \quad h'^k= \mu_A F^{kA} - h^k \quad ,
\end{eqnarray}
so that eqs. $(\ref{2bis.1})_{1,2}$ become
  \begin{eqnarray*}
&{}&  d \, h'= F^A d \, \mu_A   \quad , \quad d \, h'^k= F^{kA}  d \, \mu_A \quad ,
\end{eqnarray*}
which are equivalent to
  \begin{eqnarray}\label{2bis.3}
&{}&   F^A  = \frac{\partial h'}{\partial  \mu_A } \quad , \quad F^{kA} = \frac{\partial
h'^k}{\partial  \mu_A } \quad ,
\end{eqnarray}
if the Lagrange Multipliers are taken as independent variables. A nice consequence of
eqs. $(\ref{2bis.3})$ is that the field equations assume the symmetric form. \\
Other restrictions are given by
\item \textbf{the symmetry conditions}, that is the second component of $F^N$ is equal to the
first component of $F^{kN}$, the third component of $F^N$ is equal to the second
component of $F^{kN}$, the second component of $G^E$ is equal to the first component of
$G^{kE}$. Moreover, $F^{ij}$ is a symmetric tensor. \\
Thanks to eqs. (\ref{2bis.3}) these conditions assume the form
  \begin{eqnarray}\label{3.0}
&{}&  \frac{\partial h'}{\partial  \mu_i } =  \frac{\partial h'^i}{\partial  \mu } \quad
, \quad \frac{\partial h'}{\partial  \mu_{ij} } =  \frac{\partial h'^i}{\partial  \mu_j }
\quad , \quad \frac{\partial h'}{\partial  \lambda_i } =  \frac{\partial h'^i}{\partial
\lambda } \quad ,
\end{eqnarray}
where we have assumed the decomposition $\mu_A= ( \mu , \mu_i , \mu_{ij} , \lambda ,
\lambda_i )$ for the Lagrange Multipliers. Moreover $\mu_{ij}$ is a symmetric tensor. \\
Eventual supplementary symmetry conditions are those imposing the symmetry of the tensors
$F^{kij}$ and $G^{ki}$ and are motivated by the kinetic counterpart of this theory.
Thanks to eqs. (\ref{2bis.3}) these conditions may be expressed as
  \begin{eqnarray}\label{3.00}
&{}&  \frac{\partial h'^{[k}}{\partial  \mu_{i]j} } = 0 \quad , \frac{\partial
h'^{[k}}{\partial  \lambda_{i]} } = 0 \quad .
\end{eqnarray}
However, if the kinetic approach is used only to give suggestions on the form of the
balance equations, one may think to obtain a more general macroscopic theory if these
supplementary symmetry conditions are not imposed. For this reason they have been not
considered in \cite{1} and in the present article. \\
The next conditions come from
  \item \textbf{the Galilean Relativity Principle}.\\
  There are two ways to impose this principle. One of these is to decompose the variables
  $F^{A}$, $F^{kA}$, $P^{A}$, $\mu_{A}$ in their corresponding non convective parts $\hat{F}^{A}$,
  $\hat{F}^{kA}$, $\hat{P}^{A}$, $\hat{\mu}_{A}$ and in velocity dependent parts, where
  the velocity is defined by
    \begin{eqnarray}\label{3.1}
&{}&  v^i = F ^{-1}F^i \quad .
\end{eqnarray}
This decomposition can be written as
\begin{eqnarray}\label{3.2}
&{}& F^{A} = {X^A}_B( \vec{v}) \hat{F}^{B} \quad , \quad F^{kA} - v^k F^A = {X^A}_B(
\vec{v}) \hat{F}^{kB} \quad , \quad P^{A} = {X^A}_B( \vec{v}) \hat{P}^{B} \quad , \\
&{}& h' = \hat{h}' \quad , \quad h'^{k} - v^k h' =  \hat{h}'^{k} \quad , \quad
\hat{\mu}_{A} = \mu_B {X^B}_A( \vec{v})  \quad , \nonumber
\end{eqnarray}
where
\begin{eqnarray}\label{3.3}
&{}& {X^A}_B( \vec{v}) =
\begin{pmatrix}
  1 & 0 & 0 & 0 & 0 \\
  v^i & \delta^{i}_a & 0 & 0 & 0 \\
  v^i v^j & 2v^{(i} \delta^{j)}_a & \delta^i_{(a}\delta^j_{b)} & 0 & 0 \\
  v^2 & 2 v_a & 0 & 1 & 0 \\
  v^2 v^i & v^2 \delta^i_a + 2 v^i v_a & 2 \delta^i_{(a} v_{b)} & v^i & \delta^i_a
\end{pmatrix}
\end{eqnarray}
\end{enumerate}
After that, all the conditions are expressed in terms of the non convective parts of the
variables. \\
This procedure is described in \cite{2}, \cite{36} for the case considering only the
block $(\ref{2.1})_1$ and is followed in \cite{1} for the whole set (\ref{2.1}). \\
Another way to impose this principle leads to easier calculations; it is described in
\cite{37} for the case considering only the the block
$(\ref{2.1})_1$ and here we show how it is adapt also for the whole set (\ref{2.1}). \\
First of all, we need to know the transformation law of the variables between two
reference frames moving one with respect to the other with a translational motion with
constant translational velocity $ \vec{v}_\tau$. To know it, we may rewrite (\ref{3.2})
in both frames, that is
\begin{eqnarray}\label{4.1}
&{}& F^{A}_a = {X^A}_B( \vec{v}_a) \hat{F}^{B} \quad , \quad F^{kA}_a - v^k_a F^A_a =
{X^A}_B(
\vec{v}_a) \hat{F}^{kB} \quad ,  \\
&{}& h'_a = \hat{h}' \quad , \quad h'^{k}_a - v^k_a h'_a =  \hat{h}'^{k} \quad , \quad
\hat{\mu}_{A}^a = \mu_B^a {X^B}_A( \vec{v}_a)  \quad , \nonumber
\end{eqnarray}
\begin{eqnarray*}
&{}& F^{A}_r = {X^A}_B( \vec{v}_r) \hat{F}^{B} \quad , \quad F^{kA}_r - v^k_r F^A_r =
{X^A}_B(
\vec{v}_r) \hat{F}^{kB} \quad ,  \\
&{}& h'_r = \hat{h}' \quad , \quad h'^{k}_r - v^k_r h'_r =  \hat{h}'^{k} \quad , \quad
\hat{\mu}_{A}^r = \mu_B^r {X^B}_A( \vec{v}_r)  \quad ,
\end{eqnarray*}
where the index $a$ denotes quantities in the absolute reference frame and index $r$
denotes quantities in the relative one; $\hat{F}^{B}$,
  $\hat{F}^{kB}$, $\hat{h}'$, $\hat{h}'^{k}$,  $\hat{\mu}_{B}$ haven't the index $a$, nor
  the index $r$ because they are independent from the reference frame. \\
  Now we can use a property of the matrix ${X^A}_B( \vec{v})$ which is a consequence of
  its definition (\ref{3.3}) and reads
\begin{eqnarray}\label{4.2}
&{}& {X^C}_A(- \vec{v}) {X^A}_B( \vec{v}) = \delta^C_B \quad .
\end{eqnarray}
So we may contract $(\ref{4.1})_{6,7}$ with ${X^C}_A(- \vec{v}_r)$ so obtaining
\begin{eqnarray*}
&{}& \hat{F}^{C} = {X^C}_A(- \vec{v}_r) F^A_r  \quad , \quad \hat{F}^{kC} = {X^C}_A(-
\vec{v}_r) (F^{kA}_r - v^k_r F^A_r)
\end{eqnarray*}
which can be substituted in $(\ref{4.1})_{1,2}$. The result is
\begin{eqnarray}\label{4.3}
&{}& F^A_a = {X^A}_B( \vec{v}_a) {X^B}_C(- \vec{v}_r) F^C_r  \quad , \quad F^{kA}_a -
v^k_a F^A_a = {X^A}_B( \vec{v}_a) {X^B}_C(- \vec{v}_r) (F^{kC}_r - v^k_r F^C_r) \, .
\quad \quad \quad
\end{eqnarray}
Now we use another property of the matrix ${X^A}_B( \vec{v})$ which is a consequence of
  its definition (\ref{3.3}) and reads
\begin{eqnarray}\label{5.1}
&{}& {X^A}_B(\vec{u}) {X^B}_C( \vec{w}) = {X^A}_C( \vec{u} + \vec{w}) \quad .
\end{eqnarray}
Moreover, we use the well known property
\begin{eqnarray}\label{5.2}
&{}& \vec{v}_a = \vec{v}_r + \vec{v}_\tau  \quad .
\end{eqnarray}
In this way the equations (\ref{4.3}) become
\begin{eqnarray}\label{5.2}
&{}& F^A_a = {X^A}_C( \vec{v}_\tau)  F^C_r  \quad , \\
&{}& F^{kA}_a - v^k_a F^A_a = {X^A}_C( \vec{v}_\tau)  F^{kC}_r - v^k_r {X^A}_C(
\vec{v}_\tau)  F^C_r \, , \quad \quad \quad \nonumber
\end{eqnarray}
In eq. $(\ref{5.2})_2$ we can substitute ${X^A}_C( \vec{v}_\tau)  F^C_r$ from eq.
$(\ref{5.2})_1$ so that it becomes
\begin{eqnarray}\label{5.3}
&{}& F^{kA}_a - v^k_\tau F^A_a = {X^A}_C( \vec{v}_\tau)  F^{kC}_r \, . \quad \quad \quad
\end{eqnarray}
Finally, we deduce $\hat{h}'$, $\hat{h}'^{k}$ and $\hat{\mu}_A$ from
$(\ref{4.1})_{8,9,10}$ and substitute them in $(\ref{4.1})_{3,4,5}$ so obtaining
\begin{eqnarray}\label{5.4}
&{}& h'_a = h'_r \quad , \quad h'^{k}_a - v^k_\tau h' =  h'^{k}_r \quad , \quad \mu^r_{C}
= \mu_B^a {X^B}_C( \vec{v}_\tau)  \quad ,
\end{eqnarray}
where for the last one we have also used a contraction with ${X^A}_C(- \vec{v}_r)$. \\
Well, eqs. $(\ref{5.2})_1$,  $(\ref{5.3})$ and $(\ref{5.4})$ give the requested
transformation law between the two reference frames and it is very interesting that it
looks like eqs. $(\ref{3.2})$. \\
Now, if the Lagrange Multipliers are taken as independent variables, eqs. $(\ref{5.4})_3$
are only a change of independent variables from $\mu^a_B$ to $\mu^r_C$, while
$(\ref{5.2})_1$, $(\ref{5.3})$, $(\ref{5.4})_{1,2}$ are conditions because they involve
constitutive functions
\begin{eqnarray}\label{6.1}
&{}& F^A_a = F^A(\mu^a_B) \, , \, F^{kA}_a = F^{kA}(\mu^a_B) \, , \, h'_a  = h'(\mu^a_B)
\, , \, h'^k_a  = h'^k(\mu^a_B) \, , \\
&{}& F^A_r = F^A(\mu^r_B) \, , \, F^{kA}_r = F^{kA}(\mu^r_B) \, , \, h'_r  = h'(\mu^r_B)
\, , \, h'^k_r  = h'^k(\mu^a_B) \, , \nonumber
\end{eqnarray}
where the form of the functions $F^{A}$, $F^{kA}$, $ h'$, $ h'^k$ don' t depend on the
reference frame for the Galilean Relativity Principle. If we substitute $\mu^a_B$ from
eq. $(\ref{5.4})_3$ in $(\ref{6.1})_{1-4}$ and then substitute the result in
$(\ref{5.2})_1$, $(\ref{5.3})$, $(\ref{5.4})_{1,2}$, we obtain
\begin{eqnarray}\label{6.2}
&{}& F^A(\mu_C^r {X^C}_B(- \vec{v}_\tau)) = {X^A}_C( \vec{v}_\tau) F^C_r \, , \\
&{}& F^{kA}(\mu_C^r {X^C}_B(- \vec{v}_\tau)) - v^k_\tau {X^A}_C( \vec{v}_\tau) F^C_r =
{X^A}_C( \vec{v}_\tau) F^{kC}_r \, , \nonumber \\
&{}& h'(\mu_C^r {X^C}_B(- \vec{v}_\tau)) = h'_r \, , \nonumber \\
&{}& h'^k(\mu_C^r {X^C}_B(- \vec{v}_\tau)) - h' v^k_\tau = h'^k_r \, . \nonumber
\end{eqnarray}
Well, these expressions calculated in $v^i_\tau =0$ are nothing more than eqs.
$(\ref{6.1})_{5-8}$, as we expected. But, for the Galilean Relativity Principle they must
be coincident for whatever value of $v^i_\tau$; this amounts to say that the derivatives
of (\ref{6.2}) with respect to $v^i_\tau$ must hold. \\
This constraint can be written explicitly more easily if we take into account that
$\mu_C^r {X^C}_B(- \vec{v}_\tau)= \mu^a_B$ which can be written explicitly by use of
(\ref{3.3}) and reads
\begin{eqnarray}\label{7.1}
&{}& \mu^a= \mu^r - \mu_i^r v^i_\tau + \mu_{ij}^r v^i_\tau v^j_\tau + \lambda^r v^2_\tau
- \lambda_i^r  v^i_\tau v^2_\tau \, , \\
&{}& \mu^a_h= \mu^r_h - 2 \mu_{ih}^r v^i_\tau  -2  \lambda^r v_{\tau h} +
\lambda_i^r ( v^2_\tau \delta_{h}^i + 2 v^i_\tau v_{\tau h} ) \, , \nonumber\\
&{}& \mu^a_{hk}= \mu^r_{hk} - 2 \lambda_i^r v_{\tau (h} \delta_{k)}^i  \, , \nonumber\\
&{}& \lambda^a = \lambda^r - \lambda_i^r v^i_{\tau} \, , \nonumber \\
&{}& \lambda^a_h = \lambda^r_h \, , \nonumber
\end{eqnarray}
from which
\begin{eqnarray}\label{7.2}
&{}& \frac{\partial \mu^a}{\partial v^i_{\tau}}= - \mu^a_i \quad , \quad \frac{\partial
\mu^a_h}{\partial v^i_{\tau}}= - 2 \mu_{ih}^a   -2  \lambda^a \delta_{hi}  \quad , \\
&{}& \frac{\partial\mu^a_{hk}}{\partial v^i_{\tau}}= - 2 \lambda_{(h}^a \delta_{k)i}
\quad , \quad \frac{\partial \lambda^a}{\partial v^i_{\tau}} =  - \lambda_i^a \quad ,
\quad \frac{\partial \lambda^a_h}{\partial v^i_{\tau}} = 0 \, . \nonumber
\end{eqnarray}
Consequently, the derivatives of $(\ref{6.2})_{3,4}$ with respect to $ v^i_{\tau}$ become
\begin{eqnarray}\label{7.3}
&{}& \frac{\partial h'}{\partial \mu} \mu_i + \frac{\partial h'}{\partial \mu_h}(2
\mu_{ih} +2  \lambda \delta_{hi})  + 2 \frac{\partial h'}{\partial \mu_{hi}} \lambda_{h}
+ \frac{\partial h'}{\partial \lambda} \lambda_i=0 \, , \\
&{}& \frac{\partial h'^k}{\partial \mu} \mu_i + \frac{\partial h'^k}{\partial \mu_h}(2
\mu_{ih} +2  \lambda \delta_{hi})  + 2 \frac{\partial h'^k}{\partial \mu_{hi}}
\lambda_{h} + \frac{\partial h'^k}{\partial \lambda} \lambda_i + h' \delta^{ki}=0 \, ,
\nonumber
\end{eqnarray}
where we have omitted the index $a$ denoting variables in the absolute reference frame
because they remain unchanged if we change $ v^i_{\tau}$ with $- v^i_{\tau}$, that is, if
we exchange the absolute and the relative reference frames. \\
It is not necessary to impose the derivatives of $(\ref{6.2})_{1,2}$ with respect to $
v^i_{\tau}$  because they are consequences of (\ref{7.3}) and (\ref{2bis.3}). \\
Consequently, the Galilean Relativity Principle amounts simply in the 2 equations
(\ref{7.3}). \\
So we have to find the most general functions satisfying (\ref{3.0}) and (\ref{7.3}).
After that, we have to use eqs. $(\ref{2bis.3})_1$ to obtain the Lagrange Multipliers in
terms of the variables $F^A$. By substituting them in $(\ref{2bis.3})_2$ and in $h'=h'(
\mu_A)$, $h'^k=h'^k( \mu_A)$ we obtain the constitutive functions in terms of the
variables $F^A$. If we want the non convective parts of our expressions, it suffices to
calculate the left hand side of eqs. $(\ref{2bis.3})_1$ in $\vec{v}=\vec{0}$ so that they
become
\begin{eqnarray}\label{8.1}
&{}&   \hat{F}^A  = \frac{\partial h'}{\partial  \mu_A } \quad .
\end{eqnarray}
From this equation we obtain the Lagrange Multipliers in terms of $\hat{F}^A$ (Obviously,
they will be $\hat{\mu}_A$) and after that substitute them in $h'=h'( \mu_A)$,
$h'^k=h'^k( \mu_A)$ (the last of which will in effect be $\hat{h}'^k$) and into
$\hat{F}^{kA}  = \frac{\partial h'^k}{\partial  \mu_A }$, that is eq. $(\ref{2bis.3})_2$
calculated in $\vec{v}=\vec{0}$. \\
It has to be noted that from (\ref{3.1}) it follows $\hat{F}^i=0$, so that one of the
equations (\ref{8.1}) is $ 0 = \frac{\partial h'}{\partial  \mu_i }$; this doesn' t mean
that $h'$ doesn' t depend on $\mu_i$, but this is simply an implicit function defining
jointly with the other equations (\ref{8.1}) the quantities $\hat{\mu}_A$ in
terms of $\hat{F}^A$. \\
By using a procedure similar to that of the paper \cite{37}, we can prove that we obtain
the same results of the firstly described approach. \\
Now, from $(\ref{3.0})_2$ it follows $\frac{\partial h'^{[i}}{\partial  \mu_{j]} }=0$;
this equation, together with $(\ref{3.0})_1$ are equivalent to assuming the existence of
a scalar function $H$ such that
\begin{eqnarray}\label{8.2}
&{}&   h' = \frac{\partial H}{\partial  \mu } \quad , \quad h'^i = \frac{\partial
H}{\partial  \mu_i } \quad .
\end{eqnarray}
In fact, the integrability conditions for (\ref{8.2}) are exactly $(\ref{3.0})_1$ and
$\frac{\partial h'^{[i}}{\partial  \mu_{j]} }=0$. \\
Thanks to (\ref{8.2}), we can rewrite (\ref{3.0}) and (\ref{7.3}) as
\begin{eqnarray}\label{9.1}
&{}&  \frac{\partial^2 H}{\partial  \mu \partial  \mu_{ij}} = \frac{\partial^2
H}{\partial \mu_i \partial  \mu_{j}} \quad , \quad \frac{\partial^2 H}{\partial  \mu
\partial  \lambda_i} = \frac{\partial^2 H}{\partial \lambda \partial  \mu_{i}}\quad .
\end{eqnarray}
\begin{eqnarray}\label{9.2}
&{}& \frac{\partial^2 H}{\partial  \mu^2 } \mu_i + \frac{\partial^2 H}{\partial  \mu
\partial  \mu_{h}}(2 \mu_{ih} +2  \lambda \delta_{hi})  +
2 \frac{\partial^2 H}{\partial  \mu \partial  \mu_{hi}} \lambda_{h}
+ \frac{\partial^2 H}{\partial  \mu \partial  \lambda} \lambda_i=0 \, , \\
&{}& \frac{\partial^2 H}{\partial  \mu \partial \mu_k } \mu_i + \frac{\partial^2
H}{\partial \mu_h
\partial  \mu_{k}}(2 \mu_{ih} +2  \lambda \delta_{hi})  +
2 \frac{\partial^2 H}{\partial  \mu_k \partial  \mu_{hi}} \lambda_{h} + \frac{\partial^2
H}{\partial  \mu_k \partial  \lambda} \lambda_i + \frac{\partial H}{\partial  \mu}
\delta^{ki}=0 \, . \nonumber
\end{eqnarray}
We note now that the derivative of $(\ref{9.2})_1$ with respect to $\mu_k$ is equal to
the derivative of $(\ref{9.2})_2$ with respect to $\mu$; similarly, the derivative of
$(\ref{9.2})_1$ with respect to $\lambda_k$ is equal to the derivative of $(\ref{9.2})_2$
with respect to $\lambda$, as it can be seen by using also eqs. (\ref{9.1}). \\
Consequently, the left hand side of eq. $(\ref{9.2})_1$ is a vectorial function depending
only on two scalars $\mu$, $\lambda$ and on a symmetric tensor $\mu_{ij}$. For the
Representation Theorems \cite{38}-\cite{46}, it can be only zero. \\
In other words, eq. $(\ref{9.2})_1$ is a consequence of $(\ref{9.1})$ and
$(\ref{9.2})_2$, so that it has not to be imposed. By using eqs. (\ref{9.1}) we can
rewrite eq. $(\ref{9.2})_2$ as
\begin{eqnarray}\label{9.3}
&{}& \frac{\partial^2 H}{\partial  \mu \partial \mu_k } \mu_i + 2 \frac{\partial^2
H}{\partial \mu
\partial  \mu_{kj}}\mu_{ji}+ 2 \frac{\partial^2
H}{\partial \mu
\partial  \mu_{ki}}\lambda +
2 \frac{\partial^2 H}{\partial  \mu_k \partial  \mu_{ij}} \lambda_{j} + \frac{\partial^2
H}{\partial  \mu \partial  \lambda_k} \lambda_i + \frac{\partial H}{\partial  \mu}
\delta^{ki}=0 \, .
\end{eqnarray}
In an article which still needs to be written, Carrisi has found the general solution up
to whatever order with respect to equilibrium, of the conditions (\ref{9.1}), (\ref{9.3})
and also of (\ref{3.00}) which now, by use of eqs. (\ref{8.2}) can be written as
\begin{eqnarray}\label{9.4}
&{}& \frac{\partial^2 H}{\partial  \mu_{[k} \partial \mu_{i]j} } = 0 \quad , \quad
\frac{\partial^2 H}{\partial  \mu_{[k} \partial \lambda_{i]} } = 0\, .
\end{eqnarray}
For a more agreement with the article \cite{1} we want now to do the same thing without
imposing (\ref{9.4}). \\
However, although it may seem strange, with less conditions the calculations become
heavier! In fact, if it was possible to use the conditions (\ref{9.4}), we see that the
function $\frac{\partial H}{\partial  \mu_{k}}$ has all the derivatives with respect to
$\mu_i$, $\mu_{ij}$, $\lambda_i$ which are symmetric tensors, so that its Taylor ' s
expansion around equilibrium is
\begin{eqnarray}\label{10.1}
&{}& \frac{\partial H}{\partial  \mu_{k}}= \sum_{p,q,r}^{0 \cdots \infty} \frac{1}{p!}
\frac{1}{q!} \frac{1}{r!} H_{p,q,r}^{ki_1 \cdots i_ph_1k_1 \cdots h_qk_qj_1 \cdots j_r}
\mu_{i_1} \cdots \mu_{i_p}\mu_{h_1k_1} \cdots \mu_{h_qk_q} \lambda_{j_1} \cdots
\lambda_{j_r} \, ,
\end{eqnarray}
\begin{eqnarray*}
&{}& \mbox{where} \quad H_{p,q,r}^{ki_1 \cdots i_ph_1k_1 \cdots h_qk_qj_1 \cdots j_r} =
\left( \frac{\partial^{p+q+r} H}{\partial \mu_{i_1} \cdots
\partial \mu_{i_p} \cdots \partial \mu_{h_1k_1} \cdots \partial \mu_{h_qk_q} \cdots  \partial  \lambda_{j_1} \cdots
\partial  \lambda_{j_r}} \right)_{eq.}
\end{eqnarray*}
is a symmetric tensor depending only on the scalars $\mu$ and $\lambda$, so that it has
the form
\begin{eqnarray*}
&{}& H_{p,q,r}^{ki_1 \cdots i_ph_1k_1 \cdots h_qk_qj_1 \cdots j_r} = \left\{
\begin{array}{ll}
  H_{p,q,r}(\mu ,\lambda) \delta^{(ki_1 \cdots i_ph_1k_1 \cdots h_qk_qj_1 \cdots j_r)} &\mbox{if $p+r+1$ is even}  \\
  0 & \mbox{if $p+r+1$ is odd}
\end{array} \right.
\end{eqnarray*}
where $\delta^{(a_1 \cdots a_{2n})}$ denotes $\delta^{(a_1 a_2} \cdots  \delta^{a_{2n-1}
a_{2n})}$. \\
By integrating (\ref{10.1}) we obtain
\begin{eqnarray}\label{10.2}
H= \sum_{p,q}^{0 \cdots \infty} \sum_{r \in I_{p+1}} \frac{1}{(p+1)!} \frac{1}{q!}
\frac{1}{r!} H_{p,q,r} \delta^{(i_1 \cdots i_{p+1}h_1k_1 \cdots h_qk_qj_1 \cdots j_r)}
\mu_{i_1} \cdots \mu_{i_{p+1}}\mu_{h_1k_1} \cdots \mu_{h_qk_q} \lambda_{j_1} \cdots
\lambda_{j_r} + \\
+ \bar{H}(\mu , \mu_{ab} , \lambda , \lambda_c) \, , \nonumber
\end{eqnarray}
where $I_p$ denotes the set of all non negative integers $r$ such that $r+p$ is even. \\
But also $\frac{\partial H}{\partial  \mu}$ has all the derivatives which are symmetric
tensors; in fact, the derivatives of (\ref{9.4}) with respect to $\mu$ are
\begin{eqnarray}\label{10.3}
&{}& \frac{\partial^3 H}{\partial  \mu \partial  \mu_{[k} \partial \mu_{i]j} } = 0 \quad
, \quad \frac{\partial^3 H}{\partial  \mu \partial  \mu_{[k} \partial \lambda_{i]} } =
0\, .
\end{eqnarray}
The derivatives of (\ref{9.1}) with respect to $\mu_{ab}$ are
\begin{eqnarray*}
&{}&  \frac{\partial^3 H}{\partial  \mu_{ab} \partial  \mu \partial  \mu_{ij}} =
\frac{\partial^3 H}{\partial  \mu_{ab} \partial \mu_i \partial  \mu_{j}} \quad , \quad
\frac{\partial^3 H}{\partial  \mu_{ab} \partial  \mu
\partial  \lambda_i} = \frac{\partial^3 H}{\partial  \mu_{ab} \partial \lambda \partial  \mu_{i}}\quad
,
\end{eqnarray*}
whose skew-symmetric parts with respect to $b$ and $i$ are
\begin{eqnarray}\label{10.4}
&{}&  \frac{\partial^3 H}{\partial  \mu \partial  \mu_{a[b}  \partial  \mu_{i]j}} = 0
 \quad , \quad
\frac{\partial^3 H}{\partial  \mu \partial  \mu_{a[b}
\partial  \lambda_{i]}} = 0 \quad ,
\end{eqnarray}
thanks to eq. (\ref{9.4}). \\
By using (\ref{10.2}) we obtain that also $\frac{\partial \bar{H}}{\partial  \mu}$ has
all the derivatives which are symmetric tensors, so that its expansion around equilibrium
is
\begin{eqnarray}\label{11.1}
\frac{\partial \bar{H}}{\partial  \mu}= \sum_{q}^{0 \cdots \infty} \sum_{r \in I_{0}}
 \frac{1}{q!} \frac{1}{r!} \frac{\partial \bar{H}_{q,r}}{\partial \mu}(\mu , \lambda) \delta^{(h_1k_1
\cdots h_qk_qj_1 \cdots j_r)} \mu_{h_1k_1} \cdots \mu_{h_qk_q} \lambda_{j_1} \cdots
\lambda_{j_r}  \, ,
\end{eqnarray}
where the derivative of $\bar{H}_{q,r}$ has been introduced for later convenience and
without loss of generality. By integrating (\ref{11.1}) we obtain
\begin{eqnarray}\label{11.2}
\bar{H}= \sum_{q}^{0 \cdots \infty} \sum_{r \in I_{0}}
 \frac{1}{q!} \frac{1}{r!} \bar{H}_{q,r}(\mu , \lambda) \delta^{(h_1k_1
\cdots h_qk_qj_1 \cdots j_r)} \mu_{h_1k_1} \cdots \mu_{h_qk_q} \lambda_{j_1} \cdots
\lambda_{j_r} + \bar{\bar{H}}(\mu_{ab} , \lambda , \lambda_c) \, .
\end{eqnarray}
But the function $H$ is present in (\ref{9.1}), (\ref{9.2}), (\ref{9.3}) only through its
derivatives with respect to $\mu$ and $\mu_k$ so that the function $ \bar{\bar{H}}$
doesn't effect the results and, consequently, it is not restrictive to assume that its
derivatives are symmetric tensors. By substituting (\ref{11.2}) in (\ref{10.2}) we see
that the function $H$ has derivatives which are symmetric tensors and we have also its
expansion. \\
Now, in the present article, we cannot use this property because we don' t have the
constraints (\ref{9.4}). To overcome this difficulty we proceed as follows. Firstly,
\begin{enumerate}
  \item We show a particular solution of (\ref{9.1}) and (\ref{9.3}). \\
Let $\psi_n(\mu , \lambda)$ be a family of functions constrained by
\begin{eqnarray}\label{11.5}
\frac{\partial}{\partial  \mu } \psi_{n+1} = \psi_n \quad \mbox{for} \quad n \geq 0 \, .
\end{eqnarray}
Let us define the function
\begin{eqnarray}\label{11.6}
H_1= \sum_{p,q}^{0 \cdots \infty} \sum_{r \in I_{p}} \frac{1}{p!} \frac{1}{q!}
\frac{1}{r!} \frac{(p+2q+r+1)!!}{p+2q+r+1} \frac{\partial^{r+p}}{\partial \lambda^r
\partial \mu^p} \left[ \left( \frac{-1}{2 \lambda} \right)^{q+\frac{p+r}{2}}
\psi_{\frac{p+r}{2}}\right]
\cdot \\
\cdot \delta^{(i_1 \cdots i_{p}h_1k_1 \cdots h_qk_qj_1 \cdots j_r)} \mu_{i_1} \cdots
\mu_{i_{p}}\mu_{h_1k_1} \cdots \mu_{h_qk_q} \lambda_{j_1} \cdots \lambda_{j_r}  \, .
\nonumber
\end{eqnarray}
In Appendix 1 we will show that eqs. (\ref{9.1}) and (\ref{9.3})  are satisfied if $H$ is
replaced by $H_1$; in other words, $H=H_1$ is a particular solution of our conditions. \\
Moreover, $(H_1)_{eq.}= \psi_0(\mu , \lambda)$ which is an arbitrary two-variables
function, such as $H_{eq.}$. So we can identify
\begin{eqnarray}\label{11.7}
\psi_0(\mu , \lambda)=H_{eq.}
\end{eqnarray}
and define
\begin{eqnarray}\label{11.8}
\Delta H =H- H_1\, .
\end{eqnarray}
In this way the conditions (\ref{9.1}) and (\ref{9.3}) become
\begin{eqnarray}\label{11.9}
&{}&  \frac{\partial^2 \Delta H}{\partial  \mu \partial  \mu_{ij}} = \frac{\partial^2
\Delta H}{\partial \mu_i \partial  \mu_{j}} \quad , \quad \frac{\partial^2 \Delta
H}{\partial \mu
\partial  \lambda_i} = \frac{\partial^2 \Delta H}{\partial \lambda \partial  \mu_{i}} \quad
, \\
&{}& \frac{\partial^2 \Delta H}{\partial  \mu \partial \mu_k } \mu_i + 2 \frac{\partial^2
\Delta H}{\partial \mu
\partial  \mu_{kj}}\mu_{ji}+ 2 \frac{\partial^2
\Delta H}{\partial \mu
\partial  \mu_{ki}}\lambda +
2 \frac{\partial^2 \Delta H}{\partial  \mu_k \partial  \mu_{ij}} \lambda_{j} +
\frac{\partial^2 \Delta H}{\partial  \mu \partial  \lambda_k} \lambda_i + \frac{\partial
\Delta H}{\partial  \mu} \delta^{ki}=0 \nonumber
\end{eqnarray}
\begin{eqnarray}\label{11.10}
\mbox{and we have also} \quad (\Delta H)_{eq.} =0 \, .
\end{eqnarray}
An interesting consequence of (\ref{11.9}) and (\ref{11.10}) is the following \\
PROPERTY 1: " The expansion of $\Delta H$ up to order $n \geq 1$ with respect to
equilibrium is a polynomial of degree $n-1$ in the variable $\mu$." \\
Let us prove this with the iterative procedure and let $\Delta H^n$ denote the
homogeneous part of $\Delta H$ of order $n$ with respect to equilibrium. We have, \\
\begin{itemize}
  \item Case $n=1$: The equation $(\ref{11.9})_3$ at equilibrium, thanks to (\ref{11.10})
  becomes $2 \frac{\partial^2 \Delta H^1}{\partial \mu \partial  \mu_{ki}} \lambda =0$ from which we have that
  $\frac{\partial \Delta H^1}{\partial \mu }$ can depend only on $\mu$, $\mu_i$,
  $\lambda$, $\lambda_c$; but the representation theorems show that no scalar function of
  order $1$ with respect to equilibrium can depend only on these variables. It follows
  that $\frac{\partial \Delta H^1}{\partial \mu }=0$ so that $\Delta H^1$ is of degree
  zero with respect to $\mu$ and the property is verified for this case.
 \item Case $n \geq 2$: Let us suppose, for the iterative hypothesis that $\Delta H$ up to order $n \geq 1$ with respect to
equilibrium is a polynomial of degree $n-1$ in the variable $\mu$; we proceed now to
prove that this property holds also with $n+1$ instead of $n$.
\end{itemize}
In fact, eq. $(\ref{11.9})_1$ up to order $n-1$ gives \\
$\frac{\partial^2 \Delta H^n}{\partial  \mu \partial  \mu_{ij}} = \frac{\partial^2 \Delta
H^{n+1}}{\partial \mu_i \partial  \mu_{j}}$ from which we have
\begin{eqnarray}\label{p1.1}
\Delta H^{n+1} = P_{n-2} + \Delta H^{n+1}_i( \mu, \mu_{ab}, \lambda, \lambda_c) \mu^i +
\Delta H^{n+1}_0( \mu, \mu_{ab}, \lambda, \lambda_c)
\end{eqnarray}
where $P_{n-2}$ is a polynomial of degree $n-2$ in  $\mu$ and which is at least quadratic
in $\mu_j$. \\
After that, eq. $(\ref{11.9})_2$ up to order $n$ gives \\
$\frac{\partial^2 \Delta H^{n+1}}{\partial \mu
\partial  \lambda_i} = \frac{\partial^2 \Delta H^{n+1}}{\partial \lambda \partial
\mu_{i}}$ \\
which, thanks to (\ref{p1.1}) becomes
\begin{eqnarray}\label{p2.1}
\frac{\partial^2  P_{n-2}}{\partial \mu
\partial  \lambda_i}  + \frac{\partial^2  \Delta H^{n+1}_j}{\partial \mu
\partial  \lambda_i} + \frac{\partial^2  \Delta H^{n+1}_0}{\partial \mu
\partial  \lambda_i} = \frac{\partial^2  P_{n-2}}{\partial \lambda \partial
\mu_{i}} + \frac{\partial  \Delta H^{n+1}_i}{\partial \lambda } \, .
\end{eqnarray}
This relation, calculated in $\mu_j=0$ gives
\begin{eqnarray}\label{p2.2}
\frac{\partial^2  \Delta H^{n+1}_0}{\partial \mu
\partial  \lambda_i} = \frac{\partial  \Delta H^{n+1}_i}{\partial \lambda }
\end{eqnarray}
because $P_{n-2}$ is at least quadratic in $\mu_j$. \\
The derivative of (\ref{p2.1}) with respect to $\mu_j$, calculated then in $\mu_j=0$, is
\\
$\frac{\partial^2  \Delta H^{n+1}_j}{\partial \mu
\partial  \lambda_i} = \left( \frac{\partial^3  P_{n-2}}{\partial \mu_{j} \partial \lambda
\partial \mu_{i}} \right)_{ \mu_{j}=0}$ from which $\frac{\partial  \Delta H^{n+1}_j}{
\partial  \lambda_i} = P^{ij}_{n-1}$ with $P^{ij}_{n-1}$ a polynomial of degree $n-1$ in
$\mu$. By integrating this relation, we obtain \\
$\Delta H^{n+1}_i= P^{i}_{n-1} + f^{i}_{n-1}(\mu, \mu_{ab}, \lambda,)$  \\
where $P^{i}_{n-1}$ is a polynomial of degree $n-1$ in $\mu$. But, for the Representation
Theorems, a vectorial function such as $f^{i}_{n-1}$ is zero because it depends only on
scalars and on a second order tensor. It follows that
\begin{eqnarray}\label{p3.1}
\Delta H^{n+1}_i= P^{i}_{n-1} \, .
\end{eqnarray}
By using this result, eq. (\ref{p2.2}) can be integrated and gives
\begin{eqnarray}\label{p3.2}
\frac{\partial  \Delta H^{n+1}_0}{ \partial  \lambda_i} = P^{i}_{n}
\end{eqnarray}
with $P^{i}_{n}$  a polynomial of degree $n$ in $\mu$. \\
Now we impose eq. $(\ref{11.9})_3$ at order $n$ and see that its first, second, fifth and
sixth terms are of degree $n-2$ in $\mu$ so that we have \\
$2 \frac{\partial^2 \Delta H^{n+1}}{\partial \mu
\partial  \mu_{ki}}\lambda +
2 \frac{\partial^2 \Delta H^{n+1}}{\partial  \mu_k \partial  \mu_{ij}} \lambda_{j} =
Q_{n-2}$  \\
with $Q_{n-2}$  a polynomial of degree $n-2$ in $\mu$. This relation, thanks to
(\ref{p1.1}) becomes \\
$2 \lambda \frac{\partial^2 \Delta H_a^{n+1}}{\partial \mu
\partial  \mu_{ki}} \mu^a + 2 \lambda \frac{\partial^2 \Delta H_0^{n+1}}{\partial \mu
\partial  \mu_{ki}} + 2 \lambda_{j} \frac{\partial^2 P_{n-2}}{\partial  \mu_k \partial
\mu_{ij}}+ 2 \lambda_{j} \frac{\partial }{\partial  \partial \mu_{ij}} \Delta H_k^{n+1} =
Z_{n-2}$ \\
with $Z_{n-2}$  a polynomial of degree $n-2$ in $\mu$. \\
This relation,  calculated in $\mu_j=0$, thanks to (\ref{p3.1}) and to the fact that
$P_{n-2}$ is at least quadratic in $\mu_j$, gives \\
$2 \lambda \frac{\partial^2 \Delta H_0^{n+1}}{\partial \mu
\partial  \mu_{ki}} = \bar{Q}_{n-1}^{ki}$ \\
with $\bar{Q}_{n-1}^{ki}$ a polynomial of degree $n-1$ in $\mu$. It follows that \\
$\frac{\partial \Delta H_0^{n+1}}{
\partial  \mu_{ki}} = \bar{P}_{n}^{ki}$ \\
with $\bar{P}_{n}^{ki}$ a polynomial of degree $n$ in $\mu$. This result, jointly with
(\ref{p3.2}) gives that
\begin{eqnarray}\label{p4.1}
\Delta H_0^{n+1} = \tilde{P}_{n} + f(\mu, \lambda) \, .
\end{eqnarray}
But a function depending only on $\mu$ and $\lambda$ cannot be of order $n+1$ with
respect to equilibrium; it follows that $f(\mu, \lambda)=0$. \\
Consequently, (\ref{p4.1}), (\ref{p3.1}) and (\ref{p1.1}) give that $\Delta H^{n+1}$ is a
polynomial of degree $n$ in $\mu$ and this completes the proof. \\
Thanks to this property, it is not restrictive to assume for $\Delta H$ a polynomial
expansion of infty degree in the variable $\mu$; we simply expect that the equations will
stop by itself the terms with higher degree. \\
So, even if $\mu$ is not zero at equilibrium, for what concerns $\Delta H$, we can do an
expansion also around $\mu=0$; obviously, the situation is different for the particular
solution $H_1$ reported in eq. (\ref{11.6}). The next step  with which we proceed is the
following one
\item We note that $\frac{\partial^2 \Delta H}{\partial  \mu^2 }$ has symmetric tensors
  as derivatives. \\
  In fact, from the derivatives of $(\ref{11.9})_{1,2}$ with respect to $\mu_k$ we can
  take the skew-symmetric part with respect to $i$ and $k$ so obtaining
\begin{eqnarray}\label{11.3}
&{}&  \frac{\partial^3 \Delta H}{\partial  \mu \partial  \mu_{[k} \partial  \mu_{i]j}} =
0 \quad , \quad \frac{\partial^3 \Delta H}{\partial \mu \partial  \mu_{[k}
\partial  \lambda_{i]}} = 0 \, .
\end{eqnarray}
From the second derivatives of $(\ref{11.9})_{1,2}$ with respect to $\mu$ and $\mu_{ab}$,
we can take the skew-symmetric part with respect to $i$ and $b$ so obtaining
\begin{eqnarray}\label{11.4}
&{}&  \frac{\partial^4 \Delta H}{\partial  \mu^2 \partial \mu_{a[b} \partial  \mu_{i]j}}
= \frac{\partial^4 \Delta H}{\partial  \mu \partial \mu_{a[b} \partial \mu_{i]} \partial
\mu_{j}} = 0 \quad , \quad  \frac{\partial^4 \Delta H}{\partial  \mu^2 \partial \mu_{a[b}
\partial \lambda_{i]} } = \frac{\partial^4 \Delta H}{\partial  \mu \partial \lambda
\partial \mu_{a[b} \partial \mu_{i]} } =0 \quad \quad \quad \quad
\end{eqnarray}
where (\ref{11.3}) has been used in the second passage. \\
Well, (\ref{11.4}) and the derivatives of (\ref{11.3}),  with respect to $\mu$, prove our
property.
\item We now prove that $\frac{\partial \Delta H}{\partial  \mu }$ is sum of
  $H^{*0}(\mu_{ab} , \lambda , \lambda_c)$ and of a scalar function whose derivatives are
  all symmetric tensors.
\end{enumerate}
In fact, from (\ref{11.3}) we deduce that $\frac{\partial^2 \Delta H}{\partial \mu_k
\partial  \mu }$ has all symmetric derivatives so that its expansion around equilibrium
is of the type
\begin{eqnarray*}
\frac{\partial^2 \Delta H}{\partial  \mu_k \partial  \mu }= \sum_{p,q}^{0 \cdots \infty}
\sum_{r \in I_{p}} \frac{1}{p!} \frac{1}{q!} \frac{1}{r!} H^*_{p,q,r} \delta^{(ki_1
\cdots i_{p}h_1k_1 \cdots h_qk_qj_1 \cdots j_r)} \mu_{i_1} \cdots \mu_{i_{p}}\mu_{h_1k_1}
\cdots \mu_{h_qk_q} \lambda_{j_1} \cdots \lambda_{j_r}  \, .
\end{eqnarray*}
By integrating this expression with respect to $\mu_k$, we obtain
\begin{eqnarray}\label{12.1}
\frac{\partial \Delta H}{ \partial  \mu }= \sum_{p,q}^{0 \cdots \infty} \sum_{r \in
I_{p}} \frac{1}{(p+1)!} \frac{1}{q!} \frac{1}{r!} H^*_{p,q,r} \delta^{(i_1 \cdots
i_{p+1}h_1k_1 \cdots h_qk_qj_1 \cdots j_r)} \\
\mu_{i_1} \cdots \mu_{i_{p+1}}\mu_{h_1k_1} \cdots \mu_{h_qk_q} \lambda_{j_1} \cdots
\lambda_{j_r}   +H^*(\mu , \mu_{ab}, \lambda , \lambda_c) \, . \nonumber
\end{eqnarray}
The derivative of (\ref{12.1}) with respect to $\mu$ is
\begin{eqnarray}\label{12.2}
\frac{\partial^2 \Delta H}{ \partial  \mu^2 }= \sum_{p,q}^{0 \cdots \infty} \sum_{r \in
I_{p}} \frac{1}{(p+1)!} \frac{1}{q!} \frac{1}{r!} \frac{\partial H^*_{p,q,r}}{ \partial
\mu } \delta^{(i_1 \cdots
i_{p+1}h_1k_1 \cdots h_qk_qj_1 \cdots j_r)} \\
\mu_{i_1} \cdots \mu_{i_{p+1}}\mu_{h_1k_1} \cdots \mu_{h_qk_q} \lambda_{j_1} \cdots
\lambda_{j_r}   + \frac{\partial H^*}{ \partial  \mu } \, . \nonumber
\end{eqnarray}
But also $\frac{\partial^2 \Delta H}{ \partial  \mu^2 }$ has all symmetric derivatives so
that its expansion is
\begin{eqnarray}\label{12.3}
\frac{\partial^2 \Delta H}{ \partial  \mu^2 }= \sum_{p,q}^{0 \cdots \infty} \sum_{r \in
I_{p}} \frac{1}{p!} \frac{1}{q!} \frac{1}{r!} \frac{\partial \tilde{H}_{p,q,r}(\mu ,
\lambda)}{
\partial \mu } \delta^{(i_1 \cdots
i_{p}h_1k_1 \cdots h_qk_qj_1 \cdots j_r)} \\
\mu_{i_1} \cdots \mu_{i_{p}}\mu_{h_1k_1} \cdots \mu_{h_qk_q} \lambda_{j_1} \cdots
\lambda_{j_r}    \, . \nonumber
\end{eqnarray}
where $\tilde{H}_{p,q,r}$ appears trough its derivative with respect to $\mu$ for later
convenience and without loss of generality. \\
By substituting (\ref{12.3}) in (\ref{12.2}) we find an expression from which we deduce
$\frac{\partial H^*}{ \partial  \mu }$; by integrating it we obtain
\begin{eqnarray*}
H^* = \sum_{p,q}^{0 \cdots \infty} \sum_{r \in I_{p}} \frac{1}{p!} \frac{1}{q!}
\frac{1}{r!} \tilde{H}_{p,q,r}(\mu , \lambda) \delta^{(i_1 \cdots i_{p}h_1k_1 \cdots
h_qk_qj_1 \cdots j_r)} \mu_{i_1} \cdots \mu_{i_{p}}\mu_{h_1k_1} \cdots \mu_{h_qk_q}
\lambda_{j_1} \cdots
\lambda_{j_r}   + \\
- \sum_{p,q}^{0 \cdots \infty} \sum_{r \in I_{p}} \frac{1}{(p+1)!} \frac{1}{q!}
\frac{1}{r!}H^*_{p,q,r} \delta^{(i_1 \cdots i_{p+1}h_1k_1 \cdots h_qk_qj_1 \cdots j_r)}
\mu_{i_1} \cdots \mu_{i_{p+1}}\mu_{h_1k_1} \cdots \mu_{h_qk_q} \lambda_{j_1} \cdots
\lambda_{j_r}   + \\
+H^{*0}(\mu_{ab}, \lambda , \lambda_c) \, ,
\end{eqnarray*}
where $H^{*0}$ arises from an integration with respect to $\mu$ so that it doesn' t
depend on $\mu$; moreover, it doesn' t depend on $\mu_i$ because $H^{*}$ doesn' t depend
on $\mu_i$. \\
By substituting this expression in (\ref{12.1}) we find that $\frac{\partial \Delta H}{
\partial  \mu }$ is the sum of $H^{*0}(\mu_{ab}, \lambda , \lambda_c)$ and of a function
whose derivatives are all symmetric tensors; consequently, it can be written in the form
\begin{eqnarray}\label{13.1}
\frac{\partial \Delta H}{
\partial  \mu } = \sum_{p,q}^{0 \cdots \infty} \sum_{r \in I_{p}} \frac{1}{p!} \frac{1}{q!}
\frac{1}{r!} H_{p,q,r}(\mu , \lambda) \delta^{(i_1 \cdots i_{p}h_1k_1 \cdots h_qk_qj_1
\cdots j_r)} \mu_{i_1} \cdots \mu_{i_{p}}\mu_{h_1k_1} \cdots \mu_{h_qk_q} \lambda_{j_1}
\cdots \lambda_{j_r}   + \\
+H^{*0}(\mu_{ab}, \lambda , \lambda_c) \, . \nonumber
\end{eqnarray}
If we take into account the result of Property 1, we see that also $\frac{\partial \Delta
H}{\partial  \mu }$ can be expressed as a polynomial of infty degree in $\mu$, so that
(\ref{13.1}) can be written as
\begin{eqnarray}\label{13.1bis}
\frac{\partial \Delta H}{
\partial  \mu } = \sum_{p,q,s}^{0 \cdots \infty} \sum_{r \in I_{p}} \frac{1}{p!} \frac{1}{q!}
\frac{1}{r!} \frac{1}{s!} \vartheta_{p,q,r,s}(\lambda) \mu^s \delta^{(i_1 \cdots
i_{p}h_1k_1 \cdots h_qk_qj_1 \cdots j_r)} \mu_{i_1} \cdots \mu_{i_{p}}\mu_{h_1k_1} \cdots
\mu_{h_qk_q} \lambda_{j_1}
\cdots \lambda_{j_r}   + \\
+H^{*0}(\mu_{ab}, \lambda , \lambda_c) \, . \nonumber
\end{eqnarray}
Now, if we substitute in (\ref{13.1bis}) $H^{*0}$ with \\
$H^{*0N} - \sum_{q}^{0 \cdots
\infty} \sum_{r \in I_{0}}\frac{1}{q!} \frac{1}{r!} \vartheta_{0,q,r,0}(\lambda)
\delta^{(h_1k_1 \cdots h_qk_qj_1 \cdots j_r)}\mu_{h_1k_1} \cdots \mu_{h_qk_q}
\lambda_{j_1} \cdots \lambda_{j_r}   $, we note that  (\ref{13.1bis}) remains unchanged,
except that now we have $H^{*0N}$ instead of $H^{*0}$ and zero instead of
$\vartheta_{0,q,r,0}$. We conclude that we may still use eq. (\ref{13.1bis}) and assume,
without loss of generality that
\begin{eqnarray}\label{13.1biss}
\vartheta_{0,q,r,0}(\lambda)=0  \, .
\end{eqnarray}
If we calculate (\ref{13.1bis}) at equilibrium, and take into account (\ref{11.10}), we
obtain \\
$0= \sum_{s=0}^{\infty} \frac{1}{s!} \vartheta_{0,0,0,s}(\lambda) \mu^s +H^{*0}(0_{ab},
\lambda , 0_c)$. \\
Consequently we have $\vartheta_{0,0,0,s}(\lambda)=0$ for $s \geq 1$, from which and from
eq. (\ref{13.1biss}) it follows
\begin{eqnarray}\label{13.2}
\vartheta_{0,0,0,s}(\lambda)=0 \quad \mbox{for} \quad s \geq 0 \, .
\end{eqnarray}
Moreover, (\ref{11.10}) will give
\begin{eqnarray}\label{13.3}
H^{*0}(0_{ab}, \lambda , 0_c) = 0 \, .
\end{eqnarray}

\subsection{Further restrictions}
For the sequel, it will be useful to consider some consequences of eq. (\ref{11.9}). They
are
\begin{eqnarray}\label{14.1}
&{}& \frac{\partial^3 \Delta H}{\partial  \mu_j  \partial \lambda_i \partial  \mu} =
\frac{\partial^3 \Delta H}{\partial \lambda \partial \mu_{ij} \partial  \mu} \quad ,
 \\
&{}& \left[\frac{\partial^2 \Delta H}{\partial  \mu \partial \mu_k } \mu_i + 2
\frac{\partial^2 \Delta H}{\partial \mu
\partial  \mu_{kj}}\mu_{ji}+ 2 \frac{\partial^2
\Delta H}{\partial \mu
\partial  \mu_{ki}}\lambda +
 \frac{\partial
\Delta H}{\partial  \mu} \delta^{ki}\right]_{\lambda_{j}=0}=0 \quad , \nonumber \\
&{}& \frac{\partial^3 \Delta H}{\partial  \mu_a \partial \mu_k \partial  \mu} \mu_i + 2
\frac{\partial^2 \Delta H}{\partial \mu
\partial  \mu_{(k}}\delta_{a)i}+ 2 \frac{\partial^3
\Delta H}{\partial \mu_a
\partial  \mu_{kj} \partial \mu} \mu_{ji} + 2 \frac{\partial^3 \Delta H}{\partial \mu_a
\partial  \mu_{ki} \partial \mu} \lambda + \nonumber \\
&{}& \quad \quad \quad \quad \quad \quad + \frac{\partial^3 \Delta H}{\partial \mu_a
\partial  \lambda_{k} \partial \mu} \lambda_i + 2
\frac{\partial^3 \Delta H}{\partial  \mu_{ij} \partial  \mu_{ka} \partial \mu}
\lambda_{j} =0 \, , \nonumber \\
&{}& \frac{\partial^3 \Delta H}{\partial  \mu_{ki} \partial  \mu \partial \lambda } \mu_i
+ 2 \frac{\partial^3 \Delta H}{
\partial  \mu_{kj} \partial \mu \partial \lambda}\mu_{ji}+ 2 \frac{\partial^3
\Delta H}{\partial  \mu_{ki} \partial \mu \partial \lambda}\lambda + 2 \frac{\partial^2
\Delta H}{\partial  \mu_{ki} \partial \mu} + \nonumber \\
&{}&  \quad \quad \quad \quad \quad \quad + \frac{\partial^3 \Delta H}{\partial \lambda_k
\partial  \mu \partial \lambda } \lambda_i + \frac{\partial^2 \Delta H}{\partial \lambda
\partial \mu} \delta^{ki} + 2 \frac{\partial^3 \Delta H}{\partial \mu_{ij}
\partial \lambda_k
\partial  \mu } \lambda_j =0
\nonumber
\end{eqnarray}
The first one of these equations is obtained by taking the derivatives of
$(\ref{11.9})_2$ with respect to $\mu_j$ and by substituting in its right hand side
$\frac{\partial^2 \Delta H}{\partial \mu_i \partial \mu_j}$ from $(\ref{11.9})_1$; the
second one is obtained by simply calculating $(\ref{11.9})_3$ in $\lambda_{j} =0$;
similarly, $(\ref{14.1})_3$ is obtained by taking the derivative of $(\ref{11.9})_3$ with
respect to $\mu_a$ and, subsequently,  by substituting in its fourth term
$\frac{\partial^2 \Delta H}{\partial \mu_k \partial \mu_a}$ from $(\ref{11.9})_1$.
Finally, in the derivative of $(\ref{11.9})_3$ with respect to $\lambda$ we can
substitute $\frac{\partial^2 \Delta H}{
\partial \lambda
\partial \mu_k}$ from
$(\ref{11.9})_2$ in its fourth term; in this way $(\ref{14.1})_4$ is obtained. \\
We see that $(\ref{14.1})$ are conditions on $\frac{\partial \Delta H}{\partial \mu}$ so
that they may be considered a sort of integrability conditions on $\Delta H$, if
$\frac{\partial \Delta H}{\partial \mu}$ would be known. \\
In the next section, restrictions will be found to the scalar functions appearing in
(\ref{13.1bis}), by analyzing  eqs. (\ref{11.9}) and (\ref{14.1}).

\section{The expression for $\frac{\partial \Delta H}{\partial \mu}$.}
If we substitute (\ref{13.1bis}) in the derivative of $(\ref{11.9})_{1,2}$ with respect
to $\mu$, we obtain
\begin{eqnarray}\label{15.2}
\vartheta_{p,q+1,r,s+1} = \vartheta_{p+2,q,r,s} \quad , \quad \vartheta_{p,q,r+1,s+1} =
\frac{\partial}{\partial \lambda} \vartheta_{p+1,q,r,s} \, .
\end{eqnarray}
From $(\ref{15.2})_1$ we now obtain
\begin{eqnarray}\label{beta.2}
\vartheta_{p,q,r,s} = \left\{ \begin{array}{ll}
  \vartheta_{0,q+\frac{p}{2},r,s+\frac{p}{2}} & \mbox{if $p$ is even} \\
  \vartheta_{1,q+\frac{p-1}{2},r,s+\frac{p-1}{2}} & \mbox{if $p$ is odd} \, .
\end{array}
\right.
\end{eqnarray}
After that, we see that $(\ref{15.2})_1$ is satisfied as a consequence of (\ref{beta.2}).
\\
Let us focus now our attention to eq. $(\ref{15.2})_2$; for $p=0,1$ it becomes
\begin{eqnarray}\label{beta.3}
\vartheta_{0,q,r+1,s+1} = \frac{\partial}{\partial \lambda} \vartheta_{1,q,r,s} \quad ,
\quad \vartheta_{1,q,r+1,s+1} = \frac{\partial}{\partial \lambda} \vartheta_{0,q+1,r,s+1}
\, ,
\end{eqnarray}
where, for $(\ref{beta.3})_2$ we have used (\ref{beta.2}) with $p=2$. \\
After that, eq. $(\ref{15.2})_2$ with use of eq. (\ref{beta.2})
\begin{itemize}
  \item in the case with $p$ even, gives $(\ref{beta.3})_1$ with $(q+\frac{p}{2} ,
  s+\frac{p}{2})$ instead of $(q,s)$,
  \item in the case with $p$ odd, gives $(\ref{beta.3})_2$ with $(q+\frac{p-1}{2} ,
  s+\frac{p-1}{2})$ instead of $(q,s)$.
\end{itemize}
But we have now to impose the derivative of $(\ref{11.9})_3$ with respect to $\mu$, that
is
\begin{eqnarray*}
&{}& \frac{\partial^3 \Delta H}{\partial  \mu^2 \partial \mu_k } \mu_i + 2
\frac{\partial^3 \Delta H}{\partial \mu^2
\partial  \mu_{kj}}\mu_{ji}+ 2 \frac{\partial^3
\Delta H}{\partial \mu^2
\partial  \mu_{ki}}\lambda +
2 \frac{\partial^3 \Delta H}{\partial \mu \partial  \mu_k \partial  \mu_{ij}} \lambda_{j}
+ \frac{\partial^3 \Delta H}{\partial  \mu^2 \partial  \lambda_k} \lambda_i +
\frac{\partial^2 \Delta H}{\partial  \mu^2} \delta^{ki}=0 \, .
\end{eqnarray*}
To impose this relation, let us take its derivatives with respect to  $\mu_{i_1}$,
$\cdots$ , $\mu_{i_P}$, $\mu_{h_1k_1}$, $\cdots$ , $\mu_{h_Qk_Q}$, $\lambda_{j_1}$,
$\cdots$ , $\lambda_{j_R}$ and let us calculate the result at equilibrium; in this way we
obtain
\begin{eqnarray*}
0 = P \delta^{i \overline{i_1}} \delta^{(\overline{i_2 \cdots i_{P}}kh_1k_1 \cdots
h_Qk_Qj_1 \cdots j_R)} \vartheta_{P,Q,R,s+1} +2 Q  \delta^{i \overline{h_1}}
\delta^{(\overline{k_1 h_2k_2 \cdots h_Qk_Q}ki_1 \cdots
i_{P}j_1 \cdots j_R)} \vartheta_{P,Q,R,s+1}+ \\
+ 2 \lambda   \delta^{(ki h_1k_1 \cdots h_Qk_Q i_{1} \cdots i_{P}j_1 \cdots j_R)}
\vartheta_{P,Q+1,R,s+1}
+ 2 R   \delta^{(ki h_1k_1 \cdots h_Qk_Q i_{1} \cdots i_{P}j_1 \cdots j_R)} \vartheta_{P+1,Q+1,R-1,s}+ \\
+R \delta^{i \overline{j_1}} \delta^{(\overline{j_2 \cdots j_{R}}kh_1k_1 \cdots
h_Qk_Qi_{1} \cdots i_{P})} \vartheta_{P,Q,R,s+1} +\delta^{ki} \delta^{(i_{1} \cdots i_{P}
h_1k_1 \cdots h_Qk_Q j_1 \cdots j_R)} \vartheta_{P,Q,R,s+1} \, ,
\end{eqnarray*}
where overlined indexes denote symmetrization over those indexes, after that the other
one (round brackets around indexes) has been taken. (Note that, in the fourth term the
index $R-1$ appears; despite this fact, the equations holds also for $R=0$ but in this
case, this fourth term is not present as it is remembered also by the factor $R$).  \\
Now, the first, second, fifth and sixth term can be put together so that the above
expression becomes
\begin{eqnarray*}
0 = (P+2Q+R+1) \vartheta_{P,Q,R,s+1}  \delta^{i \overline{i_1}} \delta^{(\overline{i_2
\cdots i_{P}kh_1k_1 \cdots h_Qk_Qj_1 \cdots j_R})}  + \\
+ ( 2 \lambda \vartheta_{P,Q+1,R,s+1} + 2 R \vartheta_{P+1,Q+1,R-1,s})
 \delta^{(ki h_1k_1 \cdots h_Qk_Q i_{1} \cdots i_{P}j_1 \cdots j_R)} \, ,
\end{eqnarray*}
that is
\begin{eqnarray}\label{17.1}
0 = (P+2Q+R+1) \vartheta_{P,Q,R,s+1} + 2 \lambda \vartheta_{P,Q+1,R,s+1} + 2 R
\vartheta_{P+1,Q+1,R-1,s}  \, .
\end{eqnarray}
This relation, for $P=0,1$ reads
\begin{eqnarray}\label{beta.4}
&{}& 0 = (2Q+R+1) \vartheta_{0,Q,R,s+1} + 2 \lambda \vartheta_{0,Q+1,R,s+1} + 2 R
\vartheta_{1,Q+1,R-1,s}  \, , \\
&{}& 0 = (2Q+R+2) \vartheta_{1,Q,R,s+1} + 2 \lambda \vartheta_{1,Q+1,R,s+1} + 2 R
\vartheta_{0,Q+2,R-1,s+1}  \, , \nonumber
\end{eqnarray}
with the agreement that the last terms are not present in the case $R=0$.  (For eq.
$(\ref{beta.4})_2$ we have used $(\ref{beta.2})$ with $p=2$).  \\
For the other values of $P$, eq. $(\ref{17.1})$ with use of eq. (\ref{beta.2})
\begin{itemize}
  \item in the case with $p$ even, gives $(\ref{beta.4})_1$ with $(q+\frac{p}{2} ,
  s+\frac{p}{2})$ instead of $(q,s)$,
  \item in the case with $p$ odd, gives $(\ref{beta.4})_2$ with $(q+\frac{p-1}{2} ,
  s+\frac{p-1}{2})$ instead of $(q,s)$.
\end{itemize}
Summarizing the results, we have that (\ref{beta.2}) gives $\vartheta_{P,Q,R,s}$ in terms
of $\vartheta_{0,Q,R,s}$ and $\vartheta_{1,Q,R,s}$, while eqs. (\ref{beta.3}) and
(\ref{beta.4}) give restrictions on $\vartheta_{0,Q,R,s}$ and $\vartheta_{1,Q,R,s}$.

\subsection{Consequences of the further restrictions}
\begin{itemize}
  \item We want now to impose eqs. (\ref{14.1}). By substituting eq. (\ref{13.1bis}) in
$(\ref{14.1})_1$, we find
\end{itemize}
\begin{eqnarray}\label{17.2}
\frac{\partial^2 H^{*0}}{
\partial \lambda \partial  \mu_{ij} } = \sum_{p,q,s}^{0 \cdots \infty} \sum_{r \in I_{p}} \frac{1}{p!}
\frac{1}{q!} \frac{1}{r!} \frac{1}{s!}( \vartheta_{p+1,q,r+1,s} -
\frac{\partial}{\partial \lambda }\vartheta_{p,q+1,r,s}) \delta^{(iji_1 \cdots
i_{p}h_1k_1
\cdots h_qk_qj_1 \cdots j_r)} \\
\mu_{i_1} \cdots \mu_{i_{p}}\mu_{h_1k_1} \cdots \mu_{h_qk_q} \lambda_{j_1} \cdots
\lambda_{j_r}    \, . \nonumber
\end{eqnarray}
But, from eq. (\ref{beta.2}) and (\ref{beta.3}) it follows that $\vartheta_{p+1,q,r+1,s}
- \frac{\partial}{\partial \lambda }\vartheta_{p,q+1,r,s}=0$ for $p \geq 1$, so that in
(\ref{17.2}) only the term with $p=0$ survives. Moreover, from eq. $(\ref{beta.3})_2$ it
follows that $\vartheta_{1,q,r+1,s} - \frac{\partial}{\partial \lambda
}\vartheta_{0,q+1,r,s}=0$ for $s \geq 1$, so that in (\ref{17.2}) only the term with
$s=0$ survives. Consequently, (\ref{17.2}) becomes
\begin{eqnarray}\label{17.3}
\frac{\partial^2 H^{*0}}{
\partial \lambda \partial  \mu_{ij} } = \sum_{q}^{0 \cdots \infty} \sum_{r \in I_{0}}
\frac{1}{q!} \frac{1}{r!} \vartheta_{1,q,r+1,0} \delta^{(ijh_1k_1 \cdots h_qk_qj_1 \cdots
j_r)} \mu_{h_1k_1} \cdots \mu_{h_qk_q} \lambda_{j_1} \cdots \lambda_{j_r}    \, .
\end{eqnarray}
\begin{itemize}
  \item By substituting now eq. (\ref{13.1bis}) in $(\ref{14.1})_2$, we find
\end{itemize}
\begin{eqnarray}\label{be.3}
\mu_i \sum_{q,s}^{0 \cdots \infty} \sum_{p \in I_{1}} \frac{1}{p!} \frac{1}{q!}
\frac{1}{s!} \vartheta_{p+1,q,o,s}(\lambda) \mu^s \delta^{(ki_1 \cdots i_{p}h_1k_1 \cdots
h_qk_q)} \mu_{i_1} \cdots \mu_{i_{p}}\mu_{h_1k_1} \cdots \mu_{h_qk_q}  +
\end{eqnarray}
\begin{eqnarray*}
 + 2 \mu_{ji}
\left\{ \sum_{q,s}^{0 \cdots \infty} \sum_{p \in I_{0}} \frac{1}{p!}
\frac{1}{q!}\frac{1}{s!} \vartheta_{p,q+1,0,s}(\lambda) \mu^s \delta^{(kji_1 \cdots
i_{p}h_1k_1 \cdots h_qk_q)} \mu_{i_1} \cdots \mu_{i_{p}}\mu_{h_1k_1} \cdots \mu_{h_qk_q}
+\left[ \frac{\partial H^{*0}}{\partial \mu_{kj}}\right]_{\lambda_{j}=0}\right\} +  \\
+ 2 \lambda \left\{ \sum_{q,s}^{0 \cdots \infty} \sum_{p \in I_{0}} \frac{1}{p!}
\frac{1}{q!} \frac{1}{s!} \vartheta_{p,q+1,0,s}(\lambda) \mu^s \delta^{(kii_1 \cdots
i_{p}h_1k_1 \cdots h_qk_q)} \mu_{i_1} \cdots \mu_{i_{p}}\mu_{h_1k_1} \cdots \mu_{h_qk_q}
+\left[ \frac{\partial H^{*0}}{\partial \mu_{ki}}\right]_{\lambda_{j}=0} \right\} +   \\
+\delta^{ki} \left\{ \sum_{q,s}^{0 \cdots \infty} \sum_{p \in I_{0}} \frac{1}{p!}
\frac{1}{q!} \frac{1}{s!} \vartheta_{p,q,0,s}(\lambda) \mu^s \delta^{(i_1 \cdots
i_{p}h_1k_1 \cdots h_qk_q)} \mu_{i_1} \cdots \mu_{i_{p}}\mu_{h_1k_1} \cdots \mu_{h_qk_q}
+\left[ H^{*0}  \right]_{\lambda_{j}=0}\right\} =0 \quad ;
\end{eqnarray*}
this relation calculated in $\mu_{j}=0$ becomes
\begin{eqnarray*}
2 \mu_{ji} \left\{ \sum_{q,s}^{0 \cdots \infty}   \frac{1}{q!}\frac{1}{s!}
\vartheta_{0,q+1,0,s} \mu^s \delta^{(kjh_1k_1 \cdots h_qk_q)} \mu_{h_1k_1} \cdots
\mu_{h_qk_q}
+\left[ \frac{\partial H^{*0}}{\partial \mu_{kj}}\right]_{\lambda_{j}=0}\right\} +  \\
+ 2 \lambda \left\{ \sum_{q,s}^{0 \cdots \infty}  \frac{1}{q!} \frac{1}{s!}
\vartheta_{0,q+1,0,s} \mu^s \delta^{(kih_1k_1 \cdots h_qk_q)} \mu_{h_1k_1} \cdots
\mu_{h_qk_q}
+\left[ \frac{\partial H^{*0}}{\partial \mu_{ki}}\right]_{\lambda_{j}=0} \right\} +   \\
+\delta^{ki} \left\{ \sum_{q,s}^{0 \cdots \infty} \frac{1}{q!} \frac{1}{s!}
\vartheta_{0,q,0,s} \mu^s \delta^{(h_1k_1 \cdots h_qk_q)} \mu_{h_1k_1} \cdots
\mu_{h_qk_q} +\left[ H^{*0} \right]_{\lambda_{j}=0}\right\} =0 \quad
\end{eqnarray*}
whose derivative with respect to $\mu_{h_1k_1}$, $\cdots$, $\mu_{h_Qk_Q}$, calculated at
equilibrium is
\begin{eqnarray*}
0 =\sum_{s}^{0 \cdots \infty} \frac{\mu^s}{s!} \left[ 2 Q  \delta^{i \overline{h_1}}
\delta^{(\overline{k_1 h_2k_2 \cdots h_Qk_Q}k)} \vartheta_{0,Q,0,s}+ 2 \lambda
\delta^{(ki h_1k_1 \cdots h_Qk_Q )}
\vartheta_{0,Q+1,0,s} + \right.\\
+ \left. \delta^{ki}
\delta^{(h_1k_1 \cdots h_Qk_Q)} \vartheta_{0,Q,0,s} \right] + \\
+ \left\{ \frac{\partial^Q}{\partial \mu_{h_1k_1}  \cdots  \partial \mu_{h_Qk_Q}} \left[
2 \mu_{ji} \frac{\partial H^{*0}}{\partial \mu_{kj}} + 2 \lambda \frac{\partial
H^{*0}}{\partial \mu_{ki}} + H^{*0} \delta^{ki} \right] \right\}_{\lambda_{j}=0, \,
\mu_{ia}=0 } \, .
\end{eqnarray*}
where overlined indexes denote symmetrization over those indexes, after that the other
one (round brackets around indexes) has been taken. Now, the first and third  term can be
written as \\
$\sum_{s}^{0 \cdots \infty} \frac{\mu^s}{s!} (2 Q+1)
\vartheta_{0,Q,0,s}\delta^{i(h_1k_1 \cdots h_Qk_Q k)}$ so that the above expression
becomes
\begin{eqnarray}\label{bf.1}
0 =\sum_{s}^{0 \cdots \infty} \frac{\mu^s}{s!} \left[ (2 Q+1) \vartheta_{0,Q,0,s}+ 2
\lambda \vartheta_{0,Q+1,0,s} \right] \delta^{(ki h_1k_1 \cdots h_Qk_Q )}
 + \\
+ \left\{ \frac{\partial^Q}{\partial \mu_{h_1k_1}  \cdots  \partial \mu_{h_Qk_Q}} \left[
2 \mu_{ji} \frac{\partial H^{*0}}{\partial \mu_{kj}} + 2 \lambda \frac{\partial
H^{*0}}{\partial \mu_{ki}} + H^{*0} \delta^{ki} \right] \right\}_{\lambda_{j}=0, \,
\mu_{ia}=0 } \, . \nonumber
\end{eqnarray}
But the terms with $s \geq 1$ are zero for eq. $(\ref{beta.4})_1$ with $R=0$; also the
terms with $s= 0 $ are zero for eq. $(\ref{13.1biss})$, so that from $(\ref{bf.1})$ there
remains
\begin{eqnarray}\label{bl.1}
0= \left\{ \frac{\partial^Q}{\partial \mu_{h_1k_1}  \cdots  \partial \mu_{h_Qk_Q}} \left[
2 \mu_{ji} \frac{\partial H^{*0}}{\partial \mu_{kj}} + 2 \lambda \frac{\partial
H^{*0}}{\partial \mu_{ki}} + H^{*0} \delta^{ki} \right] \right\}_{\lambda_{j}=0, \,
\mu_{ia}=0 } \, .
\end{eqnarray}
But we have still to impose the derivative of (\ref{be.3}) with respect to $\mu_{i_1}$,
$\cdots$, $\mu_{i_P}$, $\mu_{h_1k_1}$, $\cdots$, $\mu_{h_Qk_Q}$ for $P \geq 1$ and
calculated at equilibrium. We obtain
\begin{eqnarray*}
0 = P \sum_{s=0}^\infty \frac{\mu^s}{s!} \delta^{i \overline{i_1}} \delta^{(\overline{i_2
\cdots i_{P}}h_1k_1 \cdots h_Qk_Qk)} \vartheta_{P,Q,0,s} +2 Q \sum_{s=0}^\infty
\frac{\mu^s}{s!} \delta^{i \overline{h_1}} \delta^{(\overline{k_1 h_2k_2 \cdots
h_Qk_Q}i_1 \cdots
i_{P}k)} \vartheta_{P,Q,0,s}+ \\
+ 2 \lambda   \sum_{s=0}^\infty \frac{\mu^s}{s!} \delta^{(ki h_1k_1 \cdots h_Qk_Q i_{1}
\cdots i_{P})} \vartheta_{P,Q+1,0,s} +\delta^{ki} \sum_{s=0}^\infty \frac{\mu^s}{s!}
\delta^{(i_{1} \cdots i_{P} h_1k_1 \cdots h_Qk_Q )} \vartheta_{P,Q,0,s} \, ,
\end{eqnarray*}
that is,
\begin{eqnarray}\label{bfl.3}
0 = (P+ 2Q+1) \vartheta_{P,Q,0,s} + 2 \lambda \vartheta_{P,Q+1,0,s}  \, , \, \forall P
\geq 1 \, , \, s \geq 0 \, .
\end{eqnarray}
Now $P$ must be even because $R=0$, so that this equation must hold $\forall P \geq 2 \,
, \, s \geq 0$; by using (\ref{beta.2}) it becomes \\
$0 = (P+ 2Q+1) \vartheta_{0,Q+ \frac{P}{2},0,s+ \frac{P}{2}} + 2 \lambda
\vartheta_{0,Q+1+ \frac{P}{2},0,s+ \frac{P}{2}}$ which is already satisfied because it is
nothing more than $(\ref{beta.4})_1$ with $R=0$ and with $(Q+ \frac{P}{2} \, , \, s+
\frac{P}{2})$ instead of $(Q \, , \, s)$.

\begin{itemize}
  \item Let us impose now eq. $(\ref{14.1})_3$. By using eq. (\ref{13.1bis}) it becomes
\end{itemize}
\begin{eqnarray}\label{ca.1}
0=  \mu_i \sum_{p,q,s}^{0 \cdots \infty} \sum_{r \in I_{p}} \frac{1}{p!} \frac{1}{q!}
\frac{1}{r!} \frac{1}{s!} \vartheta_{p+2,q,r,s} \mu^s \delta^{(aki_1 \cdots i_{p}h_1k_1
\cdots h_qk_qj_1 \cdots j_r)} \mu_{i_1} \cdots \mu_{i_{p}}\mu_{h_1k_1} \cdots
\mu_{h_qk_q} \lambda_{j_1}
\cdots \lambda_{j_r}   +  \\
+ 2 \sum_{p,q,s}^{0 \cdots \infty} \sum_{r \in I_{p+1}} \frac{1}{p!} \frac{1}{q!}
\frac{1}{r!} \frac{1}{s!} \vartheta_{p+1,q,r,s} \mu^s \delta^{i(a} \delta^{(k)i_1 \cdots
i_{p}h_1k_1 \cdots h_qk_qj_1 \cdots j_r)} \mu_{i_1} \cdots \mu_{i_{p}}\mu_{h_1k_1} \cdots
\mu_{h_qk_q} \lambda_{j_1} \cdots \lambda_{j_r} +\nonumber \\
+ 2 \mu_{ji} \sum_{p,q,s}^{0 \cdots \infty} \sum_{r \in I_{p+1}} \frac{1}{p!}
\frac{1}{q!} \frac{1}{r!} \frac{1}{s!} \vartheta_{p+1,q+1,r,s} \mu^s \delta^{(akji_1
\cdots i_{p}h_1k_1 \cdots h_qk_qj_1 \cdots j_r)} \mu_{i_1} \cdots \mu_{i_{p}}\mu_{h_1k_1}
\cdots \mu_{h_qk_q} \lambda_{j_1}
\cdots \lambda_{j_r}   + \nonumber \\
+ 2  \lambda \sum_{p,q,s}^{0 \cdots \infty} \sum_{r \in I_{p+1}} \frac{1}{p!}
\frac{1}{q!} \frac{1}{r!} \frac{1}{s!} \vartheta_{p+1,q+1,r,s} \mu^s \delta^{(akii_1
\cdots i_{p}h_1k_1 \cdots h_qk_qj_1 \cdots j_r)} \mu_{i_1} \cdots \mu_{i_{p}}\mu_{h_1k_1}
\cdots \mu_{h_qk_q} \lambda_{j_1}
\cdots \lambda_{j_r}   + \nonumber \\
+  \lambda_i \sum_{p,q,s}^{0 \cdots \infty} \sum_{r \in I_{p}} \frac{1}{p!} \frac{1}{q!}
\frac{1}{r!} \frac{1}{s!} \vartheta_{p+1,q,r+1,s} \mu^s \delta^{(kai_1 \cdots i_{p}h_1k_1
\cdots h_qk_qj_1 \cdots j_r)} \mu_{i_1} \cdots \mu_{i_{p}}\mu_{h_1k_1} \cdots
\mu_{h_qk_q} \lambda_{j_1} \cdots \lambda_{j_r}   + \nonumber \\
+ 2 \lambda_{j} \sum_{p,q,s}^{0 \cdots \infty} \sum_{r \in I_{p}} \frac{1}{p!}
\frac{1}{q!} \frac{1}{r!} \frac{1}{s!} \vartheta_{p,q+2,r,s} \mu^s \delta^{(ijkai_1
\cdots i_{p}h_1k_1 \cdots h_qk_qj_1 \cdots j_r)} \mu_{i_1} \cdots \mu_{i_{p}}\mu_{h_1k_1}
\cdots
\mu_{h_qk_q} \lambda_{j_1} \cdots \lambda_{j_r}   + \nonumber \\
+2 \lambda_{j} \frac{\partial^2 \, H^{*0}(\mu_{ab}, \lambda , \lambda_c)}{\partial
\mu_{ij} \partial \mu_{ka}} \, . \nonumber
\end{eqnarray}
This relation, calculated in $\mu_j=0$ gives
\begin{eqnarray}\label{cb.1}
0= 2 \sum_{q,s}^{0 \cdots \infty} \sum_{r \in I_{1}} \frac{1}{q!} \frac{1}{r!}
\frac{1}{s!} \vartheta_{1,q,r,s} \mu^s \delta^{i(a} \delta^{(k)h_1k_1 \cdots h_qk_qj_1
\cdots j_r)} \mu_{h_1k_1} \cdots
\mu_{h_qk_q} \lambda_{j_1} \cdots \lambda_{j_r} + \\
+ 2 \mu_{ji} \sum_{q,s}^{0 \cdots \infty} \sum_{r \in I_{1}} \frac{1}{q!} \frac{1}{r!}
\frac{1}{s!} \vartheta_{1,q+1,r,s} \mu^s \delta^{(akjh_1k_1 \cdots h_qk_qj_1 \cdots j_r)}
\mu_{h_1k_1} \cdots \mu_{h_qk_q} \lambda_{j_1}
\cdots \lambda_{j_r}   + \nonumber \\
+ 2  \lambda \sum_{q,s}^{0 \cdots \infty} \sum_{r \in I_{1}} \frac{1}{q!} \frac{1}{r!}
\frac{1}{s!} \vartheta_{1,q+1,r,s} \mu^s \delta^{(akih_1k_1 \cdots h_qk_qj_1 \cdots j_r)}
\mu_{h_1k_1} \cdots \mu_{h_qk_q} \lambda_{j_1}
\cdots \lambda_{j_r}   + \nonumber \\
+  \lambda_i \sum_{q,s}^{0 \cdots \infty} \sum_{r \in I_{0}} \frac{1}{q!} \frac{1}{r!}
\frac{1}{s!} \vartheta_{1,q,r+1,s} \mu^s \delta^{(kah_1k_1 \cdots h_qk_qj_1 \cdots j_r)}
\mu_{h_1k_1} \cdots
\mu_{h_qk_q} \lambda_{j_1} \cdots \lambda_{j_r}   + \nonumber \\
+ 2 \lambda_{j} \sum_{q,s}^{0 \cdots \infty} \sum_{r \in I_{0}}\frac{1}{q!} \frac{1}{r!}
\frac{1}{s!} \vartheta_{0,q+2,r,s} \mu^s \delta^{(ijkah_1k_1 \cdots h_qk_qj_1 \cdots
j_r)}\mu_{h_1k_1} \cdots
\mu_{h_qk_q} \lambda_{j_1} \cdots \lambda_{j_r}   + \nonumber \\
+2 \lambda_{j} \frac{\partial^2 \, H^{*0}(\mu_{ab}, \lambda , \lambda_c)}{\partial
\mu_{ij} \partial \mu_{ka}} \, . \nonumber
\end{eqnarray}
This relation, calculated in $\mu=0$ becomes
\begin{eqnarray}\label{cc.1}
0= 2 \sum_{q}^{0 \cdots \infty} \sum_{r \in I_{1}} \frac{1}{q!} \frac{1}{r!}
\vartheta_{1,q,r,0}  \delta^{i(a} \delta^{(k)h_1k_1 \cdots h_qk_qj_1 \cdots j_r)}
\mu_{h_1k_1} \cdots
\mu_{h_qk_q} \lambda_{j_1} \cdots \lambda_{j_r} + \\
+ 2 \mu_{ji} \sum_{q}^{0 \cdots \infty} \sum_{r \in I_{1}} \frac{1}{q!} \frac{1}{r!}
 \vartheta_{1,q+1,r,0}  \delta^{(akjh_1k_1 \cdots h_qk_qj_1 \cdots j_r)}
\mu_{h_1k_1} \cdots \mu_{h_qk_q} \lambda_{j_1}
\cdots \lambda_{j_r}   + \nonumber \\
+ 2  \lambda \sum_{q}^{0 \cdots \infty} \sum_{r \in I_{1}} \frac{1}{q!} \frac{1}{r!}
 \vartheta_{1,q+1,r,0}\delta^{(akih_1k_1 \cdots h_qk_qj_1 \cdots j_r)}
\mu_{h_1k_1} \cdots \mu_{h_qk_q} \lambda_{j_1}
\cdots \lambda_{j_r}   + \nonumber \\
+  \lambda_i \sum_{q}^{0 \cdots \infty} \sum_{r \in I_{0}} \frac{1}{q!} \frac{1}{r!}
 \vartheta_{1,q,r+1,0} \delta^{(kah_1k_1 \cdots h_qk_qj_1 \cdots j_r)}
\mu_{h_1k_1} \cdots
\mu_{h_qk_q} \lambda_{j_1} \cdots \lambda_{j_r}   + \nonumber \\
+ 2 \lambda_{j} \sum_{q}^{0 \cdots \infty} \sum_{r \in I_{0}}\frac{1}{q!} \frac{1}{r!}
\vartheta_{0,q+2,r,0} \delta^{(ijkah_1k_1 \cdots h_qk_qj_1 \cdots j_r)}\mu_{h_1k_1}
\cdots
\mu_{h_qk_q} \lambda_{j_1} \cdots \lambda_{j_r}   + \nonumber \\
+2 \lambda_{j} \frac{\partial^2 \, H^{*0}(\mu_{ab}, \lambda , \lambda_c)}{\partial
\mu_{ij} \partial \mu_{ka}} \, . \nonumber
\end{eqnarray}
Now, there remains to considers the terms with $ s \geq 1$ in (\ref{cb.1}). Taking of
this part the derivative with respect to $\mu_{h_1k_1}$, $\cdots$, $\mu_{h_Qk_Q}$,
$\lambda_{j_1}$, $\cdots$, $\lambda_{j_R}$ and calculating the result at equilibrium, it
becomes
\begin{eqnarray*}
&{}& 0 = 2 \delta^{i(a} \delta^{(k)h_1k_1 \cdots h_Qk_Qj_1 \cdots
j_R)}\vartheta_{1,Q,R,s} +2 Q  \delta^{i \overline{h_1}} \delta^{(\overline{k_1 h_2k_2
\cdots h_Qk_Q}kj_1 \cdots
j_Rak)} \vartheta_{1,Q,R,s}+ \\
&{}& + 2 \lambda   \delta^{(aki h_1k_1 \cdots h_Qk_Q j_1 \cdots j_R)}
\vartheta_{1,Q+1,R,s} +R \delta^{i \overline{j_1}} \delta^{(\overline{j_2 \cdots
j_{R}}h_1k_1 \cdots h_Qk_Qka)}
\vartheta_{1,Q,R,s} + \\
&{}& + 2 R   \delta^{(ika h_1k_1 \cdots h_Qk_Q j_1 \cdots j_R)} \vartheta_{0,Q+2,R-1,s}
 \, ,
\end{eqnarray*}
where overlined indexes denote symmetrization over those indexes, after that the other
one (round brackets around indexes) has been taken. (Note that, in the last term the
index $R-1$ appears; despite this fact, the equations holds also for $R=0$ but in this
case, this last term is not present as it is remembered also by the factor $R$).  \\
Now, the first, second, and fourth term can be put together so that the above expression
becomes
\begin{eqnarray*}\label{cd.1}
0 = (2Q+R+2) \vartheta_{1,Q,R,s}  \delta^{i \overline{a}} \delta^{(\overline{kh_1k_1 \cdots h_Qk_Qj_1
\cdots j_R})}  + \\
+ ( 2 \lambda \vartheta_{1,Q+1,R,s} + 2 R \vartheta_{0,Q+2,R-1,s})
 \delta^{(aki h_1k_1 \cdots h_Qk_Q j_1 \cdots j_R)} \, .
\end{eqnarray*}
Obviously, also for the terms with $s=0$ the corresponding  elements can be put together,
so that the expression (\ref{cc.1}) can be written also as
\begin{eqnarray}\label{cd.2}
0= \sum_{q}^{0 \cdots \infty} \sum_{r \in I_{1}} \frac{1}{q!} \frac{1}{r!} \left[
(2q+r+2) \vartheta_{1,q,r,0} + 2  \lambda
 \vartheta_{1,q+1,r,0} \right] \delta^{(akih_1k_1 \cdots h_qk_qj_1 \cdots j_r)}
\mu_{h_1k_1} \cdots \mu_{h_qk_q} \lambda_{j_1}
\cdots \lambda_{j_r}   +  \\
+2 \lambda_{j} \frac{\partial^2 \, H^{*0}(\mu_{ab}, \lambda , \lambda_c)}{\partial
\mu_{ij} \partial \mu_{ka}} \, , \nonumber
\end{eqnarray}
where (\ref{13.1biss}) has been used. After that, (\ref{cd.1}) becomes
\begin{eqnarray*}
0 = (2Q+R+2) \vartheta_{1,Q,R,s} + 2 \lambda \vartheta_{1,Q+1,R,s} + 2 R
\vartheta_{0,Q+2,R-1,s}  \quad \mbox{for} \quad s \geq 1 \, ,
\end{eqnarray*}
which is nothing more than $(\ref{beta.4})_2$. \\
There remains now to  take  the derivatives of  (\ref{ca.1}) with respect to $\mu_{i_1}$,
$\cdots$, $\mu_{i_P}$ $\mu_{h_1k_1}$, $\cdots$, $\mu_{h_Qk_Q}$, $\lambda_{j_1}$,
$\cdots$, $\lambda_{j_R}$ and calculate the result at equilibrium, but only for $P \geq
1$; we obtain
\begin{eqnarray*}
0 = P \delta^{i \overline{i_1}} \delta^{(\overline{i_2 \cdots i_{P}}h_1k_1 \cdots
h_Qk_Qj_1 \cdots j_Rak)} \vartheta_{P+1,Q,R,s} + 2 \delta^{i (a} \delta^{(k)i_1 \cdots
i_{P}h_1k_1 \cdots h_Qk_Qj_1 \cdots j_R)} \vartheta_{P+1,Q,R,s} + \\
+ 2 Q  \delta^{i \overline{h_1}} \delta^{(\overline{k_1 h_2k_2 \cdots h_Qk_Q}i_1 \cdots
i_{P}j_1 \cdots j_Rak)} \vartheta_{P+1,Q,R,s}+ 2 \lambda   \delta^{(aki i_{1} \cdots
i_{P} h_1k_1 \cdots h_Qk_Q j_1 \cdots j_R)} \vartheta_{P+1,Q+1,R,s} +
\\
+R \delta^{i \overline{j_1}} \delta^{(\overline{j_2 \cdots j_{R}}h_1k_1 \cdots
h_Qk_Qi_{1} \cdots i_{P}ka)} \vartheta_{P+1,Q,R,s} + 2 R   \delta^{(kia h_1k_1 \cdots
h_Qk_Q i_{1} \cdots i_{P}j_1 \cdots j_R)} \vartheta_{P,Q+2,R-1,s} \, ,
\end{eqnarray*}
or,
\begin{eqnarray*}
0 = (P+2Q+R+2) \vartheta_{P+1,Q,R,s}  \delta^{i \overline{a}} \delta^{(\overline{ki_1
\cdots i_{P}h_1k_1 \cdots h_Qk_Qj_1 \cdots j_R})}  + \\
+ ( 2 \lambda \vartheta_{P+1,Q+1,R,s} + 2 R \vartheta_{P,Q+2,R-1,s})
 \delta^{(ki ah_1k_1 \cdots h_Qk_Q i_{1} \cdots i_{P}j_1 \cdots j_R)} \, ,
\end{eqnarray*}
that is
\begin{eqnarray}\label{ce.1}
0 = (P+2Q+R+2) \vartheta_{P+1,Q,R,s} + 2 \lambda \vartheta_{P+1,Q+1,R,s} + 2 R
\vartheta_{P,Q+2,R-1,s} \quad \mbox{for} \quad P \geq 1 \, , \, s \geq 0 \, .
\end{eqnarray}
\begin{itemize}
  \item Now, in the case with $P$ even and $P \geq 2$, this relation by using (\ref{beta.2}) becomes
\begin{eqnarray*}
0 = (P+2Q+R+2) \vartheta_{1,Q+ \frac{P}{2},R,s+ \frac{P}{2}} + 2 \lambda
\vartheta_{1,Q+1+ \frac{P}{2},R,s+ \frac{P}{2}} + 2 R \vartheta_{0,Q+2+
\frac{P}{2},R-1,s+ \frac{P}{2}}  \, .
\end{eqnarray*}
which is satisfied as a consequence of $(\ref{beta.4})_2$ with $(Q+\frac{P}{2} ,
  s+\frac{P-2}{2})$ instead of $(Q,s)$.
\item Instead of this, in the case with $P$ odd and $P \geq 1$, eq. (\ref{ce.1}) by using
(\ref{beta.2}) becomes
\begin{eqnarray*}
0 = (P+2Q+R+2) \vartheta_{0,Q+ \frac{P+1}{2},R,s+ \frac{P+1}{2}} + 2 \lambda
\vartheta_{0,Q+1+ \frac{P+1}{2},R,s+\frac{P+1}{2}} +\\
+ 2 R \vartheta_{1,Q+2+ \frac{P-1}{2},R-1,s+ \frac{P-1}{2}}  \, .
\end{eqnarray*}
which is satisfied as a consequence of $(\ref{beta.4})_1$ with $(Q+\frac{P+1}{2} ,
  s+\frac{P-1}{2})$ instead of $(Q,s)$.
\end{itemize}
We may conclude that the only consequence of eq. $(\ref{14.1})_3$ is given by
$(\ref{cd.2})$.

\begin{itemize}
  \item Let us conclude by imposing  eq. $(\ref{14.1})_4$. By using eq. (\ref{13.1bis}) it becomes
\end{itemize}
\begin{eqnarray}\label{cf.1}
0= \mu_i \sum_{p,q,s}^{0 \cdots \infty} \sum_{r \in I_{p+1}} \frac{1}{p!} \frac{1}{q!}
\frac{1}{r!} \frac{1}{s!} \frac{\partial}{\partial \lambda} \vartheta_{p+1,q,r,s} \mu^s
\delta^{(ki_1 \cdots i_{p}h_1k_1 \cdots h_qk_qj_1 \cdots j_r)} \\
\mu_{i_1} \cdots
\mu_{i_{p}}\mu_{h_1k_1} \cdots \mu_{h_qk_q} \lambda_{j_1}
\cdots \lambda_{j_r}   + \nonumber \\
+ 2  \mu_{ji} \left\{ \sum_{p,q,s}^{0 \cdots \infty} \sum_{r \in I_{p}} \frac{1}{p!}
\frac{1}{q!} \frac{1}{r!} \frac{1}{s!} \frac{\partial}{\partial \lambda}
\vartheta_{p,q+1,r,s} \mu^s \delta^{(jki_1 \cdots i_{p}h_1k_1 \cdots h_qk_qj_1 \cdots
j_r)} \nonumber \right. \\
\mu_{i_1} \cdots \mu_{i_{p}}\mu_{h_1k_1} \cdots \mu_{h_qk_q}
\lambda_{j_1} \cdots \lambda_{j_r}
+ \left. \frac{\partial^2 H^{*0}}{\partial  \mu_{kj}  \partial \lambda}  \right\} + \nonumber \\
+ \sum_{p,q,s}^{0 \cdots \infty} \sum_{r \in I_{p}} \frac{1}{p!} \frac{1}{q!}
\frac{1}{r!} \frac{1}{s!} \left( 2 \lambda \frac{\partial}{\partial \lambda}
\vartheta_{p,q+1,r,s} + 2 \vartheta_{p,q+1,r,s}\right) \mu^s \delta^{(iki_1 \cdots
i_{p}h_1k_1 \cdots h_qk_qj_1 \cdots j_r)} \nonumber \\
\mu_{i_1} \cdots
\mu_{i_{p}}\mu_{h_1k_1} \cdots \mu_{h_qk_q} \lambda_{j_1} \cdots \lambda_{j_r}   +  2
\lambda \frac{\partial^2 H^{*0}}{\partial  \mu_{ki}  \partial \lambda} +
\frac{\partial H^{*0}}{\partial  \mu_{ki}} + \nonumber \\
+ \lambda_i \left\{ \sum_{p,q,s}^{0 \cdots \infty} \sum_{r \in I_{p+1}} \frac{1}{p!}
\frac{1}{q!} \frac{1}{r!} \frac{1}{s!} \frac{\partial}{\partial \lambda}
\vartheta_{p,q,r+1,s} \mu^s \delta^{(ki_1 \cdots i_{p}h_1k_1 \cdots h_qk_qj_1 \cdots
j_r)} \nonumber \right. \\
\left. \mu_{i_1} \cdots \mu_{i_{p}}\mu_{h_1k_1} \cdots \mu_{h_qk_q} \lambda_{j_1} \cdots
\lambda_{j_r}   + \frac{\partial^2
H^{*0}}{\partial  \lambda_{k}  \partial \lambda}\right\}+ \nonumber\\
+ \delta^{ki} \left\{ \sum_{p,q,s}^{0 \cdots \infty} \sum_{r \in I_{p}} \frac{1}{p!}
\frac{1}{q!} \frac{1}{r!} \frac{1}{s!} \frac{\partial}{\partial \lambda}
\vartheta_{p,q,r,s} \mu^s \delta^{(i_1 \cdots i_{p}h_1k_1 \cdots h_qk_qj_1 \cdots j_r)}
\nonumber \right. \\
\left. \mu_{i_1} \cdots \mu_{i_{p}}\mu_{h_1k_1} \cdots \mu_{h_qk_q} \lambda_{j_1} \cdots
\lambda_{j_r}  + \frac{\partial
H^{*0}}{\partial \lambda} \right\}+ \nonumber\\
+ 2  \lambda_j \left\{ \sum_{p,q,s}^{0 \cdots \infty} \sum_{r \in I_{p+1}} \frac{1}{p!}
\frac{1}{q!} \frac{1}{r!} \frac{1}{s!} \vartheta_{p,q+1,r+1,s} \mu^s \delta^{(ijki_1
\cdots
i_{p}h_1k_1 \cdots h_qk_qj_1 \cdots j_r)} \nonumber \right.\\
\left. \mu_{i_1} \cdots \mu_{i_{p}}\mu_{h_1k_1} \cdots \mu_{h_qk_q} \lambda_{j_1} \cdots
\lambda_{j_r}   + \frac{\partial^2 H^{*0}}{\partial \mu_{ij}  \partial \lambda_k}
\right\} \, . \nonumber
\end{eqnarray}
Let us consider the part of this expression which doesn' t involve $H^{*0}$ and let us
take its derivatives with respect to $\mu_{i_1}$, $\cdots$, $\mu_{i_P}$ $\mu_{h_1k_1}$,
$\cdots$, $\mu_{h_Qk_Q}$, $\lambda_{j_1}$, $\cdots$, $\lambda_{j_R}$ calculated at
equilibrium; it is equal to
\begin{eqnarray*}
0 = \sum_{s=0}^{\infty} \frac{\mu^s}{s!} \left[ P \delta^{i \overline{i_1}}
\delta^{(\overline{i_2 \cdots i_{P}}h_1k_1 \cdots h_Qk_Qj_1 \cdots j_Rk)}
\frac{\partial}{\partial \lambda} \vartheta_{P,Q,R,s} + \right.\\
+ 2 Q  \delta^{i \overline{h_1}} \delta^{(\overline{k_1 h_2k_2 \cdots h_Qk_Q}i_1 \cdots
i_{P}j_1 \cdots j_Rk)} \frac{\partial}{\partial \lambda} \vartheta_{P,Q,R,s}+ \\
+ \left( 2 \lambda \frac{\partial}{\partial \lambda} \vartheta_{P,Q+1,R,s}+ 2
\vartheta_{P,Q+1,R,s} \right) \delta^{(ki i_{1} \cdots i_{P} h_1k_1 \cdots h_Qk_Q j_1
\cdots j_R)} +
\\
+ R \delta^{i \overline{j_1}} \delta^{(\overline{j_2 \cdots j_{R}}h_1k_1 \cdots
h_Qk_Qi_{1} \cdots i_{P}k)} \frac{\partial}{\partial \lambda} \vartheta_{P,Q,R,s} +
\delta^{ki} \delta^{(i_{1} \cdots i_{P} h_1k_1 \cdots h_Qk_Q j_1 \cdots j_R)}
\frac{\partial}{\partial \lambda} \vartheta_{P,Q,R,s} + \\
+\left. 2 R \delta^{(kih_1k_1 \cdots h_Qk_Q i_{1} \cdots i_{P}j_1 \cdots j_R)}
\vartheta_{P,Q+1,R,s} \right] \, ,
\end{eqnarray*}
which is equal to
\begin{eqnarray*}
0 = \sum_{s=0}^{\infty} \frac{\mu^s}{s!} \left[ (P+2Q+R+1) \delta^{i \overline{k}}
\delta^{(\overline{i_1 \cdots i_{P}h_1k_1 \cdots h_Qk_Qj_1 \cdots j_R})}
\frac{\partial}{\partial \lambda} \vartheta_{P,Q,R,s} + \right.\\
+ \left. \left( 2 \lambda \frac{\partial}{\partial \lambda} \vartheta_{P,Q+1,R,s}+ 2
\vartheta_{P,Q+1,R,s} + 2 R \vartheta_{P,Q+1,R,s} \right) \delta^{(ki i_{1} \cdots i_{P}
h_1k_1 \cdots h_Qk_Q j_1 \cdots j_R)}  \right] = \\
= \sum_{s=0}^{\infty} \frac{\mu^s}{s!} \left[ (P+2Q+R+1) \frac{\partial}{\partial
\lambda} \vartheta_{P,Q,R,s} + 2 \lambda \frac{\partial}{\partial \lambda}
\vartheta_{P,Q+1,R,s} + 2 (R+1) \vartheta_{P,Q+1,R,s} \right] \\
\delta^{(ki i_{1} \cdots i_{P} h_1k_1 \cdots h_Qk_Q j_1 \cdots j_R)} \, .
\end{eqnarray*}
Now, in this expression, the terms with $s \geq 1$ are zero, thanks to derivative of
$(\ref{17.1})$ with respect to $\lambda$ with an aid from eq. $(\ref{15.2})_2$.
Consequently, there remains
\begin{eqnarray*}
\left[ (P+2Q+R+1) \frac{\partial}{\partial \lambda} \vartheta_{P,Q,R,0} + 2 \lambda
\frac{\partial}{\partial \lambda} \vartheta_{P,Q+1,R,0} + 2 (R+1) \vartheta_{P,Q+1,R,0}
\right] \delta^{(ki i_{1} \cdots i_{P} h_1k_1 \cdots h_Qk_Q j_1 \cdots j_R)} \, .
\end{eqnarray*}
This result allows to rewrite eq. (\ref{cf.1}) as
\begin{eqnarray}\label{da.1}
0= 2  \mu_{ji}\frac{\partial^2 H^{*0}}{\partial  \mu_{kj}  \partial \lambda} + 2 \lambda
\frac{\partial^2 H^{*0}}{\partial  \mu_{ki}  \partial \lambda} + 2 \frac{\partial
H^{*0}}{\partial  \mu_{ki}} + \lambda_i \frac{\partial^2 H^{*0}}{\partial  \lambda_{k}
\partial \lambda}
+ \delta^{ki} \frac{\partial H^{*0}}{ \partial \lambda} + 2  \lambda_j \frac{\partial^2
H^{*0}}{\partial \mu_{ij}  \partial \lambda_k} +  \\
+ \sum_{p,q}^{0 \cdots \infty} \sum_{r \in I_{p}} \frac{1}{p!} \frac{1}{q!} \frac{1}{r!}
\left[ (p+2q+r+1) \frac{\partial}{\partial \lambda} \vartheta_{p,q,r,0} + 2 \lambda
\frac{\partial}{\partial \lambda} \vartheta_{p,q+1,r,0} + \right. \nonumber \\
+ \left. 2 (r+1) \vartheta_{p,q+1,r,0} \right] \delta^{(ki i_{1} \cdots i_{p} h_1k_1
\cdots h_qk_q j_1 \cdots j_r)} \mu_{i_1} \cdots \mu_{i_{p}}\mu_{h_1k_1} \cdots
\mu_{h_qk_q} \lambda_{j_1} \cdots \lambda_{j_r}
 \, . \nonumber
\end{eqnarray}
Now we note that, if $p=0$, thanks to (\ref{13.1biss}), the term of (\ref{da.1}) inside
the square brackets is zero. \\
Moreover, if $p$ is even and $p \geq 2$, the term of (\ref{da.1}) inside the square
brackets can be written with use of eq. (\ref{beta.2}) as
\begin{eqnarray*}
(p+2q+r+1) \frac{\partial}{\partial \lambda} \vartheta_{0,q+\frac{p}{2},r,\frac{p}{2}} +
2 \lambda \frac{\partial}{\partial \lambda} \vartheta_{0,q+1+\frac{p}{2},r,\frac{p}{2}} +
2 (r+1) \vartheta_{0,q+1+\frac{p}{2},r,\frac{p}{2}}
\end{eqnarray*}
which is zero, thanks to derivative of $(\ref{17.1})$ with respect to $\lambda$ with an
aid from eq. $(\ref{15.2})_2$. \\
Similarly, if $p$ is odd and $p \geq 3$, the term of (\ref{da.1}) inside the square
brackets can be written with use of eq. (\ref{beta.2}) as
\begin{eqnarray*}
(p+2q+r+1) \frac{\partial}{\partial \lambda}
\vartheta_{1,q+\frac{p-1}{2},r,\frac{p-1}{2}} + 2 \lambda \frac{\partial}{\partial
\lambda} \vartheta_{1,q+1+\frac{p-1}{2},r,\frac{p-1}{2}} + 2 (r+1)
\vartheta_{1,q+1+\frac{p-1}{2},r,\frac{p-1}{2}}
\end{eqnarray*}
which is zero, thanks to derivative of $(\ref{17.1})$ with respect to $\lambda$ with an
aid from eq. $(\ref{15.2})_2$. \\
Finally, if $p=1$ the term of (\ref{da.1}) inside the square brackets is
\begin{eqnarray*}
(2q+r+2) \frac{\partial}{\partial \lambda} \vartheta_{1,q,r,0} + 2 \lambda
\frac{\partial}{\partial \lambda} \vartheta_{1,q+1,r,0} + 2 (r+1) \vartheta_{1,q+1,r,0} =
\\
= (2q+r+2)  \vartheta_{0,q,r+1,1} + 2 \lambda \vartheta_{0,q+1,r+1,1} + 2 (r+1)
\vartheta_{1,q+1,r,0} \, ,
\end{eqnarray*}
where, in the second passage $(\ref{beta.3})_1$ has been used. The result is zero, thanks
to $(\ref{beta.4})_1$ with $Q=q$, $R=r+1$, $s=0$. \\
Thanks to these results, eq. (\ref{da.1}) becomes
\begin{eqnarray}\label{da.11}
0= 2  \mu_{ji}\frac{\partial^2 H^{*0}}{\partial  \mu_{kj}  \partial \lambda} + 2 \lambda
\frac{\partial^2 H^{*0}}{\partial  \mu_{ki}  \partial \lambda} + 2 \frac{\partial
H^{*0}}{\partial  \mu_{ki}} + \lambda_i \frac{\partial^2 H^{*0}}{\partial  \lambda_{k}
\partial \lambda}
+ \delta^{ki} \frac{\partial H^{*0}}{ \partial \lambda} + 2  \lambda_j \frac{\partial^2
H^{*0}}{\partial \mu_{ij}  \partial \lambda_k}
 \, .
\end{eqnarray}
Summarizing the results of this subsection, we have found the restrictions (\ref{17.3}),
(\ref{bl.1}), (\ref{cd.2}) and (\ref{da.11}) on $H^{*0}$.

\section{The expression for $\Delta H$.}
By integrating (\ref{13.1bis}) with respect to $\mu$, we obtain
\begin{eqnarray}\label{da.2}
\Delta H = \sum_{p,q,s}^{0 \cdots \infty} \sum_{r \in I_{p}} \frac{1}{p!} \frac{1}{q!}
\frac{1}{r!} \frac{1}{(s+1)!} \vartheta_{p,q,r,s}(\lambda) \mu^{s+1} \delta^{(i_1 \cdots
i_{p}h_1k_1 \cdots h_qk_qj_1 \cdots j_r)} \\
\mu_{i_1} \cdots \mu_{i_{p}}\mu_{h_1k_1}
\cdots \mu_{h_qk_q} \lambda_{j_1} \cdots \lambda_{j_r}  +\mu H^{*0}(\mu_{ab}, \lambda ,
\lambda_c) + \tilde{\tilde{H}}(\mu_{a}, \mu_{bc}, \lambda , \lambda_d) \, . \nonumber
\end{eqnarray}
By substituting $\Delta H$ from here into $(\ref{11.9})_1$ we find
\begin{eqnarray*}
\sum_{p,q,s}^{0 \cdots \infty} \sum_{r \in I_{p}} \frac{1}{p!} \frac{1}{q!} \frac{1}{r!}
\frac{1}{s!} \vartheta_{p,q+1,r,s} \mu^{s} \delta^{(iji_1 \cdots
i_{p}h_1k_1 \cdots h_qk_qj_1 \cdots j_r)} \\
\mu_{i_1} \cdots \mu_{i_{p}}\mu_{h_1k_1} \cdots \mu_{h_qk_q} \lambda_{j_1} \cdots
\lambda_{j_r}  +  \frac{\partial H^{*0}}{\partial \mu_{ij}} = \\
= \sum_{p,q,s}^{0 \cdots \infty} \sum_{r \in I_{p}} \frac{1}{p!} \frac{1}{q!}
\frac{1}{r!} \frac{1}{s!} \vartheta_{p+2,q,r,s-1} \mu^{s} \delta^{(iji_1 \cdots
i_{p}h_1k_1 \cdots h_qk_qj_1 \cdots j_r)} \\
\mu_{i_1} \cdots \mu_{i_{p}}\mu_{h_1k_1} \cdots \mu_{h_qk_q} \lambda_{j_1} \cdots
\lambda_{j_r}  +  \frac{\partial^2 \tilde{\tilde{H}}}{\partial \mu_{i} \partial \mu_{j}}
\, ,
\end{eqnarray*}
which can be rewritten by taking into account that, from (\ref{beta.2}) it follows
$\vartheta_{p+2,q,r,s-1}=\vartheta_{p,q+1,r,s}$ for $s \geq 1$; so it becomes
\begin{eqnarray*}
\frac{\partial^2 \tilde{\tilde{H}}}{\partial \mu_{i} \partial \mu_{j}} = \sum_{p,q}^{0
\cdots \infty} \sum_{r \in I_{p}} \frac{1}{p!} \frac{1}{q!} \frac{1}{r!}
\vartheta_{p,q+1,r,0} \delta^{(iji_1 \cdots
i_{p}h_1k_1 \cdots h_qk_qj_1 \cdots j_r)} \\
\mu_{i_1} \cdots \mu_{i_{p}}\mu_{h_1k_1} \cdots \mu_{h_qk_q} \lambda_{j_1} \cdots
\lambda_{j_r}  +  \frac{\partial H^{*0}}{\partial \mu_{ij}} \, ,
\end{eqnarray*}
which can be integrated and gives
\begin{eqnarray}\label{db.2}
 \tilde{\tilde{H}} = \sum_{p,q}^{0
\cdots \infty} \sum_{r \in I_{p}} \frac{1}{(p+2)!} \frac{1}{q!} \frac{1}{r!}
\vartheta_{p,q+1,r,0} \delta^{(i_1 \cdots
i_{p+2}h_1k_1 \cdots h_qk_qj_1 \cdots j_r)} \\
\mu_{i_1} \cdots \mu_{i_{p+2}}\mu_{h_1k_1} \cdots \mu_{h_qk_q} \lambda_{j_1} \cdots
\lambda_{j_r}  + \frac{1}{2} \mu_{i} \mu_{j} \frac{\partial H^{*0}}{\partial \mu_{ij}} +
\tilde{\tilde{H}}^i (\mu_{ab}, \lambda , \lambda_c) \mu_i + \tilde{\tilde{H}}^0
(\mu_{ab}, \lambda , \lambda_c) \, , \nonumber
\end{eqnarray}
where $\tilde{\tilde{H}}^i$ and $\tilde{\tilde{H}}^0$ arise from the integration. \\
By substituting $\Delta H$ from (\ref{da.2}) into $(\ref{11.9})_2$, and by taking into
account (\ref{db.2}), we find
\begin{eqnarray}\label{dc.1}
\sum_{p,q,s}^{0 \cdots \infty} \sum_{r \in I_{p+1}} \frac{1}{p!} \frac{1}{q!}
\frac{1}{r!} \frac{1}{s!} \vartheta_{p,q,r+1,s} \mu^{s} \delta^{(ii_1 \cdots
i_{p}h_1k_1 \cdots h_qk_qj_1 \cdots j_r)} \\
\mu_{i_1} \cdots \mu_{i_{p}}\mu_{h_1k_1} \cdots \mu_{h_qk_q} \lambda_{j_1} \cdots
\lambda_{j_r}  +  \frac{\partial H^{*0}}{\partial \lambda_{i}} = \nonumber \\
= \sum_{p,q,s}^{0 \cdots \infty} \sum_{r \in I_{p+1}} \frac{1}{p!} \frac{1}{q!}
\frac{1}{r!} \frac{1}{(s+1)!} \frac{\partial }{\partial \lambda} \vartheta_{p+1,q,r,s}
\mu^{s+1} \delta^{(ii_1 \cdots
i_{p}h_1k_1 \cdots h_qk_qj_1 \cdots j_r)}  \nonumber  \\
\mu_{i_1} \cdots \mu_{i_{p}}\mu_{h_1k_1} \cdots \mu_{h_qk_q} \lambda_{j_1} \cdots
\lambda_{j_r}  +   \nonumber  \\
+ \sum_{p,q}^{0 \cdots \infty} \sum_{r \in I_{p}} \frac{1}{(p+1)!} \frac{1}{q!}
\frac{1}{r!}  \frac{\partial }{\partial \lambda} \vartheta_{p,q+1,r,0} \delta^{(ii_1
\cdots
i_{p+1}h_1k_1 \cdots h_qk_qj_1 \cdots j_r)}  \nonumber  \\
\mu_{i_1} \cdots \mu_{i_{p+1}}\mu_{h_1k_1} \cdots \mu_{h_qk_q} \lambda_{j_1} \cdots
\lambda_{j_r}  +  \mu_j  \frac{\partial^2 H^{*0}}{\partial \mu_{ij} \partial \lambda}
+\frac{\partial \tilde{\tilde{H}}^i}{\partial \lambda} \, .  \nonumber
\end{eqnarray}
Now, thanks to $(\ref{15.2})_2$, the terms with degree greater than zero in $\mu$ cancel
each other so that (\ref{dc.1}) becomes
\begin{eqnarray}\label{dd.1}
\sum_{p,q}^{0 \cdots \infty} \sum_{r \in I_{p+1}} \frac{1}{p!} \frac{1}{q!} \frac{1}{r!}
 \vartheta_{p,q,r+1,0}  \delta^{(ii_1 \cdots
i_{p}h_1k_1 \cdots h_qk_qj_1 \cdots j_r)} \\
\mu_{i_1} \cdots \mu_{i_{p}}\mu_{h_1k_1} \cdots \mu_{h_qk_q} \lambda_{j_1} \cdots
\lambda_{j_r}  +  \frac{\partial H^{*0}}{\partial \lambda_{i}} = \nonumber \\
=  \sum_{p,q}^{0 \cdots \infty} \sum_{r \in I_{p}} \frac{1}{(p+1)!} \frac{1}{q!}
\frac{1}{r!}  \frac{\partial }{\partial \lambda} \vartheta_{p,q+1,r,0} \delta^{(ii_1
\cdots
i_{p+1}h_1k_1 \cdots h_qk_qj_1 \cdots j_r)}  \nonumber  \\
\mu_{i_1} \cdots \mu_{i_{p+1}}\mu_{h_1k_1} \cdots \mu_{h_qk_q} \lambda_{j_1} \cdots
\lambda_{j_r}  +  \mu_j  \frac{\partial^2 H^{*0}}{\partial \mu_{ij} \partial \lambda}
+\frac{\partial \tilde{\tilde{H}}^i}{\partial \lambda} \, .  \nonumber
\end{eqnarray}
This relation, calculated in $\mu_j=0$, thanks to (\ref{13.1biss}), becomes
\begin{eqnarray}\label{dd.2}
\frac{\partial H^{*0}}{\partial \lambda_{i}} = \frac{\partial
\tilde{\tilde{H}}^i}{\partial \lambda} \, .
\end{eqnarray}
The terms in (\ref{dd.1}) which are linear in $\mu_j$ cancel each other thanks to
(\ref{17.3}) and (\ref{13.1biss}). Moreover, from eq. (\ref{beta.2}) and (\ref{beta.3})
it follows that $\vartheta_{p+2,q,r+1,s} - \frac{\partial}{\partial \lambda
}\vartheta_{p+1,q+1,r,s}=0$ for $p \geq 0$, so that in (\ref{dd.1}) all the term of
degree greater than 1 in $\mu_j$ cancel each other. We may conclude that $(\ref{11.9})_2$
has only (\ref{dd.2}) as a consequence. \\
Let us substitute now $\Delta H$ from (\ref{da.2}) into $(\ref{11.9})_3$ and  take into
account (\ref{db.2}); we obtain a relation whose derivative with respect to $\mu_a$ has
already been imposed because this was the way in which we obtained $(\ref{14.1})_3$.
Consequently, there remains now to impose only its value in $\mu_a=0$, which is
\begin{eqnarray}\label{de.1}
2 \mu_{ji} \left\{  \sum_{q,s}^{0 \cdots \infty} \sum_{r \in I_{0}} \frac{1}{q!}
\frac{1}{r!} \frac{\mu^{s}}{s!} \vartheta_{0,q+1,r,s}  \mu^{s} \delta^{(kjh_1k_1 \cdots
h_qk_qj_1 \cdots j_r)} \mu_{h_1k_1} \cdots \mu_{h_qk_q} \lambda_{j_1} \cdots
\lambda_{j_r}  + \frac{\partial H^{*0} }{\partial \mu_{kj}} \right\} +
\end{eqnarray}
\begin{eqnarray*}
+ 2\lambda  \left\{   \sum_{q,s}^{0 \cdots \infty} \sum_{r \in I_{0}} \frac{1}{q!}
\frac{1}{r!} \frac{\mu^{s}}{s!} \vartheta_{0,q+1,r,s}  \delta^{(kih_1k_1 \cdots h_qk_qj_1
\cdots j_r)}\mu_{h_1k_1} \cdots \mu_{h_qk_q} \lambda_{j_1} \cdots \lambda_{j_r}
+ \frac{\partial H^{*0} }{\partial \mu_{ki}} \right\} +  \\
+ \lambda_i  \left\{   \sum_{q,s}^{0 \cdots \infty} \sum_{r \in I_{1}} \frac{1}{q!}
\frac{1}{r!} \frac{\mu^{s}}{s!} \vartheta_{0,q,r+1,s}  \delta^{(kh_1k_1 \cdots h_qk_qj_1
\cdots j_r)} \mu_{h_1k_1} \cdots \mu_{h_qk_q} \lambda_{j_1} \cdots \lambda_{j_r}  +
\frac{\partial H^{*0} }{\partial \lambda_{k}} \right\} + \\
+ \delta^{ki} \left\{   \sum_{q,s}^{0 \cdots \infty} \sum_{r \in I_{0}} \frac{1}{q!}
\frac{1}{r!} \frac{\mu^{s}}{s!} \vartheta_{0,q,r,s}  \delta^{(h_1k_1 \cdots h_qk_qj_1
\cdots j_r)}
\mu_{h_1k_1} \cdots \mu_{h_qk_q} \lambda_{j_1} \cdots \lambda_{j_r}  + H^{*0} \right\} +  \\
+ 2 \lambda_{j}   \left\{     \sum_{q,s}^{0 \cdots \infty} \sum_{r \in I_{1}}
\frac{1}{q!} \frac{1}{r!} \frac{\mu^{s+1}}{(s+1)!} \vartheta_{1,q+1,r,s}
\delta^{(ijkh_1k_1 \cdots h_qk_qj_1 \cdots j_r)} \mu_{h_1k_1} \cdots \mu_{h_qk_q}
\lambda_{j_1} \cdots \lambda_{j_r}  + \frac{\partial \tilde{\tilde{H}}^k }{\partial
\mu_{ij}} \right\} =0 \, .
\end{eqnarray*}
Let us consider the part of the right hand side of this expression which doesn' t involve
$H^{*0}$, nor $\tilde{\tilde{H}}^k$ and let us take its derivatives with respect to
$\mu_{h_1k_1}$, $\cdots$, $\mu_{h_Qk_Q}$, $\lambda_{j_1}$, $\cdots$, $\lambda_{j_R}$
calculated at equilibrium; it is equal to
\begin{eqnarray*}
\sum_{s=0}^{\infty} \frac{\mu^s}{s!} \left[ 2 Q  \delta^{i \overline{h_1}}
\delta^{(\overline{k_1 h_2k_2 \cdots h_Qk_Q}j_1 \cdots j_Rk)} \vartheta_{0,Q,R,s} \right.
+ 2 \lambda \vartheta_{0,Q+1,R,s} \delta^{(ki h_1k_1 \cdots h_Qk_Q j_1 \cdots j_R)} +
\\
\left. + R \delta^{i \overline{j_1}} \delta^{(\overline{j_2 \cdots j_{R}}h_1k_1 \cdots
h_Qk_Qk)} \vartheta_{0,Q,R,s} + \delta^{ki} \delta^{(h_1k_1 \cdots h_Qk_Q j_1 \cdots
j_R)}
\vartheta_{0,Q,R,s}\right] + \\
+ \sum_{s=0}^{\infty} \frac{\mu^{s+1}}{(s+1)!}   2 R \delta^{(kih_1k_1 \cdots h_Qk_Q j_1
\cdots j_R)} \vartheta_{1,Q+1,R-1,s}  \, .
\end{eqnarray*}
Now the first, third and fourth term of this expression can be put together so that it
becomes
\begin{eqnarray*}
\sum_{s=0}^{\infty} \frac{\mu^s}{s!} \left[  (2 Q+R+1)   \delta^{i \overline{h_1}}
\delta^{(\overline{k_1 h_2k_2 \cdots h_Qk_Qj_1 \cdots j_Rk})} \vartheta_{0,Q,R,s} + 2
\lambda \vartheta_{0,Q+1,R,s} \delta^{(ki h_1k_1 \cdots h_Qk_Q j_1 \cdots j_R)}\right]  +
\\ + \sum_{s=0}^{\infty} \frac{\mu^{s+1}}{(s+1)!}   2 R \delta^{(kih_1k_1 \cdots h_Qk_Q
j_1 \cdots j_R)} \vartheta_{1,Q+1,R-1,s}  \, .
\end{eqnarray*}
But the polynomial terms in $\mu$ with degree greater than zero elide one another, thanks
to eq. $(\ref{beta.4})_1$; the remaining part is zero, thanks to eq. $(\ref{13.1biss})$.
Consequently, from $(\ref{de.1})$ there remains
\begin{eqnarray}\label{dl.1}
2 \mu_{ji}  \frac{\partial H^{*0} }{\partial \mu_{kj}} + 2\lambda \frac{\partial H^{*0}
}{\partial \mu_{ki}} + \lambda_i \frac{\partial H^{*0} }{\partial \lambda_{k}}  +
\delta^{ki} H^{*0} + 2 \lambda_{j}   \frac{\partial \tilde{\tilde{H}}^k }{\partial
\mu_{ij}} =0 \, .
\end{eqnarray}
So the situation is now that eqs. $(\ref{11.9})$ are equivalent to (\ref{db.2}) (which
gives $ \tilde{\tilde{H}}$ in terms of $\tilde{\tilde{H}}^i (\mu_{ab}, \lambda ,
\lambda_c)$ and of $\tilde{\tilde{H}}^0 (\mu_{ab}, \lambda , \lambda_c)$) and to the
conditions (\ref{dd.2}) and (\ref{dl.1}) on $\tilde{\tilde{H}}^k$, while
$\tilde{\tilde{H}}^0$ remains arbitrary, as it was obvious because in (\ref{11.9}) it
appears only trough its derivatives with respect to $\mu$ and $\mu_k$ which are zero. We
note also that eq. (\ref{bl.1}) is a particular case of (\ref{dl.1}) , when this last one
is calculated in $\lambda_j=0$. \\
Is there some integrability conditions on (\ref{dd.2}) and (\ref{dl.1}) to determine
$\tilde{\tilde{H}}^k$? Well, in the derivative of (\ref{dl.1}) with respect to $\lambda$
we can substitute $\frac{\partial \tilde{\tilde{H}}^ik}{\partial \lambda}$ from
(\ref{dd.2}); but we obtain an integrability condition which have been already taken into
account and that is (\ref{da.11}). \\
Another type of integrability condition can be obtained in the following way: Let us take
the derivative of (\ref{dl.1}) with respect to $\mu_{ab}$, let us contract the result
with $\lambda_b$ and let us take from the resulting equation the skew-symmetric part wit
respect to $i$ and $a$. In this way we obtain
\begin{eqnarray}\label{ea.1}
0=2 \mu_{j[i}  \frac{\partial^2 H^{*0} }{\partial \mu_{a]b} \partial \mu_{kj}} \lambda_b
+ 2\lambda \frac{\partial^2 H^{*0} }{\partial \mu_{k[i} \partial \mu_{a]b}} \lambda_b +
\lambda_{[i} \frac{\partial^2 H^{*0} }{\partial \mu_{a]b} \partial \lambda_{k}} \lambda_b
+ \delta^{k[i} \frac{\partial H^{*0}}{\partial \mu_{a]b}}  \lambda_b \, .
\end{eqnarray}
To conclude this section, we can say that we have to impose the conditions (\ref{17.3}),
(\ref{bl.1}), (\ref{cd.2}), (\ref{da.11}) and (\ref{ea.1}) on $H^{*0}(\mu_{ab}, \lambda ,
\lambda_c)$. \\
After that, (\ref{dd.2}) and (\ref{dl.1}) will give $\tilde{\tilde{H}}^i$; we will see
that a further little integrability condition will be necessary to this end.

\section{Solution of the conditions on $H^{*0}$. }
To solve (\ref{17.3}), it is useful to define $\psi_{0,q+1,r,0}$ and $\tilde{H}^{*0}$
from
\begin{eqnarray}\label{eb.1}
\vartheta_{1,q,r+1,0}= \frac{\partial }{\partial \lambda} \psi_{0,q+1,r,0} \, ,
\end{eqnarray}
\begin{eqnarray}\label{eb.2}
H^{*0}= \tilde{H}^{*0} + \sum_{q=0}^{\infty} \sum_{r \in I_{0}} \frac{1}{(q+1)!}
 \frac{1}{r!}   \psi_{0,q+1,r,0}
\delta^{(h_1k_1 \cdots h_{q+1}k_{q+1}j_1 \cdots j_r)} \mu_{h_1k_1} \cdots
\mu_{h_{q+1}k_{q+1}} \lambda_{j_1} \cdots \lambda_{j_r} \, .
\end{eqnarray}
By using them, we see that (\ref{17.3}) becomes $\frac{\partial^2 \tilde{H}^{*0}
}{\partial \mu_{ij} \partial \lambda} =0$, that is, $\tilde{H}^{*0}$ is sum of a function
not depending on $\lambda$ and of a function not depending on $\mu_{ij}$, that is
\begin{eqnarray*}
\tilde{H}^{*0} = \tilde{H}^{*01}(\lambda , \lambda_j) + \tilde{H}^{*02}(\mu_{ij} ,
\lambda_k)= \sum_{r \in I_{0}}
 \frac{1}{r!}   \psi_{0,0,r,0}(\lambda) \delta^{(j_1 \cdots j_r)}\lambda_{j_1} \cdots \lambda_{j_r} + \tilde{H}^{*02}(\mu_{ij} ,
\lambda_k)\, ,
\end{eqnarray*}
where we have introduced the expansion of $\tilde{H}^{*01}$ around equilibrium and called
$\psi_{0,0,r,0}$ the coefficients. We may also assume, without loss of generality, that
\begin{eqnarray}\label{eb.5}
\tilde{H}^{*02}(0 , \lambda_k) =0 \, ,
\end{eqnarray}
because its eventual non zero value can be enclosed in $\tilde{H}^{*01}$. By substituting
the result in (\ref{eb.2}) this is transformed in
\begin{eqnarray}\label{eb.3}
H^{*0}= \tilde{H}^{*02}(\mu_{ij} , \lambda_k) + \sum_{q=0}^{\infty} \sum_{r \in I_{0}}
\frac{1}{q!} \frac{1}{r!}   \psi_{0,q,r,0} \delta^{(h_1k_1 \cdots h_qk_qj_1 \cdots j_r)}
\mu_{h_1k_1} \cdots \mu_{h_qk_q} \lambda_{j_1} \cdots \lambda_{j_r} \, .
\end{eqnarray}
This is the solution of (\ref{17.3}). By substituting it in (\ref{bl.1}), we obtain
\begin{eqnarray}\label{eb.4}
0= \left[ (2Q+1) \psi_{0,Q,0,0}(\lambda) + 2 \lambda \psi_{0,Q+1,0,0}(\lambda) \right]
\delta^{(kih_1k_1 \cdots h_Qk_Q)} + \\
\left\{ \frac{\partial^Q}{\partial \mu_{h_1k_1}  \cdots  \partial \mu_{h_Qk_Q}} \left[ 2
\mu_{ji} \frac{\partial \tilde{H}^{*02}}{\partial \mu_{kj}} + 2 \lambda \frac{\partial
\tilde{H}^{*02}}{\partial \mu_{ki}} + \tilde{H}^{*02} \delta^{ki} \right]
\right\}_{\lambda_{j}=0, \, \mu_{ia}=0 } \, . \nonumber
\end{eqnarray}
The derivative of this relation with respect to $\lambda$ gives the information
\begin{eqnarray}\label{ec.1}
(2Q+1) \psi_{0,Q,0,0}(\lambda) + 2 \lambda \psi_{0,Q+1,0,0}(\lambda) = \phi_{0,Q+1,0,0}
\, ,
\end{eqnarray}
with $\phi_{0,Q+1,0,0}$ constant. After that, (\ref{eb.4}) becomes
\begin{eqnarray}\label{ec.2}
0= \left\{ \frac{\partial^Q}{\partial \mu_{h_1k_1}  \cdots  \partial \mu_{h_Qk_Q}} \left[
2 \mu_{ji} \frac{\partial \tilde{H}^{*02}}{\partial \mu_{kj}} + 2 \lambda \frac{\partial
\tilde{H}^{*02}}{\partial \mu_{ki}} + \tilde{H}^{*02} \delta^{ki} \right]
\right\}_{\lambda_{j}=0, \, \mu_{ia}=0 } + \\
+ \phi_{0,Q+1,0,0} \delta^{(kih_1k_1 \cdots h_Qk_Q)} \, . \quad \quad \quad \quad \quad
\quad \quad \quad \quad  \nonumber
\end{eqnarray}
Let us consider now (\ref{cd.2}); thanks to (\ref{eb.3}) it is transformed in
\begin{eqnarray*}
0= 2 \lambda_{j} \frac{\partial^2 \, \tilde{H}^{*02}}{\partial \mu_{ij} \partial
\mu_{ka}} + \sum_{Q=0}^{\infty} \sum_{R \in I_{1}} \frac{1}{Q!} \frac{1}{R!}
\left[ 2R \psi_{0,Q+2,R-1,0}+ (2Q+R+2) \vartheta_{1,Q,R,0} + \right. \\
\left. + 2  \lambda
 \vartheta_{1,Q+1,R,0} \right] \delta^{(akih_1k_1 \cdots h_Qk_Qj_1 \cdots j_R)}
\mu_{h_1k_1} \cdots \mu_{h_Qk_Q} \lambda_{j_1} \cdots \lambda_{j_R} \, ,
\end{eqnarray*}
which is equivalent to
\begin{eqnarray}\label{ec.3}
2R \psi_{0,Q+2,R-1,0}+ (2Q+R+2) \vartheta_{1,Q,R,0}  + 2  \lambda
 \vartheta_{1,Q+1,R,0}  = \phi_{1,Q,R,0} \quad \mbox{for} \quad R \geq 1
\end{eqnarray}
and to
\begin{eqnarray}\label{ec.4}
0= 2 \lambda_{j} \frac{\partial^2 \, \tilde{H}^{*02}}{\partial \mu_{ij} \partial
\mu_{ka}} + \sum_{Q=0}^{\infty} \sum_{R \in I_{1}} \frac{1}{Q!} \frac{1}{R!}
\phi_{1,Q,R,0} \delta^{(akih_1k_1 \cdots h_Qk_Qj_1 \cdots j_R)} \mu_{h_1k_1} \cdots
\mu_{h_Qk_Q} \lambda_{j_1} \cdots \lambda_{j_R} \, ,
\end{eqnarray}
with $\phi_{1,Q,R,0}$ constant. \\
We want now to see how (\ref{da.11}) is transformed by use of (\ref{eb.3}); to this end,
let us firstly substitute (\ref{eb.3}) except for the term $\tilde{H}^{*02}$. Of the
resulting expression, let us take the derivatives with respect to $\mu_{h_1k_1}$,
$\cdots$, $\mu_{h_Qk_Q}$, $\lambda_{j_1}$, $\cdots$, $\lambda_{j_R}$ calculated at
equilibrium; it is equal to
\begin{eqnarray*}
2 Q  \delta^{i \overline{h_1}} \delta^{(\overline{k_1 h_2k_2 \cdots h_Qk_Q}j_1 \cdots j_Rk)}
\frac{\partial}{\partial \lambda} \psi_{0,Q,R,0}+ \\
+ \left( 2 \lambda \frac{\partial}{\partial \lambda} \psi_{0,Q+1,R,0}+ 2 \psi_{0,Q+1,R,0}
\right) \delta^{(ki h_1k_1 \cdots h_Qk_Q j_1 \cdots j_R)} +
\\
+ R \delta^{i \overline{j_1}} \delta^{(\overline{j_2 \cdots j_{R}}h_1k_1 \cdots h_Qk_Qk)}
\frac{\partial}{\partial \lambda} \psi_{0,Q,R,0} + \delta^{ki} \delta^{(h_1k_1 \cdots
h_Qk_Q j_1 \cdots j_R)}
\frac{\partial}{\partial \lambda} \psi_{0,Q,R,0} + \\
+ 2 R \delta^{(kih_1k_1 \cdots h_Qk_Q j_1 \cdots j_R)} \psi_{0,Q+1,R,0} = \\
= (2 Q +R+1) \delta^{i \overline{h_1}} \delta^{(\overline{k_1 h_2k_2 \cdots h_Qk_Qj_1
\cdots j_Rk})} \frac{\partial}{\partial \lambda} \psi_{0,Q,R,0}+ \\
+ \left( 2 \lambda \frac{\partial}{\partial \lambda} \psi_{0,Q+1,R,0}+ 2 \psi_{0,Q+1,R,0}
+ 2R \psi_{0,Q+1,R,0} \right) \delta^{(ki h_1k_1 \cdots h_Qk_Q j_1 \cdots j_R)} \, .
\end{eqnarray*}
By using this result, we see that (\ref{da.11}) is transformed in
\begin{eqnarray*}
0= 2 \frac{\partial \tilde{H}^{*02}}{\partial  \mu_{ki}} + 2  \lambda_j \frac{\partial^2
\tilde{H}^{*02}}{\partial \mu_{ij}  \partial \lambda_k} + \\
+ \sum_{q=0}^{\infty} \sum_{r \in I_{0}} \frac{1}{q!} \frac{1}{r!} \left[ (2 q +r+1)
\frac{\partial}{\partial \lambda} \psi_{0,q,r,0} + 2 \lambda \frac{\partial}{\partial
\lambda} \psi_{0,q+1,r,0}+ 2(r+1) \psi_{0,q+1,r,0}  \right]
\cdot \\
 \delta^{(kih_1k_1 \cdots h_qk_qj_1 \cdots j_r)} \mu_{h_1k_1} \cdots
\mu_{h_qk_q} \lambda_{j_1} \cdots \lambda_{j_r}
 \, .
\end{eqnarray*}
which is equivalent to
\begin{eqnarray}\label{ee.1}
(2 q +r+1) \frac{\partial}{\partial \lambda} \psi_{0,q,r,0} + 2 \lambda
\frac{\partial}{\partial \lambda} \psi_{0,q+1,r,0}+ 2(r+1) \psi_{0,q+1,r,0}  =
\phi'_{0,q+1,r,0}
\end{eqnarray}
and to
\begin{eqnarray}\label{ee.2}
0= 2 \frac{\partial \tilde{H}^{*02}}{\partial  \mu_{ki}} + 2  \lambda_j \frac{\partial^2
\tilde{H}^{*02}}{\partial \mu_{ij}  \partial \lambda_k}  + \sum_{q=0}^{\infty} \sum_{r
\in I_{0}} \frac{1}{q!} \frac{1}{r!} \phi'_{0,q+1,r,0}
 \delta^{(kih_1k_1 \cdots h_qk_qj_1 \cdots j_r)} \mu_{h_1k_1} \cdots
\mu_{h_qk_q} \lambda_{j_1} \cdots \lambda_{j_r}
 \, .
\end{eqnarray}
with $\phi'_{0,q+1,r,0}$ constant. \\
Finally, let us substitute (\ref{eb.3}) in (\ref{ea.1}), so obtaining
\begin{eqnarray}\label{ef.1}
0=2 \mu_{j[i}  \frac{\partial^2 \tilde{H}^{*02} }{\partial \mu_{a]b} \partial \mu_{kj}}
\lambda_b + 2\lambda \frac{\partial^2 \tilde{H}^{*02} }{\partial \mu_{k[i} \partial
\mu_{a]b}} \lambda_b + \lambda_{[i} \frac{\partial^2 \tilde{H}^{*02} }{\partial \mu_{a]b}
\partial \lambda_{k}} \lambda_b + \delta^{k[i} \frac{\partial \tilde{H}^{*02}}{\partial \mu_{a]b}}
\lambda_b + \\
+ 2 \mu_{j[i} \sum_{q=0}^{\infty} \sum_{r \in I_{0}} \frac{1}{q!} \frac{1}{r!}
\psi_{0,q+2,r,0} \delta^{(a]bkjh_1k_1 \cdots h_qk_qj_1 \cdots j_r)} \mu_{h_1k_1} \cdots
\mu_{h_qk_q} \lambda_{j_1} \cdots \lambda_{j_r}   \lambda_b + \nonumber \\
+ \lambda_{[i} \sum_{q=0}^{\infty} \sum_{r \in I_{1}} \frac{1}{q!} \frac{1}{r!}
\psi_{0,q+1,r+1,0} \delta^{(a]bkh_1k_1 \cdots h_qk_qj_1 \cdots j_r)} \mu_{h_1k_1} \cdots
\mu_{h_qk_q} \lambda_{j_1} \cdots \lambda_{j_r} \lambda_b + \nonumber  \\
+ \delta^{k[i} \sum_{q=0}^{\infty} \sum_{r \in I_{0}} \frac{1}{q!} \frac{1}{r!}
\psi_{0,q+1,r,0} \delta^{(a]bh_1k_1 \cdots h_qk_qj_1 \cdots j_r)} \mu_{h_1k_1} \cdots
\mu_{h_qk_q} \lambda_{j_1} \cdots \lambda_{j_r}    \lambda_b   \, .  \nonumber
\end{eqnarray}
Of the expression in the right hand side, let us consider the part not involving
$\tilde{H}^{*02}$ and without skew-symmetrization; its part linear in $\lambda_{j}$ is
\begin{eqnarray*}
 2 \mu_{ji} \sum_{q=0}^{\infty} \frac{1}{q!}
\psi_{0,q+2,0,0} \delta^{(abkjh_1k_1 \cdots h_qk_q)} \mu_{h_1k_1} \cdots
\mu_{h_qk_q}    \lambda_b +  \\
 + \delta^{ki} \sum_{q=0}^{\infty}  \frac{1}{q!}
\psi_{0,q+1,0,0} \delta^{(abh_1k_1 \cdots h_qk_q)} \mu_{h_1k_1} \cdots \mu_{h_qk_q}
   \lambda_b
\end{eqnarray*}
whose derivatives with respect to $\mu_{h_1k_1}$, $\cdots$, $\mu_{h_Qk_Q}$, $\lambda_{b}$
calculated at equilibrium is equal to
\begin{eqnarray*}
2 Q  \delta^{i \overline{h_1}} \delta^{(\overline{k_1 h_2k_2 \cdots h_Qk_Q}akb)}
\psi_{0,Q+1,0,0}+ \delta^{ki} \delta^{(abh_1k_1 \cdots h_Qk_Q )} \psi_{0,Q+1,0,0} = \\
= (2 Q+3)  \delta^{i \overline{h_1}} \delta^{(\overline{k_1 h_2k_2 \cdots h_Qk_Qakb})}
\psi_{0,Q+1,0,0} - \psi_{0,Q+1,0,0} \left[ \delta^{ia} \delta^{(h_1k_1 \cdots h_Qk_Q kb)}
+ \delta^{ib} \delta^{(h_1k_1 \cdots h_Qk_Q ka)} \right] \, .
\end{eqnarray*}
So we can write this linear part in $\lambda_{j}$ as
\begin{eqnarray*}
\sum_{q=0}^{\infty} \frac{1}{q!} \psi_{0,q+1,0,0} \left[ (2 q+3) \delta^{(ih_1k_1 h_2k_2
\cdots h_qk_qakb)} - \delta^{ia} \delta^{(h_1k_1 \cdots h_qk_q kb)} - \delta^{ib}
\delta^{(h_1k_1 \cdots h_qk_q ka)} \right] \\
\mu_{h_1k_1} \cdots \mu_{h_qk_q}
   \lambda_b
\end{eqnarray*}
whose skew-symmetric part with respect to $i$ and $a$ is
\begin{eqnarray}\label{ef.10}
-\sum_{q=0}^{\infty} \frac{1}{q!} \psi_{0,q+1,0,0} \lambda^{[i} \delta^{(a]kh_1k_1 \cdots
h_qk_q )} \mu_{h_1k_1} \cdots \mu_{h_qk_q}  \, .
\end{eqnarray}
This is the linear part in $\lambda_{j}$ of the expression (\ref{ef.1}) without the terms
in $\tilde{H}^{*02}$. In order to consider the terms with degree in $\lambda_{j}$ greater
than 1, let us change index in the summations of the parts with $\mu_{j[1} \cdots$ and
$\delta^{k[i} \cdots$; the change of index is $r=R+1$. In this wy we obtain the
skew-symmetric part with respect to $i$ and $a$ of
\begin{eqnarray}\label{ef.11}
 2 \mu_{ji} \sum_{q=0}^{\infty} \sum_{R \in I_{1}} \frac{1}{q!} \frac{1}{(R+1)!}
\psi_{0,q+2,R+1,0} \delta^{(akjh_1k_1 \cdots h_qk_qj_1 \cdots j_{R+2})} \mu_{h_1k_1}
\cdots
\mu_{h_qk_q} \lambda_{j_1} \cdots \lambda_{j_{R+2}}  +  \\
+ \delta^{ij_{R+2}} \sum_{q=0}^{\infty} \sum_{R \in I_{1}} \frac{1}{q!} \frac{1}{R!}
\psi_{0,q+1,R+1,0} \delta^{(akh_1k_1 \cdots h_qk_qj_1 \cdots j_{R+1})} \mu_{h_1k_1}
\cdots
\mu_{h_qk_q} \lambda_{j_1} \cdots \lambda_{j_{R+2}}  + \nonumber  \\
+ \delta^{ki} \sum_{q=0}^{\infty} \sum_{R \in I_{1}} \frac{1}{q!} \frac{1}{(R+1)!}
\psi_{0,q+1,R+1,0} \delta^{(ah_1k_1 \cdots h_qk_qj_1 \cdots j_{R+2})} \mu_{h_1k_1} \cdots
\mu_{h_qk_q} \lambda_{j_1} \cdots \lambda_{j_{R+2}}  \, ,  \nonumber
\end{eqnarray}
where we have put also  $b=j_{R+2}$. The derivatives of this expression with respect to
$\mu_{h_1k_1}$, $\cdots$, $\mu_{h_Qk_Q}$,  calculated in $\mu_{ab}=0$ is equal to
\begin{eqnarray*}
\sum_{R \in I_{1}}  \frac{1}{(R+1)!} \left[ 2 Q  \delta^{i \overline{h_1}}
\delta^{(\overline{k_1 h_2k_2 \cdots h_Qk_Q}akj_1 \cdots j_{R+2})} \psi_{0,Q+1,R+1,0}+ \right. \\
+ (R+1) \delta^{i \overline{j_1}} \delta^{(\overline{j_2 \cdots j_{R+2}}akh_1k_1 \cdots
h_Qk_Q)}  \psi_{0,Q+1,R+1,0} + \\
\left. + \delta^{ki} \delta^{(a h_1k_1 \cdots h_Qk_Q j_1 \cdots j_{R+2})}
\psi_{0,Q+1,R+1,0} \right] \lambda_{j_1} \cdots \lambda_{j_{R+2}} = \\
=  \sum_{R \in I_{1}}  \frac{1}{(R+1)!} \psi_{0,Q+1,R+1,0} \left[ (2 Q+R+4)  \delta^{i
\overline{h_1}}
\delta^{(\overline{k_1 h_2k_2 \cdots h_Qk_Qakj_1 \cdots j_{R+2}})} + \right. \\
\left. - \delta^{i \overline{j_1}} \delta^{(\overline{j_2 \cdots j_{R+2}}akh_1k_1 \cdots
h_Qk_Q)} - \delta^{ia} \delta^{(kh_1k_1 \cdots h_Qk_Qj_1 \cdots j_{R+2})} \right]
\lambda_{j_1} \cdots \lambda_{j_{R+2}} \, ,
\end{eqnarray*}
so that (\ref{ef.11}) is equal to
\begin{eqnarray*}
\sum_{q=0}^{\infty} \sum_{R \in I_{1}} \frac{1}{q!} \frac{1}{(R+1)!}\psi_{0,q+1,R+1,0}
\left[ (2 q+R+4)
\delta^{(ih_1k_1 h_2k_2 \cdots h_qk_qakj_1 \cdots j_{R+2})} + \right. \\
\left. - \delta^{i \overline{j_1}} \delta^{(\overline{j_2 \cdots j_{R+2}}akh_1k_1 \cdots
h_qk_q)} - \delta^{ia} \delta^{(kh_1k_1 \cdots h_qk_qj_1 \cdots j_{R+2})} \right]
\lambda_{j_1} \cdots \lambda_{j_{R+2}}\mu_{h_1k_1} \cdots \mu_{h_qk_q}
\end{eqnarray*}
whose skew-symmetric part with respect to $i$ and $a$ is
\begin{eqnarray*}
-\sum_{q=0}^{\infty} \sum_{R \in I_{1}} \frac{1}{q!} \frac{1}{(R+1)!}\psi_{0,q+1,R+1,0}
 \lambda^{[i} \delta^{(a]kh_1k_1 \cdots
h_qk_qj_1 \cdots j_{R+1})}  \lambda_{j_1} \cdots \lambda_{j_{R+1}}\mu_{h_1k_1} \cdots
\mu_{h_qk_q} \, .
\end{eqnarray*}
This result, jointly with (\ref{ef.10}), allows to rewrite (\ref{ef.1}) as
\begin{eqnarray*}
0=2 \mu_{j[i}  \frac{\partial^2 \tilde{H}^{*02} }{\partial \mu_{a]b} \partial \mu_{kj}}
\lambda_b + 2\lambda \frac{\partial^2 \tilde{H}^{*02} }{\partial \mu_{k[i} \partial
\mu_{a]b}} \lambda_b + \lambda_{[i} \frac{\partial^2 \tilde{H}^{*02} }{\partial \mu_{a]b}
\partial \lambda_{k}} \lambda_b + \delta^{k[i} \frac{\partial \tilde{H}^{*02}}{\partial \mu_{a]b}}
\lambda_b + \\
-\sum_{q=0}^{\infty} \frac{1}{q!} \psi_{0,q+1,0,0} \lambda^{[i} \delta^{(a]kh_1k_1 \cdots
h_qk_q )} \mu_{h_1k_1} \cdots \mu_{h_qk_q}  + \\
-\sum_{q=0}^{\infty} \sum_{R \in I_{1}} \frac{1}{q!} \frac{1}{(R+1)!}\psi_{0,q+1,R+1,0}
 \lambda^{[i} \delta^{(a]kh_1k_1 \cdots
h_qk_qj_1 \cdots j_{R+1})}  \lambda_{j_1} \cdots \lambda_{j_{R+1}}\mu_{h_1k_1} \cdots
\mu_{h_qk_q} \, .
\end{eqnarray*}
But the last 2 terms of this expression can be written in a more compact form, so that
the whole expression becomes
\begin{eqnarray}\label{ef.2}
0=2 \mu_{j[i}  \frac{\partial^2 \tilde{H}^{*02} }{\partial \mu_{a]b} \partial \mu_{kj}}
\lambda_b + 2\lambda \frac{\partial^2 \tilde{H}^{*02} }{\partial \mu_{k[i} \partial
\mu_{a]b}} \lambda_b + \lambda_{[i} \frac{\partial^2 \tilde{H}^{*02} }{\partial \mu_{a]b}
\partial \lambda_{k}} \lambda_b + \delta^{k[i} \frac{\partial \tilde{H}^{*02}}{\partial \mu_{a]b}}
\lambda_b + \\
-\sum_{q=0}^{\infty} \sum_{R \in I_{0}} \frac{1}{q!} \frac{1}{R!}\psi_{0,q+1,R,0}
 \lambda^{[i} \delta^{(a]kh_1k_1 \cdots
h_qk_qj_1 \cdots j_{R})}  \lambda_{j_1} \cdots \lambda_{j_{R}}\mu_{h_1k_1} \cdots
\mu_{h_qk_q} \, . \nonumber
\end{eqnarray}
whose derivative with respect to $\lambda$ gives
\begin{eqnarray}\label{ef.3}
\frac{\partial}{\partial \lambda} \psi_{0,q+1,R,0} =0 \, .
\end{eqnarray}
So there remains now to impose the conditions (\ref{ec.2}), (\ref{ec.4}), (\ref{ee.2})
and (\ref{ef.2}) on $\tilde{H}^{*02}( \mu_{ab} , \lambda_c)$.

\subsection{The conditions on $\tilde{H}^{*02}$.}
\begin{itemize}
  \item Let us begin with (\ref{ec.4}).
\end{itemize}
Let us define $\tilde{H}^{*03}$ from
\begin{eqnarray}\label{el.1}
\tilde{H}^{*02}= \tilde{H}^{*03} - \frac{1}{2} \sum_{q=0}^{\infty} \sum_{r \in I_{0}}
\frac{1}{(q+2)!} \frac{1}{(r+1)!}   \phi_{1,q,r+1,0} \delta^{(h_1k_1 \cdots
h_{q+2}k_{q+2}j_1 \cdots j_r)} \mu_{h_1k_1} \cdots \mu_{h_{q+2}k_{q+2}} \lambda_{j_1}
\cdots \lambda_{j_r} \, .
\end{eqnarray}
After that, eq. (\ref{ec.4}) becomes
\begin{eqnarray}\label{el.2}
0= 2 \lambda_{j} \frac{\partial^2 \, \tilde{H}^{*03}}{\partial \mu_{ij} \partial
\mu_{ka}}  \, .
\end{eqnarray}
\begin{itemize}
  \item Instead of this, (\ref{ee.2}) (by writing explicitly the term with $q=0$ and changing
index in the remaining part according to $q=Q+1$) becomes
\end{itemize}
\begin{eqnarray}\label{el.3}
0= 2 \frac{\partial \tilde{H}^{*03}}{\partial  \mu_{ki}} + 2  \lambda_j \frac{\partial^2
\tilde{H}^{*03}}{\partial \mu_{ij}  \partial \lambda_k}  + \sum_{r \in I_{0}}
\frac{1}{r!} \phi'_{0,1,r,0}  \delta^{(kij_1 \cdots j_r)}  \lambda_{j_1} \cdots
\lambda_{j_r} + \\
- \sum_{q=0}^{\infty} \sum_{r \in I_{0}} \frac{1}{(q+1)!} \frac{1}{(r+1)!} \left(
\phi_{1,q,r+1,0} + r \phi_{1,q,r,0} - (r+1) \phi'_{0,q+2,r,0} \right)
 \delta^{(kih_1k_1 \cdots h_{q+1}k_{q+1}j_1 \cdots j_r)} \nonumber \\
 \mu_{h_1k_1} \cdots
\mu_{h_{q+1}k_{q+1}} \lambda_{j_1} \cdots \lambda_{j_r}
 \, .  \nonumber
\end{eqnarray}
Now, if we contract this expression with $\lambda_{i}$, we will obtain another expression
whose first 2 terms don' t depend on $\mu_{ab}$ for (\ref{el.2}), the third term
explicitly doesn' t depend on $\mu_{ab}$, the last one will be at least linear in
$\mu_{ab}$, except if the coefficients are zero. In other words, we have
\begin{eqnarray}\label{fa.3}
\phi_{1,q,r+1,0} + r \phi_{1,q,r,0} - (r+1) \phi'_{0,q+2,r,0} =0
 \, .
\end{eqnarray}
After this result, what remains of (\ref{el.3}) shows that $\frac{\partial
\tilde{H}^{*03}}{\partial  \mu_{ki}}$ depends only on $\lambda_{j}$; consequently,
\begin{eqnarray*}
\frac{\partial \tilde{H}^{*03}}{\partial  \mu_{ki}}+ \frac{1}{2} \sum_{r \in I_{0}}
\frac{1}{(r+1)!} \phi'_{0,1,r,0}  \delta^{(kij_1 \cdots j_r)}  \lambda_{j_1} \cdots
\lambda_{j_r}
\end{eqnarray*}
is a second order symmetric tensor depending only on $\lambda_{j}$. By applying the
Representation Theorems, we can say that it is a linear combination, trough scalar
coefficients, of $\delta^{ij}$ and of $\lambda^i \lambda^j$; writing the expansion of the
coefficients around $\lambda_{j}=0$, we obtain that
\begin{eqnarray}\label{fa.6}
\frac{\partial \tilde{H}^{*03}}{\partial  \mu_{ki}}=- \frac{1}{2} \sum_{r \in I_{0}}
\frac{1}{(r+1)!} \phi'_{0,1,r,0}  \delta^{(kij_1 \cdots j_r)}  \lambda_{j_1} \cdots
\lambda_{j_r} + \sum_{r=0}^\infty \frac{1}{r!} (\lambda_a \lambda^a)^r (\alpha_r
\delta^{ik} + \beta_r \lambda^i \lambda^k) \, ,
\end{eqnarray}
with $\alpha_r$ and $\beta_r$ constants. By using this, (\ref{el.3}) becomes
\begin{eqnarray}\label{fa.7}
\alpha_0=0 \quad , \quad \alpha_{r+1} = - (r+1) \beta_{r}
\end{eqnarray}
and (\ref{el.2}) is an identity. \\
Now, from (\ref{eb.5}) and (\ref{el.1}) it follows $\tilde{H}^{*03}(0 , \lambda_k)=0$.
This result, jointly with (\ref{fa.6}) allows us to obtain
\begin{eqnarray}\label{fa.8}
\tilde{H}^{*03}=\mu_{ki} \left[ - \frac{1}{2} \sum_{r \in I_{0}} \frac{1}{(r+1)!}
\phi'_{0,1,r,0} \delta^{(kij_1 \cdots j_r)}  \lambda_{j_1} \cdots \lambda_{j_r} +
\sum_{r=0}^\infty \frac{1}{r!} (\lambda_a \lambda^a)^r \beta_r (\lambda^i \lambda^k -
\lambda_a \lambda^a \delta^{ik}) \right] \, ,
\end{eqnarray}
so that (\ref{el.1}) now becomes
\begin{eqnarray}\label{fb.1}
\tilde{H}^{*02}= \mu_{ki} \left[ - \frac{1}{2} \sum_{r \in I_{0}} \frac{1}{(r+1)!}
\phi'_{0,1,r,0} \delta^{(kij_1 \cdots j_r)}  \lambda_{j_1} \cdots \lambda_{j_r} +
\sum_{r=0}^\infty \frac{1}{r!} (\lambda_a \lambda^a)^r \beta_r (\lambda^i \lambda^k -
\lambda_a \lambda^a \delta^{ik}) \right] + \\
- \frac{1}{2} \sum_{q=0}^{\infty} \sum_{r \in I_{0}} \frac{1}{(q+2)!} \frac{1}{(r+1)!}
\phi_{1,q,r+1,0} \delta^{(h_1k_1 \cdots h_{q+2}k_{q+2}j_1 \cdots j_r)} \mu_{h_1k_1}
\cdots \mu_{h_{q+2}k_{q+2}} \lambda_{j_1} \cdots \lambda_{j_r} \, . \nonumber
\end{eqnarray}
\begin{itemize}
  \item Let us substitute this expression in (\ref{ec.2}). For $Q=0$, $Q=1$ and $Q \geq 2$  we
  obtain, respectively
\end{itemize}
\begin{eqnarray}\label{fb.2}
&{}& \phi_{0,1,0,0} - \lambda \phi'_{0,1,0,0} = 0 \, , \\
&{}& \phi_{0,2,0,0} - \lambda \phi_{1,0,1,0} - \frac{3}{2} \phi'_{0,1,0,0} = 0 \, , \nonumber \\
&{}& \phi_{0,Q+1,0,0} - \lambda \phi_{1,Q-1,1,0} - \frac{1}{2}(2Q+1) \phi_{1,Q-2,1,0} = 0
\, . \nonumber
\end{eqnarray}
But the quantities $\phi_{\cdots}$, $\phi'_{\cdots}$ don' t depend on $\lambda$ so that
(\ref{fb.2}) now becomes
\begin{eqnarray}\label{fc.1}
\phi_{1,Q,1,0} = 0 \quad \mbox{for} \, Q \geq 0 \quad , \quad \phi'_{0,1,0,0} = 0 \quad ,
\quad \phi_{0,Q,0,0} = 0 \quad \mbox{for} \, Q \geq 1 \, .
\end{eqnarray}
\begin{itemize}
  \item Let us substitute now (\ref{fb.1}) in (\ref{ef.2}). We find
\end{itemize}
\begin{eqnarray*}
0=2 \mu_{j[i} \left( - \frac{1}{2} \sum_{q=0}^{\infty} \sum_{r \in I_{0}} \frac{1}{q!}
\frac{1}{(r+1)!} \phi_{1,q,r+1,0} \delta^{(a]bkjh_1k_1 \cdots h_{q}k_{q}j_1 \cdots j_r)}
\mu_{h_1k_1} \cdots \mu_{h_{q}k_{q}} \lambda_{b} \lambda_{j_1} \cdots \lambda_{j_r}
\right) +
\end{eqnarray*}
\begin{eqnarray}\label{fc.2}
{} \hspace{8cm}
\end{eqnarray}
\begin{eqnarray*}
 +\lambda_{[i} \left[ - \frac{1}{2} \sum_{r \in I_{0}}  \frac{r}{(r+1)!}
\phi'_{0,1,r,0} \delta^{(a]bkj_1 \cdots j_{r-1})}  \lambda_{b} \lambda_{j_1} \cdots
\lambda_{j_{r-1}} + \right.
\\
- \frac{1}{2} \sum_{q=0}^{\infty} \sum_{r \in I_{0}} \frac{1}{(q+1)!} \frac{r}{(r+1)!}
\phi_{1,q,r+1,0} \delta^{(a]bkh_1k_1 \cdots h_{q+1}k_{q+1}j_1 \cdots j_{r-1})}
\mu_{h_1k_1} \cdots \mu_{h_{q+1}k_{q+1}} \lambda_{b} \lambda_{j_1} \cdots
\lambda_{j_{r-1}} + \\
\left. + \sum_{r=0}^{\infty} \frac{1}{r!} (\lambda^c \lambda_c)^r \beta_r \left(
\delta^{a]k} \lambda^b \lambda_b - \lambda^{a]} \lambda^k \right)
\right] + \\
+ \delta^{k[i} \left( - \frac{1}{2} \sum_{r \in I_{0}}  \frac{1}{(r+1)!}
\phi'_{0,1,r,0}
\delta^{(a]bj_1 \cdots j_r)}  \lambda_{b} \lambda_{j_1} \cdots \lambda_{j_r} + \right. \\
\left. - \frac{1}{2} \sum_{q=0}^{\infty} \sum_{r \in I_{0}} \frac{1}{(q+1)!}
\frac{1}{(r+1)!} \phi_{1,q,r+1,0} \delta^{(a]bh_1k_1 \cdots h_{q+1}k_{q+1}j_1 \cdots
j_r)} \mu_{h_1k_1} \cdots \mu_{h_{q+1}k_{q+1}} \lambda_{b} \lambda_{j_1} \cdots
\lambda_{j_r} \right) + \\
-\sum_{q=0}^{\infty} \sum_{r \in I_{0}} \frac{1}{q!} \frac{1}{r!}\psi_{0,q+1,r,0}
 \lambda^{[i} \delta^{(a]kh_1k_1 \cdots
h_qk_qj_1 \cdots j_{r})}  \lambda_{j_1} \cdots \lambda_{j_{r}}\mu_{h_1k_1} \cdots
\mu_{h_qk_q} \, .
\end{eqnarray*}
But now we have
\begin{eqnarray*}
 r \lambda_{i} \delta^{(abkj_1 \cdots j_{r-1})}  \lambda_{b} \lambda_{j_1} \cdots
\lambda_{j_{r-1}} + \delta^{ki} \delta^{(abj_1 \cdots j_r)} \lambda_{b} \lambda_{j_1}
\cdots \lambda_{j_r} = \\
=\lambda_{b} \lambda_{j_1} \cdots \lambda_{j_r} [  r \delta^{j_ri} \delta^{(abkj_1 \cdots
j_{r-1})} + \delta^{ki} \delta^{(abj_1 \cdots j_r)}] = \\
= \lambda_{b} \lambda_{j_1} \cdots \lambda_{j_r} [  r \delta^{i\overline{j_r}}
\delta^{(\overline{bj_1 \cdots j_{r-1}} ka)} + \delta^{ki} \delta^{(abj_1 \cdots j_r)} +
\delta^{ia} \delta^{(kbj_1 \cdots j_r)}- \delta^{ia} \delta^{(kbj_1 \cdots j_r)}]= \\
= \lambda_{b} \lambda_{j_1} \cdots \lambda_{j_r} [  (r+3) \delta^{i\overline{j_r}}
\delta^{(\overline{bj_1 \cdots j_{r-1}ka} )} - \delta^{i\overline{j_r}}
\delta^{(\overline{bj_1 \cdots j_{r-1}} ka)} - \delta^{ia} \delta^{(kbj_1 \cdots j_r)}]=
\\
= \lambda_{b} \lambda_{j_1} \cdots \lambda_{j_r} [  (r+3) \delta^{(ikabj_1 \cdots j_{r}
)} - \delta^{i\overline{j_r}} \delta^{(\overline{bj_1 \cdots j_{r-1}} ka)} - \delta^{ia}
\delta^{(kbj_1 \cdots j_r)}] \, .
\end{eqnarray*}
By using this identity in (\ref{fc.2}) calculated in $\mu_{ab}=0$, we find
\begin{eqnarray*}
0= \frac{1}{2} \sum_{r \in I_{0}}  \frac{1}{(r+1)!} \phi'_{0,1,r,0} \lambda^{[i}
\delta^{(a]kj_1 \cdots j_{r})}   \lambda_{j_1} \cdots \lambda_{j_{r}} +
\sum_{r=0}^{\infty} \frac{1}{r!} (\lambda^c \lambda_c)^{r+1} \beta_r \lambda^{[i}
\delta^{a]k}- \psi_{0,1,0,0} \lambda^{[i} \delta^{a]k} + \\
-  \sum_{R \in I_{1}} \frac{1}{(R+1)!} \psi_{0,1,R+1,0} \lambda^{[i} \delta^{(a]kj_1
\cdots j_{R+1})} \lambda_{j_1} \cdots \lambda_{j_{R+1}} \, .
\end{eqnarray*}
This relation, by changing index in the last term according to $R=r-1$, transforms itself
in
\begin{eqnarray}\label{fd.0}
0= \frac{1}{2} \sum_{r \in I_{0}}  \frac{1}{(r+1)!} [ \phi'_{0,1,r,0} - 2(r+1)
\psi_{0,1,r,0} ] \lambda^{[i} \delta^{(a]kj_1 \cdots j_{r})}   \lambda_{j_1} \cdots
\lambda_{j_{r}} + \\
+ \sum_{r=0}^{\infty} \frac{1}{r!} (\lambda^c \lambda_c)^{r+1} \beta_r \lambda^{[i}
\delta^{a]k}- \psi_{0,1,0,0} \lambda^{[i} \delta^{a]k} \, . \nonumber
\end{eqnarray}
The linear part in $\lambda^{j}$ of this equation gives
\begin{eqnarray}\label{fd.1}
\phi'_{0,1,0,0} = 2 \psi_{0,1,0,0} \, .
\end{eqnarray}
For the remaining part, we use the identity
\begin{eqnarray*}
\delta^{(akj_1 \cdots j_{r})} = \delta^{a(kj_1 \cdots j_{r})} = \frac{1}{r+1} (
\delta^{ak} \delta^{(j_1 \cdots j_{r})} + r \delta^{a(j_1 \cdots j_{r})k}
\end{eqnarray*}
from which it follows
\begin{eqnarray*}
\delta^{(akj_1 \cdots j_{r})} \lambda_{j_1} \cdots \lambda_{j_{r}} = \frac{1}{r+1} \left[
\delta^{ak} (\lambda^c \lambda_c)^{\frac{r}{2}} + r \lambda^a \lambda^k (\lambda^c
\lambda_c)^{\frac{r-2}{2}} \right] \, .
\end{eqnarray*}
By using this identity, eq. (\ref{fd.0}) becomes
\begin{eqnarray*}
0= \frac{1}{2} \sum_{r \in I_{0}}  \frac{1}{(r+1)!} [ \phi'_{0,1,r,0} - 2(r+1)
\psi_{0,1,r,0} ] \frac{1}{r+1} \lambda^{[i} \delta^{a]k} (\lambda^c
\lambda_c)^{\frac{r}{2}}   + \sum_{R=0}^{\infty} \frac{1}{R!} (\lambda^c \lambda_c)^{R+1}
\beta_R \lambda^{[i} \delta^{a]k} \, .
\end{eqnarray*}
or, equivalently,
\begin{eqnarray}\label{fd.2}
\phi'_{0,1,2R+2,0} = 2(2R+3) \psi_{0,1,2R+2,0} -  \frac{(2R+3)!}{R!} 2 (2R+3)\beta_R \, .
\end{eqnarray}
After this result, what remains in the right hand side of eq. (\ref{fc.2}), without
skew-symmetrization and by putting $b=j_{r+1}$, becomes
\begin{eqnarray*}
- \frac{1}{2} \sum_{q=0}^{\infty} \sum_{r \in I_{0}} \frac{1}{(q+1)!} \frac{1}{(r+1)!}
\phi_{1,q,r+1,0} \left[ 2(q+1) \delta^{i \overline{h_{q+1}}} \delta^{(\overline{k_{q+1}
h_1k_1 \cdots
h_{q}k_{q}}j_1 \cdots j_{r+1} ak)} + \right. \\
+ \left. r \delta^{i \overline{j_{r+1}}} \delta^{(\overline{j_1 \cdots j_{r}} h_1k_1
\cdots h_{q+1}k_{q+1} ak)} + \delta^{i k} \delta^{(j_1 \cdots j_{r+1} h_1k_1 \cdots
h_{q+1}k_{q+1}a)} \right] \mu_{h_1k_1} \cdots \mu_{h_{q+1}k_{q+1}}
\lambda_{j_1} \cdots \lambda_{j_{r+1}} + \\
-\sum_{Q=0}^{\infty} \sum_{r \in I_{0}} \frac{1}{(Q+1)!} \frac{1}{r!}\psi_{0,Q+2,r,0}
 \lambda^{[i} \delta^{(a]kh_1k_1 \cdots
h_{Q+1}k_{Q+1}j_1 \cdots j_{r})}  \lambda_{j_1} \cdots \lambda_{j_{r}}\mu_{h_1k_1} \cdots
\mu_{h_{Q+1}k_{Q+1}} = \\
=- \frac{1}{2} \sum_{q=0}^{\infty} \sum_{r \in I_{0}} \frac{1}{(q+1)!} \frac{1}{(r+1)!}
\phi_{1,q,r+1,0} \left[ (2q+r+5)) \delta^{i \overline{h_{q+1}}}
\delta^{(\overline{k_{q+1} h_1k_1 \cdots
h_{q}k_{q}j_1 \cdots j_{r+1} ak})} + \right. \\
- \left. \delta^{i \overline{j_{r+1}}} \delta^{(\overline{j_1 \cdots j_{r}} h_1k_1 \cdots
h_{q+1}k_{q+1} ak)} - \delta^{i a} \delta^{(j_1 \cdots j_{r+1} h_1k_1 \cdots
h_{q+1}k_{q+1}k)} \right] \mu_{h_1k_1} \cdots \mu_{h_{q+1}k_{q+1}} \lambda_{j_1} \cdots
\lambda_{j_{r+1}} + \\-\sum_{Q=0}^{\infty} \sum_{r \in I_{0}} \frac{1}{(Q+1)!}
\frac{1}{r!}\psi_{0,Q+2,r,0}
 \lambda^{[i} \delta^{(a]kh_1k_1 \cdots
h_{Q+1}k_{Q+1}j_1 \cdots j_{r})}  \lambda_{j_1} \cdots \lambda_{j_{r}}\mu_{h_1k_1} \cdots
\mu_{h_{Q+1}k_{Q+1}} \, .
\end{eqnarray*}
Now (\ref{fc.2}) says that the skew-symmetric part of this expression, with respect to
$i$ and $a$, is zero, that is,
\begin{eqnarray*}
0 =\sum_{q=0}^{\infty} \sum_{r \in I_{0}} \frac{1}{(q+1)!} \frac{1}{(r+1)!} \left[
\frac{1}{2} \phi_{1,q,r+1,0} -(r+1) \phi_{0,q+2,r,0} \right]
 \lambda^{[i} \delta^{(a]kh_1k_1 \cdots
h_{Q+1}k_{Q+1}j_1 \cdots j_{r})} \\
 \lambda_{j_1} \cdots \lambda_{j_{r}}\mu_{h_1k_1}
\cdots \mu_{h_{Q+1}k_{Q+1}} \, ,
\end{eqnarray*}
from which
\begin{eqnarray}\label{fe.1}
\phi_{1,q,r+1,0} = 2(r+1) \phi_{0,q+2,r,0} \, .
\end{eqnarray}
By using (\ref{fd.1}), (\ref{fd.2}) and (\ref{fe.1}), the expression (\ref{fb.1}) becomes
\begin{eqnarray*}
\tilde{H}^{*02}=  -  \sum_{r \in I_{0}} \frac{1}{r!} \psi_{0,1,r,0}
\delta^{(ijj_1 \cdots j_r)} \mu_{ij} \lambda_{j_1} \cdots \lambda_{j_r} + \\
- \sum_{q=0}^{\infty} \sum_{r \in I_{0}} \frac{1}{(q+2)!} \frac{1}{r!} \psi_{0,q+2,r,0}
\delta^{(h_1k_1 \cdots h_{q+2}k_{q+2}j_1 \cdots j_r)} \mu_{h_1k_1} \cdots
\mu_{h_{q+2}k_{q+2}} \lambda_{j_1} \cdots \lambda_{j_r} + \\
+\sum_{r=0}^\infty \frac{1}{r!} (\lambda_a \lambda^a)^r \beta_r  \mu_{ik}(\lambda^i
\lambda^k - \lambda_b \lambda^b \delta^{ik}) + \sum_{r=0}^\infty \frac{(2r+3)!}{r!}
\beta_r \delta^{(ijj_1 \cdots j_{2r+2})}\mu_{ij} \lambda_{j_1} \cdots \lambda_{j_{2r+2}}
\, .
\end{eqnarray*}
But $\delta^{(ijj_1 \cdots j_{2r+2})}= \delta^{i(jj_1 \cdots j_{2r+2})}=
\frac{1}{2r+3}\left( \delta^{ij} \delta^{(j_1 \cdots j_{2r+2})} + (2r+2) \delta^{i(j_1
\cdots j_{2r+2})j} \right)$ so that the last 2 terms in the above expression are equal to
$\sum_{r=0}^\infty \frac{2r+3}{r!} (\lambda_a \lambda^a)^r \beta_r  \mu_{ik}\lambda^i
\lambda^k$. This allows to rewrite the total expression as
\begin{eqnarray}\label{fe.2}
\tilde{H}^{*02}=  -  \sum_{r \in I_{0}} \frac{1}{r!} \psi_{0,1,r,0}
\delta^{(ijj_1 \cdots j_r)} \mu_{ij} \lambda_{j_1} \cdots \lambda_{j_r} + \\
- \sum_{q=0}^{\infty} \sum_{r \in I_{0}} \frac{1}{(q+2)!} \frac{1}{r!} \psi_{0,q+2,r,0}
\delta^{(h_1k_1 \cdots h_{q+2}k_{q+2}j_1 \cdots j_r)} \mu_{h_1k_1} \cdots
\mu_{h_{q+2}k_{q+2}} \lambda_{j_1} \cdots \lambda_{j_r} +  \nonumber \\
+ \sum_{r=0}^\infty \frac{2r+3}{r!} (\lambda_a \lambda^a)^r \beta_r  \mu_{ik}\lambda^i
\lambda^k \, . \nonumber
\end{eqnarray}
Thanks to this result, eq. (\ref{eb.3}) now becomes
\begin{eqnarray}\label{ff.2}
H^{*0}=\sum_{r \in I_{0}} \frac{1}{r!}   \psi_{0,0,r,0} \delta^{(j_1 \cdots j_r)}
\lambda_{j_1} \cdots \lambda_{j_r} + \sum_{r=0}^\infty \frac{2r+3}{r!} (\lambda_a
\lambda^a)^r \beta_r  \mu_{ik}\lambda^i \lambda^k \, .
\end{eqnarray}
Let us resume now some of the conditions found on the coefficients. \\
Eq. (\ref{fe.1}) says that
\begin{eqnarray}\label{ff.3}
\phi_{1,q,r+1,0} = 2(r+1) \phi_{0,q+2,r,0} \, .
\end{eqnarray}
Eqs. (\ref{fd.1}) and Eqs. (\ref{fd.2}) give
\begin{eqnarray}\label{ff.3bis}
\phi'_{0,1,0,0} = 2 \psi_{0,1,0,0} \quad , \quad \phi'_{0,1,2R+2,0} = 2(2R+3)
\psi_{0,1,2R+2,0} -  \frac{(2R+3)!}{R!} 2 (2R+3)\beta_R \, .
\end{eqnarray}
Eq. (\ref{fc.1}), thanks also to Eq. (\ref{ff.3}) and Eq. $(\ref{ff.3bis})_1$ says that
\begin{eqnarray}\label{ff.4}
&{}& \psi_{0,q+2,0,0} = 0 \, ; \, \psi_{0,1,0,0} = 0 \, (\mbox{which can be compacted in}
\, \psi_{0,q+1,0,0} = 0) \, ; \\
&{}& \phi_{0,q+1,0,0} = 0 \, . \nonumber
\end{eqnarray}
Eq. (\ref{fa.3})
\begin{eqnarray}\label{ff.4bis}
&{}& \mbox{for} \quad r= 0 \quad \mbox{gives} \quad 2 \psi_{0,q+2,0,0} = \phi'_{0,q+2,0,0}  \, , \\
&{}& \mbox{with} \, r+1 \, \mbox{instead of} \, r, \, \mbox{gives} \quad 2(r+2)
\psi_{0,q+2,r+1,0} + 2(r+1)^2 \psi_{0,q+2,r,0} - (r+1) \phi'_{0,q+2,r+1,0} =0
 \, . \nonumber
\end{eqnarray}
Eq. (\ref{ef.3}) is
\begin{eqnarray}\label{ff.5}
\frac{\partial}{\partial \lambda} \psi_{0,q+1,R,0} =0 \, .
\end{eqnarray}
Eq. (\ref{ec.3}) with $R=r+1$ and with  aid by (\ref{ff.3}), is
\begin{eqnarray}\label{ff.5bis}
(2q+r+3) \vartheta_{1,q,r+1,0}  + 2  \lambda \vartheta_{1,q+1,r+1,0}  = 0 \, .
\end{eqnarray}
Eq. (\ref{eb.1}),  thanks to (\ref{ff.5}), says that
\begin{eqnarray}\label{ff.5ter}
\vartheta_{1,q,r+1,0} = 0  \, ,
\end{eqnarray}
which implies (\ref{ff.5bis}) as a particular case. \\
Eq. (\ref{ec.1}) is a consequence of $(\ref{ff.4})_{3,4}$, except for $Q=0$; in this case
it gives
\begin{eqnarray}\label{ff.6}
\psi_{0,0,0,0} = 0  \, .
\end{eqnarray}
Eq. (\ref{ee.1}) with $q+1$ instead of $q$ gives
\begin{eqnarray}\label{fl.1}
2(r+1) \psi_{0,q+2,r,0}  = \phi'_{0,q+2,r,0} \, ,
\end{eqnarray}
where (\ref{ff.5}) has been used.  \\
Eq. (\ref{ee.1}) with $q=0$, thanks to (\ref{ff.5}) and $(\ref{ff.3bis})_1$ gives
\begin{eqnarray*}
(r+1) \frac{\partial}{\partial \lambda} \psi_{0,0,r,0} + 2(r+1) \psi_{0,1,r,0}  =
\phi'_{0,1,r,0} \, ,
\end{eqnarray*}
which for $r=0$ gives again (\ref{ff.6}), while written with $2r+2$ instead of $r$ gives
\begin{eqnarray*}
\frac{\partial}{\partial \lambda} \psi_{0,0,2r+2,0}   = -2 \frac{(2r+3)!}{r!}  \beta_r
\, ,
\end{eqnarray*}
where we have used $(\ref{ff.3bis})_2$. By integrating the result, we obtain
\begin{eqnarray}\label{fl.2}
\psi_{0,0,2r+2,0}   = -2 \lambda \frac{(2r+3)!}{r!}  \beta_r + \psi_{0,0,2r+2,0,0} \, ,
\end{eqnarray}
where $\psi_{0,0,2r+2,0,0}$ is a constant arising from integration. \\
Eq. (\ref{ff.2}), thanks to (\ref{ff.6}) and (\ref{fl.2}) becomes
\begin{eqnarray}\label{fl.3}
H^{*0}=\sum_{r \in I_{0}} \frac{1}{(r+2)!}   \psi_{0,0,r,0,0} \delta^{(j_1 \cdots
j_{r+2})} \lambda_{j_1} \cdots \lambda_{j_{r+2}} - \sum_{r=0}^\infty 2 \lambda
\frac{2r+3}{r!} (\lambda_a \lambda^a)^{r+1} \beta_r + \\
+ \sum_{r=0}^\infty \frac{2r+3}{r!} (\lambda_a \lambda^a)^r \beta_r  \mu_{ik}\lambda^i
\lambda^k \, . \nonumber
\end{eqnarray}
It is interesting that this expressions determines $H^{*0}$ in terms of two arbitrary
sets of constants $\psi_{0,0,r,0,0}$ and $\beta_r$ and that it satisfies eq. (\ref{13.3}).\\
The doubt may arise that the other equations on the coefficients $\phi_{\cdots}$,
$\phi'_{\cdots}$, $\psi_{\cdots}$, are too much restrictive. The doubt may be eliminate
by direct analysis or, equivalently, noting that they appeared only in the present
section; only the $\vartheta_{\cdots}$ were present also in previous section and here we
have found that they are further restricted  by (\ref{ff.5ter}). This can be rewritten
also as
\begin{eqnarray}\label{fl.4}
\vartheta_{1,q,r,0} = 0  \, ,
\end{eqnarray}
because the sum of the first and third index must be an even number, so that obviously we
must have $r \geq 1$ in $\vartheta_{1,q,r,0}$. \\
Well, we may eliminate the above mentioned doubt by verifying directly that $H^{*0}$
given by (\ref{fl.3}) satisfies, for whatever value of the constants $\psi_{0,0,r,0,0}$
and $\beta_r$, the conditions described at the end of the previous section, that is
(\ref{17.3}), (\ref{bl.1}), (\ref{cd.2}), (\ref{da.11}) and (\ref{ea.1}). To this end we
may use only the restriction (\ref{fl.4}) on $\vartheta_{\cdots}$, jointly with those
found in the previous sections. \\

\subsection{Verifying the conditions on $H^{*0}$.}
\begin{itemize}
  \item It is to verify equation (\ref{17.3}), because $H^{*0}$ given by (\ref{fl.3}) is
  sum of a function not depending on $\lambda$ and of a function not depending on
  $\mu_{ij}$; moreover the right hand side of eq. (\ref{17.3}) is zero, thanks to
  (\ref{fl.4}).
\item It is to verify equation (\ref{bl.1}), because $H^{*0}$ given by (\ref{fl.3}) becomes zero when
calculated in $\lambda_i=0$.
\item It is to verify equation (\ref{cd.2}), thanks to  (\ref{fl.4}) and because $H^{*0}$ given by
(\ref{fl.3}) is linear in $\mu_{ij}=0$.
\item Let us verify equation (\ref{da.11}). By a substitution of $H^{*0}$ from
(\ref{fl.3}) it becomes
\end{itemize}
\begin{eqnarray*}
0= 2 \sum_{r=0}^\infty \frac{2r+3}{r!} \beta_r (\lambda_a \lambda^a)^r \lambda^k
\lambda^i + \lambda_i \frac{\partial}{\partial \lambda_k} \left[- \sum_{r=0}^\infty 2
\frac{2r+3}{r!} \beta_r (\lambda_a \lambda^a)^{r+1} \right] + \\
+ \delta^{ki} \left[- \sum_{r=0}^\infty 2 \frac{2r+3}{r!} \beta_r (\lambda_a
\lambda^a)^{r+1} \right]+ 2 \lambda_j \frac{\partial}{\partial \lambda_k} \left[
\sum_{r=0}^\infty  \frac{2r+3}{r!} \beta_r (\lambda_a \lambda^a)^{r} \lambda^i \lambda^j
\right] \, ,
\end{eqnarray*}
which is true because the sum of $1^{th}$ and $4^{th}$ term is equal to
$\frac{\partial}{\partial \lambda_k} \left[2 \sum_{r=0}^\infty  \frac{2r+3}{r!} \beta_r
(\lambda_a \lambda^a)^{r} \lambda^i \lambda^j \lambda_j \right] $, while the sum of
$2^{th}$ and $3^{th}$ term is equal to $\frac{\partial}{\partial \lambda_k} \left[ -
\lambda_i \sum_{r=0}^\infty 2 \frac{2r+3}{r!} \beta_r (\lambda_a \lambda^a)^{r+1}
\right]$.
\begin{itemize}
\item Let us verify equation (\ref{ea.1}). By a substitution of $H^{*0}$ from
(\ref{fl.3}) it becomes
\end{itemize}
\begin{eqnarray*}
0= \lambda_b \lambda^{[i} \frac{\partial}{\partial \lambda_k} \left[\sum_{r=0}^\infty
\frac{2r+3}{r!} \beta_r (\lambda_c \lambda^c)^{r} \lambda^{a]} \lambda^b \right] +
\lambda_b \delta^{k[i} \left[\sum_{r=0}^\infty \frac{2r+3}{r!} \beta_r (\lambda_c
\lambda^c)^{r} \lambda^{a]} \lambda^b \right] \, .
\end{eqnarray*}
In the first term, when we don' t take the derivative of $\lambda^a$ with respect to
$\lambda_k$, we obtain zero for the identity $\lambda^{[i} \lambda^{a]}=0$; when we take
the derivative of $\lambda^a$ with respect to $\lambda_k$, we obtain $\lambda_b
\lambda^{[i} \delta^{a]k} \sum_{r=0}^\infty \frac{2r+3}{r!} \beta_r (\lambda_c
\lambda^c)^{r} \lambda^b$ which is the opposite of the second term! \\
This completes our verification.

\section{Solution of the conditions on $\tilde{\tilde{H}}^i$. }
Let us firstly change unknown function, from $\tilde{\tilde{H}}^k$ to
$\tilde{\tilde{H}}^{*k}$ defined by
\begin{eqnarray}\label{fl.5}
&{}& \tilde{\tilde{H}}^k = \tilde{\tilde{H}}^{*k} + \\
&{}& + \sum_{r \in I_{0}} \frac{1}{(r+1)!} \left[ \lambda  \delta^{(kj_1 \cdots j_{r+1})}
- \frac{1}{2} \frac{r+3}{r+2} \delta^{(kijj_1 \cdots j_{r+1})} \mu_{ij} \right]
\psi_{0,0,r,0,0}  \lambda_{j_1} \cdots
\lambda_{j_{r+1}} + \nonumber \\
&{}& + \frac{\partial}{\partial \lambda_k} \left[ \sum_{r=0}^\infty \frac{2r+3}{r!}
\beta_r (\lambda_a \lambda^a)^r \left( \lambda  \mu_{bc} \lambda^b \lambda^c - \lambda^2
\lambda_b \lambda^b \right) \right] + \nonumber \\
&{}& -  \mu^{kd} \lambda_d (\mu_{bc} \lambda^b \lambda^c) \sum_{r=2}^\infty
\frac{2r+3}{r!} \beta_r (\lambda_a \lambda^a)^{r-1} - \frac{1}{4} \lambda^k (\mu_{bc}
\lambda^b \lambda^c)^2 \sum_{r=2}^\infty (2r-3) \frac{2r+3}{r!} \beta_r (\lambda_a
\lambda^a)^{r-2} + \nonumber \\
&{}& - \lambda^k (\mu_{bd} \mu_{dc} \lambda^b \lambda^c) \sum_{r=2}^\infty
\frac{2r+3}{r!} \beta_r (\lambda_a \lambda^a)^{r-1}  \, . \nonumber
\end{eqnarray}
By substituting $\tilde{\tilde{H}}^k$ from (\ref{fl.5}) and $H^{*0}$ from (\ref{fl.3}) in
(\ref{dd.2}) and (\ref{dl.1}) these equations are transformed respectively in
\begin{eqnarray}\label{fl.6}
\frac{\partial}{\partial \lambda} \tilde{\tilde{H}}^{*i}= 0  \, ,
\end{eqnarray}
\begin{eqnarray}\label{fl.7}
&{}& 0= 2 \mu_{ji} \left\{ \sum_{r=0}^\infty \frac{2r+3}{r!} \beta_r (\lambda_a
\lambda^a)^r \lambda^k \lambda^j \right\} +  2 \lambda \left\{ \sum_{r=0}^\infty
\frac{2r+3}{r!} \beta_r (\lambda_a \lambda^a)^r
\lambda^k \lambda^i \right\} + \quad \quad \quad \quad \quad \quad \quad \quad \\
&{}& + \lambda_i \left\{ \sum_{r \in I_{0}} \frac{1}{(r+1)!} \psi_{0,0,r,0,0}
\delta^{(kj_1
\cdots j_{r+1})}\lambda_{j_1} \cdots \lambda_{j_{r+1}} + \right. \nonumber \\
&{}& + \left. \frac{\partial}{\partial \lambda_k} \left[ \sum_{r=0}^\infty
\frac{2r+3}{r!} \beta_r (\lambda_a \lambda^a)^r \left( \mu_{bc} \lambda^b \lambda^c - 2
\lambda \lambda_b \lambda^b \right) \right] \right\} + \nonumber
\end{eqnarray}
\begin{eqnarray*}
+ \delta^{ki} \left[ \sum_{r \in I_{0}} \frac{1}{(r+2)!} \psi_{0,0,r,0,0} \delta^{(j_1
\cdots j_{r+2})}\lambda_{j_1} \cdots \lambda_{j_{r+2}} +\sum_{r=0}^\infty \frac{2r+3}{r!}
\beta_r (\lambda_a \lambda^a)^r \left( \mu_{bc} \lambda^b \lambda^c - 2 \lambda \lambda_b
\lambda^b \right) \right] +
\end{eqnarray*}
\begin{eqnarray*}
&{}& + 2 \lambda_j \left\{ \frac{\partial \tilde{\tilde{H}}^{*k}}{\partial \mu_{ij}} -
\frac{1}{2} \sum_{r \in I_{0}} \frac{r+3}{(r+2)!} \psi_{0,0,r,0,0} \delta^{(kijj_1 \cdots
j_{r+1})}\lambda_{j_1} \cdots \lambda_{j_{r+1}} + \right. \quad \quad \quad \quad \quad \quad \quad \quad \quad \quad \quad \quad \quad \\
&{}& + \frac{\partial}{\partial \lambda_k} \left[ \sum_{r=0}^\infty \frac{2r+3}{r!}
\beta_r
(\lambda_a \lambda^a)^r \lambda \lambda^i \lambda^j \right] + \\
&{}& - \sum_{r=2}^\infty \frac{2r+3}{r!} \beta_r (\lambda_a \lambda^a)^{r-1}  \left[
\lambda^i \lambda^j  \mu^{kd} \lambda_d + (\mu_{bc} \lambda^b \lambda^c) \delta^{k(i}
\lambda^{j)}
\right]  +  \\
&{}& - \frac{1}{4} \sum_{r=2}^\infty (2r-3) \frac{2r+3}{r!} \beta_r (\lambda_a
\lambda^a)^{r-2} \, 2 (\mu_{bc} \lambda^b \lambda^c) \lambda^i \lambda^j \lambda^k + \\
&{}& - \left.  \sum_{r=2}^\infty  \frac{2r+3}{r!} \beta_r (\lambda_a \lambda^a)^{r-1}
\lambda^k (\lambda^i \mu^{jb} \lambda_b + \lambda^j \mu^{ib} \lambda_b) \right\} \, .
\end{eqnarray*}
Now, in eq. (\ref{fl.7}), the second and third term containing $\lambda$ can be written
together as \\
$\frac{\partial}{\partial \lambda_k} \left[ \sum_{r=0}^\infty \frac{2r+3}{r!} \beta_r
(\lambda_a \lambda^a)^r ( -2\lambda ) \lambda_i (\lambda^b  \lambda_b) \right]$, \\
while the first and fourth term can be written
together as \\
$2 \frac{\partial}{\partial \lambda_k} \left\{ \left[ \sum_{r=0}^\infty \frac{2r+3}{r!}
\beta_r (\lambda_a \lambda^a)^r \lambda \lambda^i \lambda^j \right] \lambda_j \right\}$; \\
consequently the terms containing $\lambda$ elide each other. This result could be
expected for eq. (\ref{fl.6}). \\
Furthermore,  the coefficient of $\psi_{0,0,r,0,0}$ in the third  term containing it, can
be written as
\begin{eqnarray*}
- \sum_{r \in I_{0}} \frac{r+3}{(r+2)!}  \delta^{i(kjj_1 \cdots j_{r+1})} \lambda_{j}
\lambda_{j_1} \cdots \lambda_{j_{r+1}} = - \sum_{r \in I_{0}} \frac{r+3}{(r+2)!}
\delta^{i(kj_1 \cdots j_{r+2})}\lambda_{j_1} \cdots \lambda_{j_{r+2}} = \\
= - \sum_{r \in I_{0}} \frac{1}{(r+2)!} \left[ \delta^{ik} \delta^{(j_1 \cdots j_{r+2})}
+ (r+2) \delta^{i(j_1 \cdots j_{r+2})k} \right] \lambda_{j_1} \cdots \lambda_{j_{r+2}}=
\\
= - \sum_{r \in I_{0}} \frac{1}{(r+2)!} \left[ \delta^{ik} \delta^{(j_1 \cdots j_{r+2})}
\lambda_{j_1} \cdots \lambda_{j_{r+2}} + (r+2) \lambda^i \delta^{(j_1 \cdots j_{r+1})k}
\lambda_{j_1} \cdots \lambda_{j_{r+1}} \right] \, ,
\end{eqnarray*}
so that it and the other terms in eq. (\ref{fl.7}) containing $\psi_{0,0,r,0,0}$ elide
each other. After that, of eq. (\ref{fl.7}) there remains
\begin{eqnarray*}
&{}& 0= 2 \mu_{ji} \sum_{r=0}^\infty \frac{2r+3}{r!} \beta_r (\lambda_a \lambda^a)^r
\lambda^k \lambda^j  + \lambda_i \frac{\partial}{\partial \lambda_k} \left[
\sum_{r=0}^\infty \frac{2r+3}{r!} \beta_r (\lambda_a \lambda^a)^r ( \mu_{bc} \lambda^b
\lambda^c) \right] + \\
&{}& + \delta^{ki} \sum_{r=0}^\infty \frac{2r+3}{r!} \beta_r (\lambda_a \lambda^a)^r (
\mu_{bc} \lambda^b \lambda^c) +
 2 \lambda_j \frac{\partial \tilde{\tilde{H}}^{*k}}{\partial \mu_{ij}} + \\
&{}& - \sum_{r=2}^\infty \frac{2r+3}{r!} \beta_r (\lambda_a \lambda^a)^{r-1}  \left[ 2
(\lambda_c \lambda^c) \lambda^i  \mu^{kd} \lambda_d + (\mu_{bc} \lambda^b \lambda^c)
(\lambda_d \lambda^d)  \delta^{ki} + (\mu_{bc} \lambda^b \lambda^c) \lambda^k \lambda^i
\right]  +  \\
&{}& - \sum_{r=2}^\infty (2r-3) \frac{2r+3}{r!} \beta_r (\lambda_a
\lambda^a)^{r-2} (\mu_{bc} \lambda^b \lambda^c) (\lambda_d \lambda^d) \lambda^i  \lambda^k + \\
&{}& - 2 \sum_{r=2}^\infty  \frac{2r+3}{r!} \beta_r (\lambda_a \lambda^a)^{r-1} \lambda^k
\left[  \lambda^i  (\mu_{bc} \lambda^b \lambda^c) +  (\lambda_d \lambda^d) \mu^{ib}
\lambda_b) \right] \, ,
\end{eqnarray*}
or,
\begin{eqnarray*}
&{}& 0=  2 \lambda_j \frac{\partial \tilde{\tilde{H}}^{*k}}{\partial \mu_{ij}} + \\
&{}& + \delta^{ki} \left[ \sum_{r=0}^\infty \frac{2r+3}{r!} \beta_r (\lambda_a
\lambda^a)^r ( \mu_{bc} \lambda^b \lambda^c) - \sum_{r=2}^\infty \frac{2r+3}{r!} \beta_r
(\lambda_a \lambda^a)^r ( \mu_{bc} \lambda^b \lambda^c) \right] + \\
&{}& + \lambda^k \mu^{ij} \lambda_j \left[ 2 \sum_{r=0}^\infty \frac{2r+3}{r!} \beta_r
(\lambda_a \lambda^a)^r  - 2 \sum_{r=2}^\infty \frac{2r+3}{r!} \beta_r (\lambda_a
\lambda^a)^r \right] + \\
&{}& + \lambda^i \mu^{kj} \lambda_j \left[ 2 \sum_{r=0}^\infty \frac{2r+3}{r!} \beta_r
(\lambda_a \lambda^a)^r  - 2 \sum_{r=2}^\infty \frac{2r+3}{r!} \beta_r (\lambda_a
\lambda^a)^r \right] + \\
&{}& + \lambda^i \lambda^k \left[ 2 \sum_{r=0}^\infty r \frac{2r+3}{r!} \beta_r
(\lambda_a \lambda^a)^{r-1} ( \mu_{bc} \lambda^b \lambda^c) - \sum_{r=2}^\infty
\frac{2r+3}{r!} \beta_r (\lambda_a \lambda^a)^{r-1} ( \mu_{bc} \lambda^b \lambda^c) + \right. \\
&{}& - \left.  \sum_{r=2}^\infty (2r-3) \frac{2r+3}{r!} \beta_r (\lambda_a
\lambda^a)^{r-1} ( \mu_{bc} \lambda^b \lambda^c) - 2 \sum_{r=2}^\infty \frac{2r+3}{r!}
\beta_r (\lambda_a \lambda^a)^{r-1} ( \mu_{bc} \lambda^b \lambda^c) \right] \, .
\end{eqnarray*}
But in this expression, all the terms with $r \geq 2$ elide each other, so that there
remain only the terms with $r=0,1$, that is
\begin{eqnarray}\label{fl.8}
0=  2 \lambda_j \frac{\partial \tilde{\tilde{H}}^{*k}}{\partial \mu_{ij}} + \left[
\delta^{ki} ( \mu_{bc} \lambda^b \lambda^c) + 4 \lambda^{(k} \mu^{i)j} \lambda_j \right]
\left( 3 \beta_0 + 5 \beta_1 \lambda_a \lambda^a \right) + 10 \lambda^i \lambda^k (
\mu_{bc} \lambda^b \lambda^c) \beta_1 \, .
\end{eqnarray}
A further refinement of the situation can be obtained with another  change of unknown
function, from $\tilde{\tilde{H}}^{*k}$ to $\tilde{\tilde{H}}^{**k}$ defined by
\begin{eqnarray}\label{fl.9}
&{}& \tilde{\tilde{H}}^{*k}= \tilde{\tilde{H}}^{**k} - \frac{5}{4} \beta_1 \left\{ 4 (
\mu_{bc} \lambda^b \lambda^c)  \mu^{kd} \lambda_d + \lambda^k \left[ (\mu_{bc} \mu^{bc})
(\lambda_a \lambda^a) + 2 \lambda^a \lambda^b  \mu_{ac} \mu_{cb} \right] \right\}
  \, .
\end{eqnarray}
By using this, eq. (\ref{fl.8}) becomes
\begin{eqnarray}\label{fl.10}
0=  2 \lambda_j \frac{\partial \tilde{\tilde{H}}^{**k}}{\partial \mu_{ij}} + 3 \beta_0
\left[ \delta^{ki} ( \mu_{bc} \lambda^b \lambda^c) + 4 \lambda^{(k} \mu^{i)j} \lambda_j
\right]  \, .
\end{eqnarray}
We can now prove that, as a consequence of this equation, we have
\begin{eqnarray}\label{fl.11}
\beta_0 =0 \, .
\end{eqnarray}

\subsection{Solution of the conditions on $\tilde{\tilde{H}}^{**k}$. }

Let us consider the Taylor expansion of $\tilde{\tilde{H}}^{**k}$ around the state with
$\mu^{ij}=0$; eq. (\ref{fl.10}) at the order 1 with respect to this state is
\begin{eqnarray}\label{fl.12}
0=  2 \lambda_j \frac{\partial \tilde{\tilde{H}}_2^{**k}}{\partial \mu_{ij}} + 3 \beta_0
\left[ \delta^{ki} ( \mu_{bc} \lambda^b \lambda^c) + 4 \lambda^{(k} \mu^{i)j} \lambda_j
\right]  \, ,
\end{eqnarray}
where $\tilde{\tilde{H}}_2^{**k}$ is the homogeneous part of $\tilde{\tilde{H}}^{**k}$ of
second degree with respect to $\mu^{ij}$. Thanks to the Representation Theorems, it has
the form
\begin{eqnarray*}
\tilde{\tilde{H}}_2^{**k} =  f_1(G_0) \mu^{ka} \mu^{ab} \lambda_b + f_2(G_0) \mu^{ll}
\mu^{ka} \lambda_a  + f_3(G_0) ( \mu_{bc} \lambda^b \lambda^c) \mu^{ka} \lambda_a + \quad
\quad \quad \quad \quad \quad \quad \quad \quad \quad \quad \quad \quad
\\
+ \lambda^k \left[ f_4(G_0) (\mu^{ll})^2 + f_5(G_0) ( \mu_{bc} \lambda^b \lambda^c)^2 +
f_6(G_0) ( \mu_{bc} \lambda^b \lambda^c) \mu^{ll} + f_7(G_0) \mu_{bc} \mu^{bc} + f_8(G_0)
 \mu_{ba}  \mu_{ac} \lambda^b \lambda^c) \right] \, ,
\end{eqnarray*}
where $G_0=\lambda_a \lambda^a$. By substituting this in (\ref{fl.12}), we obtain
\begin{eqnarray*}
0= 2 \lambda_j \left\{ f_1 \delta^{k(i} \mu^{j)b} \lambda_b + f_1 \mu^{k(i} \lambda^{j)}
+ f_2 \delta^{ij} \mu^{kb} \lambda_b + f_2 \mu^{ll} \delta^{k(i} \lambda^{j)} + f_3
\lambda^i \lambda^j \mu^{kb} \lambda_b + f_3  ( \mu_{bc} \lambda^b \lambda^c)
\delta^{k(i}
\lambda^{j)} + \right. \\
+ \lambda^k \left.  \left[ 2 f_4 \mu^{ll} \delta^{ij}  + 2 f_5 ( \mu_{bc} \lambda^b
\lambda^c) \lambda^i \lambda^j + f_6  \lambda^i \lambda^j \mu^{ll} + f_6 ( \mu_{bc}
\lambda^b \lambda^c) \delta^{ij} + 2 f_7 \mu^{ij} + 2 f_8 \lambda^{(i} \mu^{j)b}
\lambda_b \right] \right\} + \\
+ 3 \beta_0 \left[ \delta^{ki} ( \mu_{bc} \lambda^b \lambda^c) + 4 \lambda^{(k} \mu^{i)j}
\lambda_j \right] \, ,
\end{eqnarray*}
that is,
\begin{eqnarray*}
0=  f_1 \delta^{ki} ( \mu_{bc} \lambda^b \lambda^c) + 2f_1 \lambda^{(k} \mu^{i)b}
\lambda_b + f_1 \mu^{ki} G_0 +2 f_2 \lambda^{i } \mu^{kb} \lambda_b + f_2 \mu^{ll}
\delta^{ki} G_0 + f_2 \mu^{ll} \lambda^k \lambda^i + \\
+ 2f_3 G_0 \lambda^i \mu^{kb} \lambda_b + f_3  ( \mu_{bc} \lambda^b \lambda^c)
\delta^{ki} G_0 +
f_3  ( \mu_{bc} \lambda^b \lambda^c)  \lambda^k \lambda^i + \\
+ \lambda^k   \left[ 4 f_4 \mu^{ll} \lambda^i  + 4 f_5 ( \mu_{bc} \lambda^b \lambda^c)
G_0 \lambda^i  + 2 f_6  G_0 \lambda^i \mu^{ll} + 2 f_6 ( \mu_{bc} \lambda^b \lambda^c)
\lambda^i + 4 f_7 \mu^{ij} \lambda_j + 2 f_8 \lambda^{i} ( \mu_{bc} \lambda^b \lambda^c)
+ \right. \\
+ \left. 2 f_8 G_0 \mu^{ib} \lambda_b \right] + 3 \beta_0 \left[ \delta^{ki} ( \mu_{bc}
\lambda^b \lambda^c) + 4 \lambda^{(k} \mu^{i)j} \lambda_j \right] \, .
\end{eqnarray*}
The skew-symmetric part of this relation, with respect to $i$ and $k$ is
\begin{eqnarray*}
0=  \lambda^{[k}  \mu^{i]b} \lambda_b (-2f_2 - 2f_3 G_0 + 4 f_7 + 2 f_8 G_0)
 \, ,
\end{eqnarray*}
from which we obtain
\begin{eqnarray}\label{fl.13}
4 f_7 = 2f_2 + 2f_3 G_0 -2 f_8 G_0 \, .
\end{eqnarray}
By taking into account this value of $f_7$, the remaining part of our condition becomes
\begin{eqnarray*}
0=  \delta^{ki} \left[ f_2 \mu^{ll} G_0 +( f_1 + f_3 G_0 +3 \beta_0) ( \mu_{bc} \lambda^b
\lambda^c) \right] + f_1 \mu^{ki} G_0 + \\
+ \lambda^{(k} \mu^{i)b} \lambda_b ( 2 f_1 + 4f_2+4f_3G_0+ 12 \beta_0) + \\
+ \lambda^k \lambda^i \left[ \mu^{ll} (f_2 + 4f_4+2f_6G_0)+ ( \mu_{bc} \lambda^b
\lambda^c) ( f_3 +4f_5G_0+ 2f_6  +2 f_8 ) \right]
 \, .
\end{eqnarray*}
In this relation, the coefficients of $\mu^{ki}$ and $ \delta^{ki} \mu^{ll}$ give
respectively $f_1=0$ and $f_2=0$. After that, the coefficient of $ \delta^{ki} ( \mu_{bc}
\lambda^b \lambda^c)$ gives $f_3 G_0 + 3 \beta_0=0$, which calculated in $\lambda_j=0$
gives the above mentioned eq. (\ref{fl.11}). \\
This result transforms eq. (\ref{fl.10}) into
\begin{eqnarray}\label{fl.14}
0=  2 \lambda_j \frac{\partial \tilde{\tilde{H}}^{**k}}{\partial \mu_{ij}}  \, .
\end{eqnarray}
Now we proceed to find the general solution of this last equation and we prove that it is
\begin{eqnarray}\label{fl.15}
\tilde{\tilde{H}}^{**k}=  \lambda^k F(G_0 \, , \, G_1 \, , \, G_2)  \, ,
\end{eqnarray}
where
\begin{eqnarray*}
G_1= G_0  \delta^{bc} \mu_{bc} - \mu_{bc} \lambda^b \lambda^c \, , \quad G_2= G_0
\mu^{bc} \mu_{bc} - 2 \mu_{bc} \mu_{ca} \lambda^b \lambda^a + 2 (\delta^{bc} \mu_{bc})
(\mu_{bc} \lambda^b \lambda^c) - G_0 (\delta^{bc} \mu_{bc})^2
\end{eqnarray*}
and $F$ is an arbitrary function of its variables. \\
In fact, if $\lambda_j=0$, from the Representation Theorems we know that
$\tilde{\tilde{H}}^{**k}=0$, just as in (\ref{fl.15}) and (\ref{fl.14}) is an identity.
\\
If $\lambda_j \neq 0$, we can define the projector into the subspace orthogonal to
$\lambda_j$, that is
\begin{eqnarray}\label{lb.4}
h^{ij}= \delta^{ij} - \frac{1}{G_0} \lambda^i \lambda^j \, ,
\end{eqnarray}
from which it follows $h^{ij} \lambda_j =0$, as obvious. By taking as independent
variables \\
$\lambda^i$, $\tilde{\mu}= \mu_{bc} \lambda^b \lambda^c$, $\tilde{\mu}^i= h^{ij} \mu_{ja}
\lambda^a$, $\tilde{\mu}^{ij}= h^{ia} \mu_{ab } h^{bj}$, \\
eq. (\ref{fl.14}) becomes
\begin{eqnarray*}
0=  2 \lambda_j \left( \frac{\partial \tilde{\tilde{H}}^{**k}}{\partial \tilde{\mu}}
\lambda^i \lambda^j + \frac{\partial \tilde{\tilde{H}}^{**k}}{\partial \tilde{\mu}^b}
h^{b(j} \lambda^{i)} + \frac{\partial \tilde{\tilde{H}}^{**k}}{\partial
\tilde{\mu}^{ab}} h^{a(i} h^{j)b} \right) = \\
= 2 G_0 \frac{\partial \tilde{\tilde{H}}^{**k}}{\partial \tilde{\mu}} \lambda^i
 + \frac{\partial \tilde{\tilde{H}}^{**k}}{\partial \tilde{\mu}^b} G_0 h^{bi}
 \, .
\end{eqnarray*}
By contracting this relation with $\lambda_i$ and with $h_{ia}$, we obtain   respectively
\begin{eqnarray}\label{lc.1}
\frac{\partial \tilde{\tilde{H}}^{**k}}{\partial \tilde{\mu}} =0 \quad , \quad
\frac{\partial \tilde{\tilde{H}}^{**k}}{\partial \tilde{\mu}^a}=0  \, .
\end{eqnarray}
It follows that $\tilde{\tilde{H}}^{**k}$ may depend only on $ \lambda^i$ and
$\tilde{\mu}^{ij}$. But $\tilde{\mu}^{ij} \lambda_j=0$ so that, for the Representation
Theorems we have that $\tilde{\tilde{H}}^{**k}$ is proportional to $\lambda^k$ as in eq.
(\ref{fl.15}); moreover, the coefficient $F$ can be a scalar function of $G_0$, $Q_1=
\delta_{ij} \tilde{\mu}^{ij}$, $Q_2= \delta_{ij}\tilde{\mu}^{ia}\tilde{\mu}^{aj}$. \\
Now we have
\begin{eqnarray*}
Q_1= \delta_{ij} \tilde{\mu}^{ij}= h_{ij} \mu^{ij}=  \delta_{ij} \mu^{ij} -
\frac{1}{G_0} \lambda_i \lambda_j  \mu^{ij} \, , \\
Q_2= \delta_{ij} h^{ib} \mu_{bc} h^{ca}  h^{ad} \mu_{de}   h^{ej} = h_{be} \mu^{bc}
h_{cd} \mu^{de} = \delta_{be} \mu^{ec} \delta_{cd} \mu^{de} - \frac{2}{G_0} \delta_{be}
\mu^{bc} \lambda_c \lambda_d \, \mu^{de}  + \left( \frac{1}{G_0} \right)^2 \left(\mu_{bc}
\lambda^b \lambda^c\right)^2 \, .
\end{eqnarray*}
But an arbitrary function of $G_0$, $Q_1$, $Q_2$ is also an arbitrary function of $G_0$,
$Q_1$ and of
\begin{eqnarray*}
Q_2 - (Q_1)^2 = \mu^{ec} \mu_{ec} - \frac{2}{G_0} \mu^{de} \mu^{ec} \lambda_d \lambda_c
+\frac{2}{G_0} (\delta_{ij} \mu^{ij}) (\mu_{ab} \lambda^a \lambda^b) -  (\delta_{ij}
\mu^{ij})^2
\end{eqnarray*}
and an arbitrary function of $G_0$, $Q_1$, $Q_2- (Q_1)^2 $ is also an arbitrary function
of \\
$G_0$, $Q_1G_0$, $[Q_2- (Q_1)^2] G_0 $ and this completes the proof of eq.
$(\ref{fl.15})_{2,3}$. These last passages have been done with the end to have a function
defined also in $\lambda^j=0$, without going too much far from equilibrium.

\section{Conclusions.}
We can collect now all our results. By substituting $\tilde{\tilde{H}}_2^{**k}$ from
(\ref{fl.15}) into (\ref{fl.9}) we obtain $\tilde{\tilde{H}}^{*k}$; by substituting this
and (\ref{fl.11}) into (\ref{fl.5}), we obtain

\begin{eqnarray}\label{lc.2}
 \tilde{\tilde{H}}^k = \lambda^k F(G_0 \, , \, G_1 \, , \, G_2) - \frac{5}{4} \beta_1
\left\{ 4 ( \mu_{bc} \lambda^b \lambda^c)  \mu^{kd} \lambda_d + \lambda^k \left[
(\mu_{bc} \mu^{bc}) (\lambda_a \lambda^a) + 2 \lambda^a \lambda^b \mu_{ac} \mu_{cb}
\right] \right\} +
\end{eqnarray}
\begin{eqnarray*}
&{}& + \sum_{r \in I_{0}} \frac{1}{(r+1)!} \left[ \lambda \delta^{(kj_1 \cdots j_{r+1})}
- \frac{1}{2} \frac{r+3}{r+2} \delta^{(kijj_1 \cdots j_{r+1})} \mu_{ij} \right]
\psi_{0,0,r,0,0}  \lambda_{j_1} \cdots
\lambda_{j_{r+1}} + \\
&{}& + \frac{\partial}{\partial \lambda_k} \left[ \sum_{r=1}^\infty \frac{2r+3}{r!}
\beta_r (\lambda_a \lambda^a)^r \left( \lambda  \mu_{bc} \lambda^b \lambda^c - \lambda^2
\lambda_b \lambda^b \right) \right] +  \\
&{}& -  \mu^{kd} \lambda_d (\mu_{bc} \lambda^b \lambda^c) \sum_{r=2}^\infty
\frac{2r+3}{r!} \beta_r (\lambda_a \lambda^a)^{r-1} - \frac{1}{4} \lambda^k (\mu_{bc}
\lambda^b \lambda^c)^2 \sum_{r=2}^\infty (2r-3) \frac{2r+3}{r!} \beta_r (\lambda_a
\lambda^a)^{r-2} +  \\
&{}& - \lambda^k (\mu_{bd} \mu_{dc} \lambda^b \lambda^c) \sum_{r=2}^\infty
\frac{2r+3}{r!} \beta_r (\lambda_a \lambda^a)^{r-1}  \, .
\end{eqnarray*}
Thanks to this expression and of (\ref{fl.3}), taking also into account (\ref{fl.11}), we
can rewrite the expression for  $\tilde{\tilde{H}}$ in (\ref{db.2}); finally, we can
substitute this new expression and that of (\ref{fl.3}) for $H^{*0}$ in (\ref{da.2}). In
this way we obtain
\begin{eqnarray}\label{lc.3}
\Delta H = \sum_{p,q,s}^{0 \cdots \infty} \sum_{r \in I_{p}} \frac{1}{p!} \frac{1}{q!}
\frac{1}{r!} \frac{1}{(s+1)!} \vartheta_{p,q,r,s}(\lambda) \mu^{s+1} \delta^{(i_1 \cdots
i_{p}h_1k_1 \cdots h_qk_qj_1 \cdots j_r)} \\
\mu_{i_1} \cdots \mu_{i_{p}}\mu_{h_1k_1} \cdots \mu_{h_qk_q} \lambda_{j_1} \cdots
\lambda_{j_r}   + \nonumber
\end{eqnarray}
\begin{eqnarray*}
+ \mu \left\{\sum_{r \in I_{0}} \frac{1}{(r+2)!}   \psi_{0,0,r,0,0} \delta^{(j_1 \cdots
j_{r+2})} \lambda_{j_1} \cdots \lambda_{j_{r+2}} - \sum_{r=1}^\infty 2 \lambda
\frac{2r+3}{r!} (\lambda_a \lambda^a)^{r+1} \beta_r + \right.\\
+ \left. \sum_{r=1}^\infty \frac{2r+3}{r!} (\lambda_a \lambda^a)^r \beta_r
\mu_{ik}\lambda^i \lambda^k \right\} +
\end{eqnarray*}
\begin{eqnarray*}
+ \sum_{p,q}^{0 \cdots \infty} \sum_{r \in I_{p}} \frac{1}{(p+2)!} \frac{1}{q!}
\frac{1}{r!} \vartheta_{p,q+1,r,0} \delta^{(i_1 \cdots
i_{p+2}h_1k_1 \cdots h_qk_qj_1 \cdots j_r)} \\
\mu_{i_1} \cdots \mu_{i_{p+2}}\mu_{h_1k_1} \cdots \mu_{h_qk_q} \lambda_{j_1} \cdots
\lambda_{j_r}  + \frac{1}{2} \mu_{i} \mu_{j} \sum_{r=1}^\infty \frac{2r+3}{r!} (\lambda_a
\lambda^a)^r \beta_r  \lambda^i \lambda^j +   \\
+ \mu_i \left\{ \lambda^i F(G_0 \, , \, G_1 \, , \, G_2) - \frac{5}{4} \beta_1 \left[ 4 (
\mu_{bc} \lambda^b \lambda^c)  \mu^{id} \lambda_d + \lambda^i \left( (\mu_{bc} \mu^{bc})
(\lambda_a \lambda^a) + 2 \lambda^a \lambda^b \mu_{ac} \mu_{cb} \right) \right] \right. +
\end{eqnarray*}
\begin{eqnarray*}
&{}& + \sum_{r \in I_{0}} \frac{1}{(r+1)!} \left[ \lambda \delta^{(ij_1 \cdots j_{r+1})}
- \frac{1}{2} \frac{r+3}{r+2} \delta^{(ikjj_1 \cdots j_{r+1})} \mu_{kj} \right]
\psi_{0,0,r,0,0}  \lambda_{j_1} \cdots
\lambda_{j_{r+1}} + \\
&{}& + \frac{\partial}{\partial \lambda_i} \left[ \sum_{r=1}^\infty \frac{2r+3}{r!}
\beta_r (\lambda_a \lambda^a)^r \left( \, \lambda \, \, \mu_{bc} \lambda^b \lambda^c -
\lambda^2
\lambda_b \lambda^b \right) \right] +  \\
&{}& -  \mu^{id} \lambda_d (\mu_{bc} \lambda^b \lambda^c) \sum_{r=2}^\infty
\frac{2r+3}{r!} \beta_r (\lambda_a \lambda^a)^{r-1} - \frac{1}{4} \lambda^i (\mu_{bc}
\lambda^b \lambda^c)^2 \sum_{r=2}^\infty (2r-3) \frac{2r+3}{r!} \beta_r (\lambda_a
\lambda^a)^{r-2} +  \\
&{}& \left. - \lambda^i (\mu_{bd} \mu_{dc} \lambda^b \lambda^c) \sum_{r=2}^\infty
\frac{2r+3}{r!} \beta_r (\lambda_a \lambda^a)^{r-1} \right\}  + \tilde{\tilde{H}}^0
(\mu_{ab}, \lambda , \lambda_c) \, .
\end{eqnarray*}
We recall that in this expression, $F(G_0 \, , \, G_1 \, , \, G_2)$ is an arbitrary
function, $\psi_{0,0,r,0,0}$ and $\beta_r$ are arbitrary constants, while
$\vartheta_{p,q,r,s}(\lambda) $ are constrained by (\ref{beta.2}), (\ref{beta.3}),
(\ref{beta.4}), (\ref{fl.4}), (\ref{13.1biss}) and (\ref{13.2}). The presence of the
arbitrary function $\tilde{\tilde{H}}^0(\mu_{ab}, \lambda , \lambda_c)$  is obvious
because it is not constrained by  (\ref{11.9})  because it doesn' t depend on $\mu$, nor
on $\mu_k$. Consequently, it is not necessary to impose the condition
$\tilde{\tilde{H}}^0(0_{ab}, \lambda , 0_c)=0$ which comes out from eqs. (\ref{11.10}),
(\ref{13.2}) and (\ref{lc.3}).\\
The sum of the expression (\ref{lc.3}) for $\Delta H$ and of the expression (\ref{11.6})
for $H_1$ give the general solution for the unknown function $H$. Let us substitute it in
the equations
\begin{eqnarray}\label{lc.4}
&{}&   F^{kij} = \frac{\partial^2 H }{\partial  \mu_k \partial \mu_{ij}  } \quad , \quad
G^{ki} = \frac{\partial^2 H }{\partial  \mu_k \partial \lambda_i}
\end{eqnarray}
which are a subset of the equations (\ref{2bis.3}). We obtain that  $F^{kij}$ is sum of a
symmetric tensor and of
\begin{eqnarray}\label{lc.5}
\Delta   F^{kij} = \lambda^k \frac{\partial F(G_0 \, , \, G_1 \, , \, G_2) }{\partial
\mu_{ij}  } - \frac{5}{4} \beta_1 \left[  4 ( \mu_{bc} \lambda^b \lambda^c) \delta^{k(i}
\lambda^{j)} + \lambda^k \left( 2 \mu^{ij} (\lambda_a \lambda^a) - 4 \lambda^{(i}
\mu^{j)b} \lambda_b \right) \right]  +
\end{eqnarray}
\begin{eqnarray*}
&{}& + 2  \sum_{r=1}^\infty \frac{2r+3}{r!} \beta_r (\lambda_a \lambda^a)^r \, \lambda \,
\delta^{k(i}  \lambda^{j)}  - \delta^{k(i} \lambda^{j)} (\mu_{bc} \lambda^b \lambda^c)
\sum_{r=2}^\infty \frac{2r+3}{r!} \beta_r (\lambda_a \lambda^a)^{r-1} \, .
\end{eqnarray*}
Similarly, $G^{ki}$ is sum of a symmetric tensor and of
\begin{eqnarray}\label{lc.6}
\Delta   G^{ki} =
 \sum_{r=1}^\infty \frac{2r+3}{r!} (\lambda_a
\lambda^a)^r \beta_r  \lambda^k \mu^{i} + \lambda^k \frac{\partial F }{\partial G_1}
\frac{\partial G_1}{\partial \lambda_i} + \lambda^k \frac{\partial F }{\partial G_2}
\frac{\partial G_2}{\partial \lambda_i}
 - 5 \beta_1  \lambda^k  \mu^{ic} \mu^{cb}   \lambda_b   +
\end{eqnarray}
\begin{eqnarray*}
&{}& - 2 \lambda^i \mu^{kd} \lambda_d (\mu_{bc} \lambda^b \lambda^c) \sum_{r=2}^\infty
\frac{2r+3}{r!} (r-1) \beta_r (\lambda_a \lambda^a)^{r-2} +   \\
&{}&  - \lambda^k \mu^{id} \lambda_d (\mu_{bc} \lambda^b \lambda^c) \sum_{r=2}^\infty
(2r-3) \frac{2r+3}{r!} \beta_r (\lambda_a \lambda^a)^{r-2} - 2 \lambda^k \mu^{id}
\mu^{dc}  \lambda_c \sum_{r=2}^\infty \frac{2r+3}{r!} \beta_r (\lambda_a \lambda^a)^{r-1}
\, .
\end{eqnarray*}
But, from $(\ref{fl.15})_{2,3}$ we see that $\frac{\partial G_1}{\partial \mu_{ij}}$ is a
tensor at least of second order with respect to equilibrium and $\frac{\partial
G_2}{\partial \mu_{ij}}$ is a tensor at least of third order with respect to equilibrium.
Consequently, from (\ref{lc.5}) it is clear that $F^{kij}$ up to second order with
respect to equilibrium is a symmetric tensor; its eventual non symmetric parts may appear
only from the third order with respect to equilibrium. This result is different from its
counterpart in \cite{1} where a non symmetric part appeared also at first order with
respect to equilibrium. We shall see in Appendix 2 that from the equations of that paper
it follows that this non symmetric part is proportional to a constant; consequently, here
we have proved that this constant is zero and that this further constraint follows by
imposing the equations up to  order higher than one with respect to equilibrium. This is
not a problem, because the authors of paper \cite{1} assumed (for example in the first 3
lines of subsection 7.2) that the integration constants vanish and furnished reasons for
this assumption based on the kinetic theory approach.  \\
Similarly, from $(\ref{fl.15})_{2,3}$ we see that $\frac{\partial G_1}{\partial
\lambda_i}$ is a tensor  of second order with respect to equilibrium and $\frac{\partial
G_1}{\partial \lambda_i}$ is a tensor at least of third order with respect to
equilibrium. Consequently, from (\ref{lc.6}) it is clear that $G^{ki}$ up to second order
with respect to equilibrium is a symmetric tensor; its eventual non symmetric parts may
appear only from the third order with respect to equilibrium. This result agrees with its
counterpart in \cite{1}. \\
A last question that can be considered is the following: It is possible to write
(\ref{lc.3}) in a form close to (\ref{13.1bis}), that is
\begin{eqnarray}\label{lc7}
\Delta H = \sum_{p,q,s}^{0 \cdots \infty} \sum_{r \in I_{p}} \frac{1}{p!} \frac{1}{q!}
\frac{1}{r!} \frac{1}{s!} \chi_{p,q,r,s}(\lambda) \mu^s \delta^{(i_1 \cdots i_{p}h_1k_1
\cdots h_qk_qj_1 \cdots j_r)} \mu_{i_1} \cdots \mu_{i_{p}}\mu_{h_1k_1} \cdots
\mu_{h_qk_q} \lambda_{j_1}
\cdots \lambda_{j_r}   + \\
+\omega(\mu_{ab}, \lambda , \lambda_c) \, . \nonumber
\end{eqnarray}
where $\chi_{p,q,r,s}(\lambda)$ are the counterparts of the
$\vartheta_{p,q,r,s}(\lambda)$ and $\omega(\mu_{ab}, \lambda , \lambda_c)$ is the
counterpart of $H^{*0}(\mu_{ab}, \lambda , \lambda_c)$? Obviously, they must also satisfy
the counterparts of eqs. (\ref{13.1biss}), (\ref{13.2}), (\ref{13.3}), (\ref{beta.2}),
(\ref{beta.3}) and (\ref{beta.4}) because in this case we would have all the symmetries
(In fact we obtained the expression for $\frac{\partial \Delta H}{
\partial  \mu } $, just imposing all the symmetries), that is
\begin{eqnarray}\label{lc.8}
\chi_{0,q,r,0}(\lambda)=0  \, .
\end{eqnarray}
\begin{eqnarray}\label{lc.9}
\chi_{0,0,0,s}(\lambda)=0 \quad \mbox{for} \quad s \geq 0 \, .
\end{eqnarray}
\begin{eqnarray}\label{lc.10}
\omega(0_{ab}, \lambda , 0_c) = 0 \, .
\end{eqnarray}
\begin{eqnarray}\label{lc.12}
\chi_{p,q,r,s} = \left\{ \begin{array}{ll}
  \chi_{0,q+\frac{p}{2},r,s+\frac{p}{2}} & \mbox{if $p$ is even} \\
  \chi_{1,q+\frac{p-1}{2},r,s+\frac{p-1}{2}} & \mbox{if $p$ is odd} \, .
\end{array}
\right.
\end{eqnarray}
\begin{eqnarray}\label{lc.13}
\chi_{0,q,r+1,s+1} = \frac{\partial}{\partial \lambda} \chi_{1,q,r,s} \quad , \quad
\chi_{1,q,r+1,s+1} = \frac{\partial}{\partial \lambda} \chi_{0,q+1,r,s+1} \, ,
\end{eqnarray}
\begin{eqnarray}\label{lc.14}
&{}& 0 = (2Q+R+1) \chi_{0,Q,R,s+1} + 2 \lambda \chi_{0,Q+1,R,s+1} + 2 R
\chi_{1,Q+1,R-1,s}  \, , \\
&{}& 0 = (2Q+R+2) \chi_{1,Q,R,s+1} + 2 \lambda \chi_{1,Q+1,R,s+1} + 2 R
\chi_{0,Q+2,R-1,s+1}  \, , \nonumber
\end{eqnarray}

\appendix

\section{Appendix 1: The particular solution $H=H_1$.}
Let us prove that $H=H_1$, with $H_1$ given by (\ref{11.6}) and $\psi_n$ constrained by
(\ref{11.5}), is a particular solution of (\ref{9.1}) and (\ref{9.3}). \\
In fact, by substituting (\ref{11.6}) in $(\ref{9.1})_1$, we obtain \\
\begin{eqnarray*}
\frac{\partial^{r+p+1}}{\partial \lambda^r
\partial \mu^{p+1}} \left[ \left( \frac{-1}{2 \lambda} \right)^{q+1+\frac{p+r}{2}}
\psi_{\frac{p+r}{2}}\right] = \frac{\partial^{r+p+2}}{\partial \lambda^r
\partial \mu^{p+2}} \left[ \left( \frac{-1}{2 \lambda} \right)^{q+\frac{p+2+r}{2}}
\psi_{\frac{p+2+r}{2}}\right]
\end{eqnarray*}
which surely holds because $\psi_{\frac{p+r}{2}}= \frac{\partial}{\partial \mu}
\psi_{\frac{p+r}{2}+1}$, thanks to (\ref{11.5}).  \\
By substituting (\ref{11.6}) in $(\ref{9.1})_2$, we obtain \\
\begin{eqnarray*}
\frac{\partial^{r+p+2}}{\partial \lambda^{r+1}
\partial \mu^{p+1}} \left[ \left( \frac{-1}{2 \lambda} \right)^{q+\frac{p+r+1}{2}}
\psi_{\frac{p+r+1}{2}}\right] =\frac{\partial^{r+p+2}}{\partial \lambda^{r+1}
\partial \mu^{p+1}} \left[ \left( \frac{-1}{2 \lambda} \right)^{q+\frac{p+1+r}{2}}
\psi_{\frac{p+1+r+}{2}}\right]
\end{eqnarray*}
which is an evident identity. \\
It is more delicate to verify (\ref{9.3}). To do it, let us substitute (\ref{9.3}) with
its derivatives with respect to $\mu_{i_1}$, $\cdots$ , $\mu_{i_P}$, $\mu_{h_1k_1}$,
$\cdots$ , $\mu_{h_Qk_Q}$, $\lambda_{j_1}$, $\cdots$ , $\lambda_{j_R}$; let us substitute
(\ref{11.6}) in the resulting equation and let us calculate the last form at equilibrium.
We obtain
\begin{eqnarray*}
0 = P \delta^{i \overline{i_1}} \delta^{(\overline{i_2 \cdots i_{P}}kh_1k_1 \cdots
h_Qk_Qj_1 \cdots j_R)} \frac{(P+2Q+R+1)!!}{P+2Q+R+1} \frac{\partial^{R+P+1}}{\partial
\lambda^R
\partial \mu^{P+1}} \left[ \left( \frac{-1}{2 \lambda} \right)^{Q+\frac{P+R}{2}}
\psi_{\frac{P+R}{2}}\right] + \\
+2 Q  \delta^{i \overline{h_1}} \delta^{(\overline{k_1 h_2k_2 \cdots h_Qk_Q}ki_1 \cdots
i_{P}j_1 \cdots j_R)} \frac{(P+2Q+R+1)!!}{P+2Q+R+1} \frac{\partial^{R+P+1}}{\partial
\lambda^R
\partial \mu^{P+1}} \left[ \left( \frac{-1}{2 \lambda} \right)^{Q+\frac{P+R}{2}}
\psi_{\frac{P+R}{2}}\right] + \\
+ 2 \lambda   \delta^{(ki h_1k_1 \cdots h_Qk_Q i_{1} \cdots i_{P}j_1 \cdots j_R)}
(P+2Q+R+1)!! \frac{\partial^{R+P+1}}{\partial \lambda^R
\partial \mu^{P+1}} \left[ \left( \frac{-1}{2 \lambda} \right)^{Q+1+\frac{P+R}{2}}
\psi_{\frac{P+R}{2}}\right] + \\
+ 2 R   \delta^{(ki h_1k_1 \cdots h_Qk_Q i_{1} \cdots i_{P}j_1 \cdots j_R)} (P+2Q+R+1)!!
\frac{\partial^{R+P}}{\partial \lambda^{R-1}
\partial \mu^{P+1}} \left[ \left( \frac{-1}{2 \lambda} \right)^{Q+1+\frac{P+R}{2}}
\psi_{\frac{P+R}{2}}\right] + \\
+R \delta^{i \overline{j_1}} \delta^{(\overline{j_2 \cdots j_{R}}kh_1k_1 \cdots
h_Qk_Qi_{1} \cdots i_{P})} \frac{(P+2Q+R+1)!!}{P+2Q+R+1} \frac{\partial^{R+P+1}}{\partial
\lambda^R
\partial \mu^{P+1}} \left[ \left( \frac{-1}{2 \lambda} \right)^{Q+\frac{P+R}{2}}
\psi_{\frac{P+R}{2}}\right] + \\
+\delta^{ki} \delta^{(i_{1} \cdots i_{P} h_1k_1 \cdots h_Qk_Q j_1 \cdots j_R)}
\frac{(P+2Q+R+1)!!}{P+2Q+R+1} \frac{\partial^{R+P+1}}{\partial \lambda^R
\partial \mu^{P+1}} \left[ \left( \frac{-1}{2 \lambda} \right)^{Q+\frac{P+R}{2}}
\psi_{\frac{P+R}{2}}\right] \, ,
\end{eqnarray*}
where overlined indexes denote symmetrization over those indexes, after that the other
one (round brackets around indexes) has been taken. \\
Now, the first, second, fifth and sixth term can be put together so that the above
expression becomes
\begin{eqnarray*}
0 = \delta^{i \overline{i_1}} \delta^{(\overline{i_2 \cdots i_{P}kh_1k_1 \cdots h_Qk_Qj_1
\cdots j_R})} (P+2Q+R+1)!! \frac{\partial^{R+P+1}}{\partial \lambda^R
\partial \mu^{P+1}} \left[ \left( \frac{-1}{2 \lambda} \right)^{Q+\frac{P+R}{2}}
\psi_{\frac{P+R}{2}}\right] + \\
+ (P+2Q+R+1)!! \delta^{(ki h_1k_1 \cdots h_Qk_Q i_{1} \cdots i_{P}j_1 \cdots j_R)}
\left\{ 2 \lambda    \frac{\partial^{R+P+1}}{\partial \lambda^R
\partial \mu^{P+1}} \left[ \left( \frac{-1}{2 \lambda} \right)^{Q+1+\frac{P+R}{2}}
\psi_{\frac{P+R}{2}}\right] \right. + \\
+ \left. 2 R \frac{\partial^{R+P}}{\partial \lambda^{R-1}
\partial \mu^{P+1}} \left[ \left( \frac{-1}{2 \lambda} \right)^{Q+1+\frac{P+R}{2}}
\psi_{\frac{P+R}{2}}\right] \right\} \, ,
\end{eqnarray*}
which is satisfied as a consequence of the property \\
$\delta^{i \overline{i_1}} \delta^{(\overline{i_2 \cdots i_{P}kh_1k_1 \cdots h_Qk_Qj_1
\cdots j_R})} = \delta^{(ki h_1k_1 \cdots h_Qk_Q i_{1} \cdots i_{P}j_1 \cdots j_R)}$ and
of the identity
\begin{eqnarray*}
\frac{\partial^{R}}{\partial \lambda^R} \left[ \left( \frac{-1}{2 \lambda}
\right)^{Q+\frac{P+R}{2}} \psi_{\frac{P+R}{2}}\right] = \frac{\partial^{R}}{\partial
\lambda^R} \left[ - 2 \lambda \left( \frac{-1}{2 \lambda} \right)^{Q+1+ \frac{P+R}{2}}
\psi_{\frac{P+R}{2}}\right]  = \\
= - 2 \lambda \frac{\partial^{R}}{\partial \lambda^R} \left[ \left( \frac{-1}{2 \lambda}
\right)^{Q+1+ \frac{P+R}{2}} \psi_{\frac{P+R}{2}}\right] -2R
\frac{\partial^{R-1}}{\partial \lambda^{R-1}} \left[ \left( \frac{-1}{2 \lambda}
\right)^{Q+1+\frac{P+R}{2}} \psi_{\frac{P+R}{2}}\right] \, .
\end{eqnarray*}
This completes the  proof that $H=H_1$ is a particular solution of (\ref{9.1}) and
(\ref{9.3}).

\section{Appendix 2: A further integration in the framework of the initial article.}

A further integration is possible for one combination of eqs. (44) of the paper \cite{1}. \\

In fact, the integrability condition on $(44)_{1,4}$ of that paper allows us to obtain
\begin{equation}\label{eq1}
  \frac{\partial}{\partial \, \rho}\,  h_4 = - 2 T^2 \frac{\partial \varepsilon }{\partial \,
  T}\frac{\partial p}{\partial \, \rho} - 2 \frac{T^3}{\rho^2} \left( \frac{\partial p}{\partial \,
  T}\right)^2 = -2 \frac{\partial \varepsilon }{\partial \, T} \left( 2 \frac{T}{\rho} h_2 +
  \frac{5T^2p}{3 \rho} \right) \, .
\end{equation}
(Here and in the sequel, we use the notation of \cite{1}. For example, the scalars
$\beta_2$, $\beta_3$ are different from the constants with the same name the present
paper). \\
After that, by using $(44)_{2,3}$ and the present eq. (\ref{eq1}), we obtain
\begin{equation}\label{eq2}
  \frac{\partial}{\partial \, \rho} \left[ \beta_2 - \frac{5}{6} \beta_3 - \left(4 h_2+
  \frac{10}{3}pT \right) \left( \varepsilon + \frac{p}{\rho} \right) \right] =0 \, .
\end{equation}
Consequently, $ \beta_2 - \frac{5}{6} \beta_3 - \left(4 h_2+
  \frac{10}{3}pT \right) \left( \varepsilon + \frac{p}{\rho} \right)$ may depend only on
  temperature.      \\
Similarly, from $(44)_{5,6}$ and the present eq. (\ref{eq1}), we obtain
\begin{equation}\label{eq3}
  \frac{\partial}{\partial \, T} \left[ \beta_2 - \frac{5}{6} \beta_3 - \left(4 h_2+
  \frac{10}{3}pT \right) \left( \varepsilon + \frac{p}{\rho} \right) \right] = - \frac{1}{T}
  \left[ \beta_2 - \frac{5}{6} \beta_3 - \left(4 h_2+
  \frac{10}{3}pT \right) \left( \varepsilon + \frac{p}{\rho} \right) \right] \, ,
\end{equation}
which is a differential equation for the unknown  $ \beta_2 - \frac{5}{6} \beta_3 -
\left(4 h_2+
  \frac{10}{3}pT \right) \left( \varepsilon + \frac{p}{\rho} \right)$ whose solution is
\begin{equation}\label{eq4}
   \beta_2 - \frac{5}{6} \beta_3 - \left(4 h_2+
  \frac{10}{3}pT \right) \left( \varepsilon + \frac{p}{\rho} \right)  = \frac{constant}{T} \, .
\end{equation}
Consequently, we have the symmetry of $M_{ijk}$ at first order, if and only if this
constant arising from integration is zero! On the other hand $m_{ppik}$ at first order is
already symmetric; eventual its skew-symmetric parts may appear at higher orders with
respect to equilibrium.

\end{document}